\documentclass[12pt]{article}
\usepackage{amsfonts,amsthm}
\usepackage{amsmath,amssymb}
\usepackage{cite}
\usepackage{mathrsfs}
\usepackage{dsfont}
\usepackage[title]{appendix}
\usepackage{axodraw2}
\usepackage{caption}
\usepackage{graphicx}
\usepackage{comment}
\usepackage{slashed}
\usepackage{subcaption}
\usepackage[top= 3.3cm, bottom= 3.3cm, left = 3.5cm, right= 3.5cm]{geometry} 
\usepackage[hypertexnames=false,pdfencoding=auto]{hyperref}
\usepackage[all]{hypcap}        

\makeatletter
\g@addto@macro\bfseries{\boldmath}
\makeatother	

\let\Re\relax
\let\Im\relax
\DeclareMathOperator{\Re}{Re}
\DeclareMathOperator{\Im}{Im}
\DeclareMathOperator{\sgn}{sgn}
\DeclareMathOperator{\Tr}{Tr}
\newenvironment{Eqnarray}%
     {\arraycolsep 0.14em\begin{eqnarray}}{\end{eqnarray}}
\newcommand\beq{\begin{equation}}
\newcommand\eeq{\end{equation}}
\newcommand\beqa{\begin{Eqnarray}}
\newcommand\eeqa{\end{Eqnarray}}

\newcommand\Tstrut{\rule{0pt}{2.9ex}}         
\newcommand\Bstrut{\rule[-1.2ex]{0pt}{0pt}}   
\newcommand\TBstrut{\Tstrut\Bstrut}

\newcommand{\overbar}[1]{\mkern 1.5mu\overline{\mkern-2.5mu#1\mkern-0.5mu}\mkern 1.5mu}
\newcommand{\MHp}{M_{H^\pm}}

\def\<{\left\langle}
\def\>{\right\rangle}

\newcommand{\bt}{\begin{tabular}}
\newcommand{\et}{\end{tabular}}

\def\Tr{{\rm Tr}}
\def\abar{{\bar a}}
\def\bbar{{\bar b}}

\def\lam{\lambda}

\def\ifmath#1{\relax\ifmmode #1\else $#1$\fi}
\def\lsim{\mathrel{\raise.3ex\hbox{$<$\kern-.75em\lower1ex\hbox{$\sim$}}}}
\def\gsim{\mathrel{\raise.3ex\hbox{$>$\kern-.75em\lower1ex\hbox{$\sim$}}}}

\def\eq#1{eq.~(\ref{#1})}
\def\eqs#1#2{eqs.~(\ref{#1}) and (\ref{#2})}

\def\eqst#1#2{eqs.~(\ref{#1})--(\ref{#2})}

\def\Eq#1{Eq.~(\ref{#1})}

\def\mud{M_U}
\def\mdd{M_D}
\def\vev#1{\langle #1 \rangle}

\def\cbma{c_{\beta-\alpha}}

\def\sbma{s_{\beta-\alpha}}

\def\pht{\phantom{i}}
\def\phm{\phantom{-}}
\def\phaa{\phantom{AA}}

\def\beq{\begin{equation}}
\def\eeq{\end{equation}}
\def\ifmath#1{\relax\ifmmode #1\else $#1$\fi}

\def\zv{\overbar{Z}_5}
\def\zvi{\overbar{Z}_6}
\def\zvii{\overbar{Z}_7}

\def\sbmaii{s^2_{\beta-\alpha}}
\def\cbmaii{c^2_{\beta-\alpha}}
\def\sbmaiii{s^3_{\beta-\alpha}}
\def\cbmaiii{c^3_{\beta-\alpha}}

\def\mw{m_W}

\def\ls#1{\ifmath{_{\lower1.5pt\hbox{$\scriptstyle #1$}}}}
\def\lss#1{\ifmath{^{\,\lower2.5pt\hbox{$\scriptstyle #1$}}}}
\def\lsup#1{^{\lower 6pt\hbox{$\scriptstyle#1$}}}
\def\llsup#1{^{\lower 3pt\hbox{$\scriptstyle#1$}}}
\def\lasup#1{^{\lower 2pt\hbox{$\scriptstyle#1$}}}

\def\half{\ifmath{{\textstyle{\frac{1}{2}}}}}

\def\quarter{\ifmath{{\textstyle{\frac{1}{4}}}}}

\def\T{{\mathsf T}}

\def\ddel{\!\!\mathrel{\raise1.5ex\hbox{$\leftrightarrow$\kern-.85em
\lower1.7ex\hbox{$\partial$}}}}

\newcommand{\nn}{\nonumber}
\def\thefootnote{\fnsymbol{footnote}}

\allowdisplaybreaks[2]
\begin{document}

\begin{flushright}
\normalsize{
\phantom{x}\\[-50pt]
DIAS-STP-22-02 \\
SCIPP-22/01 
}
\end{flushright}
\vspace{0.5cm}

\begin{center}
\LARGE
P-even, CP-violating Signals in Scalar-Mediated Processes
\end{center}
\vspace{0.4cm}
\begin{center}

Howard E.~Haber,$^{1,}$\footnote{E-mail: {\tt haber@scipp.ucsc.edu}} 
Venus~Keus,$^{2,}$\footnote{E-mail: {\tt venus@stp.dias.ie}}  
Rui~Santos,$^{3,4,}$\footnote{E-mail: {\tt rasantos@fc.ul.pt}}\\

\vspace{0.4cm}
{\sl $^1$ Santa Cruz Institute for Particle Physics \\
 University of California, Santa Cruz, CA 95064 USA}\\[0.2cm]
\vspace{0.2cm}
{\sl $^2$ Dublin Institute for Advanced Studies, School of Theoretical Physics \\
10 Burlington road, Dublin, D04 C932, Ireland}\\[0.2cm]
\vspace{0.2cm}
{\sl $^3$ Centro de F\'{\i}sica Te\'{o}rica e Computacional, Faculdade de Ci\^{e}ncias\\
Universidade de Lisboa, Campo Grande, Edif\'{\i}cio C8, 1749-016 Lisboa, Portugal}\\[0.2cm]
\vspace{0.2cm}
{\sl $^4$ ISEL -  Instituto Superior de Engenharia de Lisboa\\
  Instituto Polit\'ecnico de Lisboa  1959-007 Lisboa, Portugal}\\[0.2cm]
\end{center}
\vspace{0.2cm}

\renewcommand{\thefootnote}{\arabic{footnote}}
\setcounter{footnote}{0}

\begin{abstract}
Most studies of Higgs sector CP violation focus on the detection of CP-violating neutral Higgs-fermion Yukawa couplings, which yield P-odd, CP-violating phenomena.  There is some literature on purely bosonic signatures of Higgs sector CP violation, where the simultaneous observation of three processes (suitably chosen) constitutes a signal of P-even CP violation.  However, in the examples previously analyzed, some of the processes are strongly suppressed in the approximate Higgs alignment limit (corresponding to the existence of a Standard Model like Higgs boson as suggested by LHC data), in which case the proposed CP-violating signals are difficult to observe in practice.    In this paper, we extend the existing literature by examining processes that do not vanish in the Higgs alignment limit and whose simultaneous observation would provide unambiguous evidence for scalar-mediated P-even CP violation.  We assess the discovery potential of such signals at various future multi-TeV lepton (and $\gamma\gamma$) colliders.  The potential for detecting loop-induced P-even, CP-violating phenomena is also considered.

\end{abstract}
\newpage

\setcounter{page}{1}

\section{Introduction}
\label{intro}

The Higgs boson of the Standard Model (SM) is a CP-even scalar.   Strictly speaking, this statement is only approximately true, since the SM Lagrangian is not CP-conserving due to the presence of an unremovable complex phase in the Cabibbo-Kobayashi-Maskawa (CKM) matrix.   This phase will generate CP-violating observables in processes involving the Higgs boson via radiative corrections.   However, in practice any such contributions are extremely tiny and can be neglected.

There have been a number of theoretical motivations advanced in the literature that suggest the existence of additional scalar degrees of freedom beyond the Higgs boson of the SM.  However, even independently of such motivations, it is striking to observe that the fermion and gauge sectors of the SM are nonminimal.   Thus, if the scalar sector of the SM follows a similar pattern, then one should expect the existence of additional Higgs scalars in nature.   Perhaps some of these scalars will possess masses that are not significantly larger than the observed Higgs boson, in which case they may be discoverable at the LHC or at future hadron and/or lepton colliders now under consideration.

Current LHC data already place nontrivial restrictions on the structure of an extended Higgs sector.   First, the observation that the electroweak $\rho$ parameter is very close to 1 suggests that any extended Higgs sector is most likely composed of hypercharge $Y=\pm 1$ doublet fields (and perhaps singlet fields) with respect to the electroweak gauge group.\footnote{We have chosen a convention, such that $Q=T_3+\half Y$ where $Q$ is the electric charge, $T_3$ is the third component of the weak isospin, and $Y$ is the hypercharge.  Note that if other scalar multiplets are included, then (in most cases) the Higgs sector parameters must be fine tuned to achieve $\rho\simeq 1$, in contrast to extended Higgs sectors consisting of $Y=0$ singlet and $Y=\pm 1$ doublet scalar fields where the tree-level value of the $\rho$-parameter is automatically equal to 1. For further details and references to the original literature, see Ref.~\cite{Gunion:1989we}.}   Second, the LHC Higgs data have already achieved a precision that implies that the properties of the observed Higgs boson closely approximate those of the SM Higgs boson 
to within an accuracy that is typically in the range of $10\%$--$20\%$ depending on the observable~\cite{ATLAS:2022vkf,CMS:2022dwd}. 
The existence of a SM-like Higgs boson in the scalar spectrum implies that the Higgs sector is close to the so-called Higgs alignment limit~\cite{Ginzburg:2001wj,Gunion:2002zf,Craig:2013hca,Asner:2013psa,Carena:2013ooa,Haber:2013mia}.

Consider an extended Higgs sector with $n$ hypercharge-one SU(2) doublet scalar fields~$\Phi_i$ and $m$ additional hypercharge-zero singlet scalar fields $\phi_i$.
After minimizing the scalar potential, we assume that only the neutral
scalar fields acquire nonzero vacuum expectation values (vevs) thereby preserving U(1)$_{\rm EM}$, with
$\vev{\Phi_i^0}=v_i/\sqrt{2}$ and $\vev{\phi_j^0}=x_j$,
where $v^2\equiv \sum_i|v_i|^2=(\sqrt{2}\,G_F)^{-1}\simeq (246~{\rm GeV})^2$ and $G_F$ is the Fermi constant of weak interactions.
One can then introduce new linear combinations, $H_i$,  
which define the so-called \textit{Higgs basis}~\cite{Georgi:1978ri,Lavoura:1994yu,Lavoura:1994fv,Botella:1994cs,Branco:1999fs,Davidson:2005cw}.
In particular,
\beq
H_1=\begin{pmatrix} H_1^+ \\ H_1^0\end{pmatrix}=\frac{1}{v}\sum_i v_i^*\Phi_i\,,\qquad\quad
\vev{H_1^0}=v/\sqrt{2}\,,
\eeq
and $H_2, H_3,\ldots, H_n$ are the other mutually orthogonal linear combinations of doublet scalar fields such that $\vev{H_i^0}=0$ (for $i\neq 1$).
That is $H_1^0$ is \textit{aligned} in field space with the direction of the scalar field vev.  
In the exact Higgs alignment limit, 
\beq
\varphi\equiv\sqrt{2}\,\Re H_1^0-v
\eeq
 is a scalar mass eigenstate such that the tree-level couplings of $\varphi$ to itself, to gauge bosons and to fermions coincide with those of the SM Higgs boson.
However, generically $\varphi$ is \textit{not} a scalar mass eigenstate due to its mixing with other neutral scalars.
An approximate Higgs alignment limit, in which the Higgs sector contains a SM-like neutral Higgs scalar, is achieved 
if at least one of the following two conditions are satisfied:
(i) the diagonal squared masses of $H_2, H_3,\ldots, H_n$ are all large compared to the squared mass of the observed Higgs boson (corresponding to the \textit{decoupling limit}~\cite{Haber:1989xc,Gunion:2002zf}), and/or (ii) the elements of the neutral scalar squared mass matrix that govern the mixing of $\varphi$ with the other neutral scalar fields are suppressed.

The Higgs alignment limit is most naturally achieved in the decoupling regime, where the observed Higgs boson [whose mass is of $\mathcal{O}(v)$] is significantly lighter than all additional scalars of the Higgs sector.  That is, a new mass parameter, $M\gg v$, exists 
that characterizes the mass scale of the additional scalars.
After integrating out all the heavy degrees of freedom at the mass scale $M$, one is left with a low-energy effective theory that consists of the SM particles, including a single neutral scalar boson, which is identified with the SM Higgs boson.  
Alternatively, one can achieve approximate Higgs alignment if the mixing of $\varphi$ with the other neutral scalars of the extended Higgs sector is suppressed.   In this case, it is possible that additional scalar states with masses of $\mathcal{O}(v)$ are present in the scalar spectrum beyond the scalar state currently identified as the SM-like Higgs boson.  This is the case of Higgs alignment without decoupling.   It is typically achieved by fune tuning the parameters of the extended Higgs sector.  However, in some special cases, exact Higgs alignment without decoupling can be implemented due to the presence of a symmetry~\cite{Dev:2014yca, Dev:2017org,Benakli:2018vqz,Lane:2018ycs,Benakli:2018vjk,Darvishi:2020teg,Draper:2020tyq,Eichten:2021qbm}.
For example, certain exact discrete symmetries can be imposed such that the mixing of the SM Higgs doublet with additional (so-called inert) scalar doublets is forbidden~\cite{Barbieri:2006dq,LopezHonorez:2006gr,Grzadkowski:2010au,Ivanov:2012hc,Keus:2014jha,Keus:2014isa,Cordero-Cid:2016krd,Azevedo:2018fmj,Aranda:2019vda,Khater:2021wcx}.
In such models, the tree-level properties of the 125 GeV Higgs boson coincide with those of the SM Higgs boson, corresponding to the exact Higgs alignment limit.  

At present, no additional Higgs-like
scalar states beyond the SM-like Higgs boson have been found at the LHC. Furthermore, the properties of the observed Higgs boson approximate those of the SM Higgs boson as noted above.
Consequently, if additional Higgs bosons exist then the Higgs alignment limit of the extended Higgs sector must be approximately realized. In this scenario, the detection of new neutral 
scalars\footnote{In our notation, $h_1\equiv h_1(125)$ is the observed Higgs boson of mass 125 GeV and the $h_i$ for $i\neq 1$ correspond to new neutral scalars of the extended Higgs sector.\label{fnh1}}
via couplings to gauge bosons is difficult due to the sum rule~\cite{Gunion:1990kf}, 
\beq \label{coupsumrule}
\sum_i \lambda_{V V h_i}^2  = \lambda_{VV h_{\rm SM}}^2\,,   \quad \text{for $VV=W^+W^-$ or $ZZ$},
\eeq 
that is satisfied in any extended Higgs sector with $\rho=1$ and no $ZW^\pm H_i^\mp$ couplings at tree level.\footnote{These conditions are automatically realized in extended Higgs sectors made up exclusively of $Y=0$ singlet and $Y=\pm 1$ doublet scalars.}   In particular, \eq{coupsumrule}
forces the couplings of $h_i$ (for $i>1$) to gauge bosons to be very small when $h_1\simeq h_{\rm SM}$ is SM-like.

In models with extended Higgs sectors, there is a potential for new sources of CP violation arising in the scalar sector.  Thus, if new scalar states are discovered at the LHC, it will be important to search for new signals of scalar-mediated CP-violating phenomena.  In the approximate Higgs alignment limit, the observed Higgs boson will continue to behave as an approximate CP-even scalar.   However, possible CP-violating phenomena associated with the additional scalar states of the Higgs sector can be present and may be observable.  In this paper, we shall discuss various possible CP-violating observables associated with the extended Higgs sector.
 
Here, we provide two examples of CP-violating observables associated with the scalar sector.  Consider the process $\gamma\gamma\to H$ mediated by a fermion loop.   If the scalar state $H$ is not an eigenstate of CP, then the $Hf\bar{f}$ vertex will be of the form
\beq \label{hffb}
H\bar{f}(a+ib\gamma_5)f\,,
\eeq
where $a$ and $b$ are real (as a consequence of hermiticity of the effective interaction).  
The one-loop diagram for $H\gamma\gamma$ then yields an effective interaction
\beq \label{hFF}
\mathscr{L}_{\rm int}=H(a'FF+b'F\widetilde{F})\,,
\eeq
where $a'$ and $b'$ are real, $F$ is the photon field strength tensor and $\widetilde{F}$ is the dual field strength tensor.  In particular, if
$ab\neq 0$ then $a'b'\neq 0$, in which case the $H\gamma\gamma$ interaction is CP-violating and $H$ is a scalar of indefinite CP quantum number.
In this case, CP violation can be experimentally detected if the initial photon polarizations can be specified.
For example, polarization asymmetries are defined in Ref.~\cite{Grzadkowski:1992sa} that would provide an unambiguous signal of CP violation in the production of $H$.  Such a process could be studied in a $\gamma\gamma$ collider mode of a future $e^+e^-$ linear collider (using Compton backscattered laser beams).  The initial photon polarizations are determined by the polarization of the initial electron and positron beams.

A second example makes direct use of the Higgs-fermion coupling given in \eq{hffb}.  For example, the energy distribution of the $\tau$ decay products in $H\to\tau^+\tau^-$ 
are correlated with the $\tau$ polarization.   The end result is a nontrivial azimuthal angle correlation of the final state decay products that provides the CP-violating signal~\cite{Grzadkowski:1995rx,Bernreuther:1997af,Berge:2011ij,Berge:2014sra,Cordero-Cid:2020yba}.
Finally, CP violation can also be probed in $\bar t t H$ production. Although the scalar and pseudoscalar amplitudes do not interfere, the cross section can be written as a function of $a^2+b^2$ and $a^2-b^2$, the latter being proportional
to the top quark mass. In order to optimize the extraction of the $(a^2-b^2)$ term, many variables have been proposed over the years; see e.g.~Refs.~\cite{Gunion:1996xu, Boudjema:2015nda, Mileo:2016mxg, AmorDosSantos:2017ayi,Goncalves:2018agy}.
 
The ATLAS and CMS collaborations have performed various studies to probe the CP nature of the 125~GeV Higgs boson in its couplings to top quarks and to $\tau$ leptons. 
Both experimental collaborations~\cite{Sirunyan:2020sum, Aad:2020ivc}  employ the two photons decay channel $H\rightarrow \gamma\gamma$ in the production process $pp\rightarrow t\bar{t}H$, and attempt to measure a CP-violating mixing angle defined as $\theta \equiv \arctan (b/a)$,
where $a$ and $b$ are given in \eq{hffb}.
The purely CP-odd hypothesis is excluded at the level of 3.9 standard deviations, and an observed (expected) exclusion upper limit at 95\% CL was obtained for the mixing angle of $\theta=43^{\circ}$~(63$^{\circ}$).
Using data collected at $\sqrt{s} =$ 13~TeV (with an integrated luminosity of 137~fb$^{-1}$), the CMS Collaboration~\cite{CMS:2021sdq} has recently measured the CP mixing angle of the tau lepton, 
$\theta=4^{\circ}$ $\pm$ 17$^{\circ}$, while setting an observed (expected) exclusion upper limit of $36^{\circ}$~(55$^{\circ}$).

In both of the examples cited above, the actual experimental observable violates P and CP while preserving C.
That is, in the case of \eq{hffb}, $\bar{f}f$ and $i\bar{f}\gamma_5 f$ are C-even bilinear covariants.  Hence, if one identifies $H$ as a C-even state then the $H\bar{f}f$ vertex is C-conserving.  Likewise, the photon is a C-odd state so that both $FF$ and $F\widetilde F$ are C-even operators, which implies that
the effective interaction given by \eq{hFF} is C-even.    Of course, C is not a good quantum number of the SM; it is 
violated by $W$ and $Z$ mediated interactions.  However, in the context of the examples presented above, the effects of the C-violating $W$ and $Z$ mediated interactions only appear via radiative corrections.  Thus, in first approximation, we may treat \eq{hFF} as a C-conserving interaction.

The examples above, where the CP-violating interactions of the scalars are attributed to approximately C-conserving, P-violating interactions, originate from the structure of the Higgs-fermion couplings.   However, there is an alternative possibility in which the CP-violating interactions of the scalars are attributed to approximately C-violating, P-conserving interactions that originate from the structure of the bosonic couplings of the Higgs bosons (these include the Higgs couplings to vector bosons and the Higgs boson self-couplings).  

In this paper,  we shall focus on this second class of CP-violating signals in the scalar sector that arise from P-even, CP-violating interactions.  
One example of this phenomena, which is well studied in the literature~\cite{Mendez:1991gp,Cvetic:1992wa,Fontes:2015xva,Keus:2015hva,LHCHiggsCrossSectionWorkingGroup:2016ypw}, involves the coupling of the $Z$ boson to a pair of neutral scalars $h_i$ via the $Zh_i h_j$ vertex ($i< j$). Since the pair of scalars that couples to a spin-one boson has 
relative orbital angular momentum equal to one, it follows that $h_i h_j$ 
must be a CP-odd state.   In the two Higgs doublet model (2HDM) where
$i=1,2,3$, if all three possible combinations of $Zh_i h_j$ for $i< j$ are observed, then CP must necessarily be violated. Experimentally, these couplings can be probed by either observing  
the three decays $h_3 \to h_2 Z$, $h_3 \to h_1 Z$ and $h_2 \to h_1 Z$ or the production processes $Z^*\to h_3 h_2$, $Z^*\to h_3 h_1$ and $Z^*\to h_2 h_1$ via $s$-channel $Z$ boson exchange.  
As shown in Refs.~\cite{Fontes:2015xva, LHCHiggsCrossSectionWorkingGroup:2016ypw}, the three scalar decays, if kinematically allowed, would be visible at future LHC runs in a significant portion of the parameter space. In contrast, the identification of the production of scalar pairs via $s$-channel $Z$ exchange is significantly more difficult at a hadron collider because 
other production mechanisms, such as gluon initiated processes via top and bottom quark loops
tend to dominate~\cite{Abouabid:2021yvw}.  
 
Moreover, the viability of the P-even, CP-violating signal discussed above
depends on appreciable $Zh_i h_j$ couplings for all possible $i<j$.  However, since
$h_1$ is identified as the observed SM-like Higgs boson, the $Zh_1 h_j$ ($j\neq 1$) couplings all vanish
in the exact Higgs alignment limit, and the CP-violating signal is completely lost.  In the approximate Higgs alignment limit, the suppression of the corresponding $Zh_1 h_j$ ($j\neq 1$) couplings limits the usefulness of the P-even, CP-violating signal.   Thus, one must either rely on a set of $Z h_i h_j$ couplings where $i\neq 1$ and $j\neq 1$ (which would require an extended Higgs sector beyond two doublets) or else search for alternative CP-violating signals that are not suppressed in the Higgs alignment limit.

Any search for scalar-mediated P-even, CP-violating  phenomena can be contaminated by fermionic P-violating contributions in the production process.
If no evidence is found for all three vertices $Zh_i h_j$ ($i< j$) by end of the high-luminosity stage of the LHC, then one may conclude that at least one (or more) of the following statements must hold:
(i) there are no new scalars beyond the SM Higgs boson that are kinematically accessible at the LHC;  (ii) some or all of the bosonic decays of $h_2$ and $h_3$ are not kinematically allowed; and/or
(iii) the departure from the Higgs alignment limit deduced from the precision Higgs data is so severe that the suppression of the three vertices precludes their observation at the LHC.

For a robust interpretation of the signal for P-even CP violation, one should ensure that there is no contamination of the P-even, CP-violating signal due to effects from the neutral Higgs-fermion interactions, either via the production process or from competing contributions to the bosonic vertices.
In this paper, we will eliminate possible contamination due to CP-violating neutral Higgs-fermion Yukawa couplings
by focusing on the production of Higgs bosons at a high energy lepton collider, with a focus on the bosonic Higgs vertices that arise at tree-level and are unsuppressed in the Higgs alignment limit.

The assertion above that the existence of three couplings of the form $Zh_i h_j$ ($i< j$) can be interpreted as a P-even, CP-violating observable requires some elucidation.  In Section~\ref{section2}, we 
demonstrate that in a CP-conserving gauge theory of scalars and gauge bosons where the fermions are excluded, C and P are separately conserved.  In this case we are free to set $P=+1$ for \textit{all} scalars, in which case the presence or absence of the $Zh_i h_j$ coupling is determined completely by the C properties of the Higgs bosons and gauge bosons.   
As an example, it is instructive to employ the 2HDM with the fermions omitted since CP-violating phenomena arising in this theory are associated with 
P-even, CP-odd observables.

In Section \ref{section3}, we examine a class of P-conserving C-violating processes involving scalars of an extended Higgs sector that can be used to identify the presence of scalar-mediated CP violation.   In this analysis, we initially focus on processes that are not suppressed in the Higgs alignment limit.  We then identify both tree-level and loop-induced processes that can provide evidence for CP violation.   In any realistic application, loop-induced processes may include ``fermion pollution''---that is, contributions from fermion loops that can complicate the interpretation of the CP-violating signal.  Thus, we focus primarily on tree-level purely bosonic P-odd, CP-violating phenomena, where the effects of the Yukawa interactions can only enter via small radiative corrections.  

In Section \ref{sec:collider}, we study the discovery potential of our proposed observables at future lepton (both $e^+e^-$ and $\mu^+\mu^-$) and photon colliders.  We calculate the relevant cross sections and estimate the total number of signal events for each observable for a given set
of Higgs sector parameters. Future lepton colliders with a center of mass energy of a few TeV are better suited for the $s$-channel processes, whereas two photon processes only become relevant at energies above 10 TeV.  

An indirect way to detect the presence of P-even CP violation is to probe the loop contributions to  the triple $Z$ form factor~\cite{Grzadkowski:2016lpv}.
CP-violating contributions to the triple gauge boson vertices $ZZZ$ and $ZW^+W^-$ vertices were first studied 
in Refs.~\cite{Hagiwara:1986vm,Gounaris:1999kf,Gounaris:2000dn,Baur:2000ae}. 
The Lorentz structure of the general $ZZZ$ and $ZW^+ W^-$ vertices contains seven independent form factors, one of which (denoted by $f_4$)
measures the P-even, CP-violating contribution to the vertex.  The possible contribution of the extended Higgs sector to $f_4$ is discussed in  
Section~\ref{sec:ZZZ}.  Loop processes are also relevant in
Section \ref{sec:photons}, where we examine P-even, CP-violating observables with multiple photons and/or $Z$ bosons in the final state (with some details relegated to Appendix~\ref{3gamma}).  Unfortunately, such signals will be very difficult to nearly impossible to detect in any future experimental program.

Although our examples are presented in the context of the Higgs alignment limit, in some cases one can tolerate some suppression factors that arise if the Higgs alignment limit is only approximately realized.
 In Section \ref{sec:beyond}, we classify additional signals of P-even, CP-odd observables that can be employed if deviations from the exact Higgs alignment limit are taken into account.
In Section \ref{sec:conclusion}, we summarize our findings and indicate some possible future directions.  The 2HDM formalism employed in this paper is reviewed and summarized in Appendix~\ref{2hdm}. 

\section{C and P symmetries in a gauge theory of spin-0 and spin-1 fields}
\label{section2}

Consider a theory of scalar fields and gauge fields.  The scalar fields transform locally and nontrivially under a gauge group $G$.  In our analysis, we assume that the gauge theory Lagrangian is renormalizable (i.e., terms of dimension five or greater are excluded).   We shall allow for all possible terms in the Lagrangian of dimension four or less, consistent with the gauge symmetry.   

It is well known that if the scalar self-interactions are turned off, then all kinetic energy terms (where derivatives are replaced by covariant derivatives) and mass terms of the Lagrangian separately conserve C, P and T. 
When scalar self-interactions are included, CP-violating interaction terms can arise.\footnote{One further potential source of CP violation can arise via a topological term of the form $\theta F\widetilde{F}$, where $F$ is the field strength tensor of the gauge field.  However, this is a P-odd, CP-odd term, which is not the focus of this work.\label{fntop}}

Moreover, we assert that even in the presence of scalar self-interactions, the theory automatically conserves parity.  That is, in all scattering processes, any parity-violating observable must vanish. To prove this statement, note that one can consistently choose the P quantum number of $+1$ for all scalar fields and $-1$ for all gauge fields.\footnote{The choice for $-1$ for the P quantum number of gauge fields is based on classical arguments.   The interaction Lagrangian of the gauge field $A_\mu$ is of the form $\mathscr{L}= -j^\mu A_\mu$, and the behavior of the current $j^\mu=\rho u^\mu$ is fixed by parity properties of the four-velocity vector $u^\mu$.}  
One can check that P will be conserved in all interactions.  For example, in a vertex with one gauge boson and two scalars, we see that parity is conserved after taking into account the $(-1)^\ell$ factor, where $\ell=1$ is the orbital angular momentum involved in the transition from a spin-1 particle to a pair of spin-0 particles.
If the theory exhibits CP violation (or equivalently T-violation via the CPT theorem) due to the presence of CP-violating scalar self-interaction terms, then it can be 
interpreted as C-violation.   

As another example that illustrates the absence of P-violation, suppose that a scalar mass eigenstate exists that does not possess a definite CP quantum number (which will arise in a theory of a CP-violating scalar sector).   In this case, consider the decay of a neutral scalar $\phi$ to two gauge bosons $VV$, which can be generated at one loop due to a loop of charged scalars.   It is clear that the effective Lagrangian that governs the $\phi\to VV$ decay will be of the form $\mathscr{L}_{\rm eff}=c \phi FF$, where $c$ is a constant that depends on the relevant couplings.
Note that the effective operator $\phi F\widetilde{F}$ 
is absent, since it is not possible to generate an
$\epsilon_{\mu\nu\alpha\beta}$ at any order of perturbation theory in a renormalizable gauge theory of scalar and gauge fields.  Consequently, there cannot be any parity-violating observable associated with the effective $\phi VV$ interaction.  Thus, we must again conclude that the presence of CP violation should be interpreted as a violation of C.   Indeed, the same argument applies to the analysis of the $ZZZ$ and $ZW^+ W^-$ form factor that is generated at one loop due to scalars circulating in the loop, which is discussed in Section~\ref{sec:ZZZ}.

It is instructive to illustrate these remarks in more detail in a simple setting. 
Thus, we examine a renormalizable theory of two hypercharge-one scalar doublets coupled to an SU(2)$\times$U(1) gauge theory.  This is the 2HDM with the fermions removed.  
The theoretical structure of the 2HDM is reviewed in Appendix~\ref{2hdm}, which also establishes our notation.

Consider first the case of a CP-conserving scalar potential and vacuum.   In this case, the Higgs basis scalar potential parameters [cf.~\eq{hbasisdef}] satisfy,
\beq \label{CPconditions}
\Im(Z_5^* Z_6^2)=\Im(Z_5^* Z_7^2)=\Im(Z_6^* Z_7)=0\,.
\eeq
If \eq{CPconditions} is satisfied, then a real Higgs basis exists---that is, one can choose a basis-dependent phase $\eta$ that appears in the definition of the Higgs basis field $\mathcal{H}_2$ [cf.~\eq{invhiggs}] such that all the coefficients $Z_i$ of the scalar potential in the Higgs basis are real. 

In the CP-conserving 2HDM with the fermions removed,
C, P and T are separately conserved, and we can assign in a consistent manner P quantum numbers to all bosonic fields and C quantum numbers to all neutral bosonic fields.\footnote{This conclusion applies in the case of a generic CP-conserving scalar potential.   However, 
if a $\mathbb{Z}_2$ symmetry is present that is not broken by the vacuum, then one cannot uniquely assign a C quantum number to $H$ and~$A$, although the C quantum numbers of $H$ and $A$ are \textit{relatively} odd.   This behavior is associated with the fact that $H$ and $A$ are odd under the $\mathbb{Z}_2$ symmetry.  In particular, the $\mathbb{Z}_2$ symmetry removes bosonic vertices of the theory that  would otherwise be used to identify the separate C quantum numbers of $H$ and~$A$. \label{fnidm}} 
In the conventional 2HDM notation, $h$ and $H$ are neutral CP-even Higgs bosons, with $m_h<m_H$, the neutral CP-odd Higgs boson is denoted by $A$,
and $H^\pm$ refers to the charged scalar fields.  Working in the $R_\xi$ gauge, one must also include the neutral CP-odd Goldstone field $G$ and the charged Goldstone scalars, $G^\pm$.
The C and P assignments were exhibited first in Ref.~\cite{Gunion:1986nh} and are reproduced in Table~\ref{candp} (see also Ref.~\cite{Cvetic:1992wa}).
The quantum numbers presentefd in Table~\ref{charges} have been obtained by examining all possible gauge invariant bosonic interaction terms of dimension 4 or less.
Finally, as discussed above, all scalars are even under parity transformations.

\begin{table}[t!]
\centering
\begin{tabular}{|c|c|c|}\hline
$\mbox{bosonic field}$ & $J^{\mathrm{PC}}$ & $J^{\mathrm{P}}$\Tstrut \\ [2mm]
\hline 
$\gamma$ & $1^{--}$ & $ $\Tstrut \\[2mm]
$Z$ & $1^{--}$ & $ $ \\[2mm]
$h, H$ & $0^{++}$ & $ $ \\[2mm]
$A, G$ & $0^{+-}$ & $ $ \\ [2mm]
$W^\pm$ & $ $ & $1^{-}$ \\ [2mm]
$H^\pm, G^\pm$ & $ $ & $0^{+}$ \\ [2mm]
\hline
\end{tabular}
\caption{\small $J^{\mathrm{PC}}$ quantum numbers of the Higgs/Goldstone scalars and gauge bosons of the 2HDM (in the absence of fermions) when the scalar potential and vacuum are CP-conserving~\cite{Gunion:1989we,Gunion:1986nh}. \label{candp}}
\label{charges}
\end{table}

We justify these C and P quantum number assignments with the following arguments.   First, the $H^+ H^-$ pair with relative orbital angular momentum $L$ has ${\rm C}={\rm P}=(-1)^L$.   In light of the existence of the $ZH^+ H^-$ coupling, where $L=1$, it follows that one must assign $J^{\mathrm{PC}}=1^{--}$ for the $Z$ boson.
Second, the existence of the $hhh_i$ vertex ($h_i=h, H$) implies the $0^{++}$ assignment for $h, H$.   Third, the existence of the $Zh_i a_j$ vertex ($a_j=A, G$) justifies the assignment of $0^{+-}$ for $A, G$.\footnote{Alternatively, note that the imaginary part of the neutral SM-Higgs doublet is the neutral Goldstone boson, $G$, which is ``eaten'' by the $Z$ boson. Like all Goldstone bosons, this field is derivatively coupled and is therefore CP-odd. Since we have argued that all scalar bosons are P-even and C and P are conserved separately, it follows that the Goldstone boson field $G$ must have ${\rm C}=-1$.}

The absence of certain Higgs couplings such as $W^+ W^-a_i$, $ZZa_i$, $Zh_i h_j$ and $Za_i a_j$ is in agreement with these C and P assignments.\footnote{In the case of $i=j$, the absence of the coupling of the $Z$ to a pair of identical scalars is forbidden by Bose symmetry.   For a pair of nonidentical scalars the coupling is only present if the two scalars have opposite CP quantum numbers as exemplified by the $Zh_i a_i$ vertex.}  
The $V Vh_i$ couplings arise from the gauge-covariant kinetic energy terms of the scalar multiplets.
Assuming that the vevs are real, no $VVa_i$ term in a CP-conserving theory arises from the covariant derivative terms since~$a_i$ derives from the imaginary part of~$\phi$.  In particular, given that the $a_i$ are CP-odd, a gauge invariant interaction with vector bosons must have the form $\epsilon^{\mu\nu\alpha\beta} F_{\mu\nu} F_{\alpha\beta}\, a_i$ which is a dimension-five term and is potentially generated at loop level.  However, in the bosonic theory, C-invariance guarantees that such terms do not arise at loop level.\footnote{Since C-invariance is broken once we include fermions in the theory, such couplings can be radiatively generated through fermion loops.}
Moreover, the covariant derivative terms do not generate an $h_i$ coupling to a massless vector boson pair (i.e., photons and gluons). These types of couplings only occur at one-loop order through the dimension-five operator $F_{\mu\nu}F^{\mu\nu} h_i$. 

In some cases, the absence of a particular Higgs coupling cannot be attributed to C and~P invariance.  For example, 
the $H^+W^-\gamma$ vertex is absent at tree level due to the conservation of the electromagnetic current.
The absence of the $H^+W^-Z$ vertex at tree level is a special feature of models with Higgs singlets and doublets and no higher scalar multiplets~\cite{Gunion:1989we}.

\section{P-even, CP-violating processes of extended Higgs sectors in the Higgs alignment limit}
\label{section3}

As noted earlier, in the absence of fermions, invariance under the parity transformation is always guaranteed.  Hence, if the scalar potential of the bosonic Lagrangian violates CP (either explicitly or spontaneously),
then any resulting CP-violating phenomena must be associated with a P-conserving C-violating
observable.  Indeed, an observable that violates both P and CP requires a term in the corresponding scattering or decay matrix element with a Lorentz-structure of the form $\epsilon_{\alpha\beta\gamma\delta} p_1^\alpha p_2^\beta p_3^\gamma p_4^\delta$, with the totally antisymmetric Levi-Civita symbol $\epsilon_{\alpha\beta\gamma\delta}$ and $p_i$ the 4-momenta of the interacting particles. Such a structure can only be generated through interactions with fermions.

In this section we begin our discussion in the context of the 2HDM and examine how P-even, CP-violating processes of the 2HDM can be observed. 
We assume that the Higgs alignment limit is exactly realized,\footnote{This assumption will be relaxed in Section~\ref{sec:beyond}.} 
which implies the following relations among Higgs basis parameters,
\beq \label{Halign}
Z_6=0, \qquad \Im (Z_5 e^{-2i\eta})=0\,,
\eeq
as noted in \eq{exact}.

The only potentially complex Higgs basis parameter that remains is $Z_7$.   In the Higgs alignment limit, the bosonic sector of the 2HDM is CP-conserving if and only if
\beq \label{AlignedCPcondition}
\Im(Z_5^* Z_7^2)= 0\,,
\eeq
in  light of \eq{CPconditions}.  In particular, \eq{AlignedCPcondition} is satisfied if and only if 
\beq \label{AlignedCPcondition2}
\Re(Z_7 e^{-i\eta})\Im(Z_7 e^{-i\eta})=0\,.
\eeq
The neutral scalar squared-mass matrix does not depend on the parameter $Z_7$.   Thus  independently of the value of $Z_7$, we can regard the neutral scalar squared mass eigenstates as eigenstates of CP, consisting of two CP-even scalars $h$ and $H$ and a CP-odd scalar $A$, where one of the CP-even states is to be identified with the observed SM-like Higgs boson.  
If CP violation is present due to $\Im(Z_5^* Z_7^2)\neq  0$, then it is more useful to relabel the neutral states as $h_1$, $h_2$ and $h_3$ (where no mass ordering is implied).   One of these three states, which we take by convention to be~$h_1$ (cf.~footnote~\ref{fnh1}), is identified as the SM-like Higgs boson, which in the exact Higgs alignment limit possesses no CP-violating interactions.  The remaining two physical neutral scalar states $h_2$ and $h_3$ are conventionally defined in the CP-conserving limit following Table~\ref{cpalign}.   Without loss of generality, if $Z_7\neq 0$ and \eq{AlignedCPcondition2} is satisfied then it is convenient to take $\Re(Z_7 e^{-i\eta})\neq 0$ and $\Im(Z_7 e^{-i\eta})=0$, since the other choice simply interchanges the roles of $h_2$ and $h_3$.   This is equivalent to declaring $h_2$ to be CP-even and $h_3$ to be CP-odd.

\begin{table}[t!]
\centering
\begin{tabular}{|c|c|c|}\hline
$\mbox{neutral scalar }$ & $\Re(Z_7 e^{-i\eta})\neq 0 $ and  $\Im(Z_7 e^{-i\eta})=0$  &  $\Re(Z_7 e^{-i\eta})= 0 $ and  $\Im(Z_7 e^{-i\eta})\neq 0$ \Tstrut \\ [2mm]
\hline 
$h_1$ & CP-even & CP-even\Tstrut \\[2mm]
$h_2$ & CP-even  & CP-odd \\[2mm]
$h_3$ & CP-odd & CP-even  \\[2mm]
$G$ & CP-odd & CP-odd \\
\hline
\end{tabular}
\caption{\small CP quantum number of the neutral scalars of the 2HDM in the exact Higgs alignment limit, where $Z_6=\Im(Z_5 e^{-2i\eta})=0$.  By definition, the tree-level couplings of $h_1$
coincide with those of the SM Higgs boson.
If $Z_7=0$ then the individual quantum numbers of $h_2$ and $h_3$ are indeterminate, although the relative sign of the CP quantum numbers of these two states is negative.
The CP-even scalar $h_1$ and the CP-odd Goldstone boson $G$ possess CP-conserving tree-level interactions even in the case of
$\Re(Z_7 e^{-i\eta})\Im(Z_7 e^{-i\eta})\neq 0$, whereas cubic and quartic scalar coupling involving $h_2$ and $h_3$ exhibit CP-violating
interactions. \label{cpalign}}
\end{table}

When $\Im(Z_5^* Z_7^2)\neq 0$, then one can regard $\Im(Z_7 e^{-i\eta})$ as a perturbation.  
In the presence of the perturbation, the interactions of $h_1$ (and $G$) remain CP-conserving in the exact Higgs alignment limit, whereas there exist cubic and quartic scalar interactions involving $h_2$ and $h_3$ that violate CP.  In particular, the following interactions are the only sources of CP violation due to a nonzero value of $\Im(Z_7 e^{-i\eta})$ in the exact Higgs alignment limit [cf.~Table~\ref{trialign}],
\beq  \label{triquad}
h_3 h_3 h_3 \quad h_3 h_2 h_2 \quad
h_3 H^+ H^-, \quad  h_3 h_3 h_3 h_1, \quad h_3 h_1 h_2 h_2, \quad
h_3 h_1 H^+ H^-, 
\eeq
as each interaction vertex involves an odd number of would-be CP-odd scalar fields.  
These interactions guide us to construct scenarios that involve sets of processes whose simultaneous observation would signal the presence of a new scalar source of CP violation.
In this section, we shall impose the following three additional requirements: (i) all processes under consideration are unsuppressed in the Higgs alignment limit; (ii) no quartic scalar couplings are involved (due to coupling and phase space suppressions of the corresponding physical processes); and (iii) the dominant contribution to the CP-violating signal is P-even.   In particular, C-even, CP-violating contributions due to the presence of neutral Higgs-fermion Yukawa couplings are assumed to be either absent or suppressed.\footnote{The SM fermions can induce a CP-violating $h_i H^+ H^-$ interaction that is proportional to the complex CKM phase.   However, this effect only enters at the three-loop level and is thus much too small to be observable.   Additional CP-violating neutral Higgs-fermion Yukawa couplings can provide sources of P-odd CP violation that would contribute at one loop to the three scalar couplings.}  

In the 2HDM, we are led to the simultaneous observation of the processes governed by the following bosonic interactions:
\beqa
&& \hspace{-1.5in}  1.~~h_2H^+ H^-\,,\quad h_3 H^+ H^-\,, \quad Zh_2 h_3\,,  \label{s1} \\
&& \hspace{-1.5in}  2.~~h_2 h_k h_k\,,\quad h_3 H^+ H^-\,, \quad Zh_2 h_3\,,\quad \text{(for $k=2$ or 3)}, \label{s2} \\
&& \hspace{-1.5in}  3.~~h_3 h_k h_k,,\quad h_2 H^+ H^-\,, \quad Zh_2 h_3\,,\quad \text{(for $k=2$ or 3)},\label{s3} \\
&& \hspace{-1.5in}  4.~~h_2 h_k h_k\,,\quad h_3h_\ell  h_\ell\,,\qquad Zh_2 h_3\,,\quad \text{(for $k, \ell=2$ or 3}). \label{s4}
\eeqa
If CP is conserved, then the observation of a $Zh_2 h_3$ interaction would imply that the CP quantum numbers of $h_2$ and $h_3$
are relatively odd, whereas the observation of the first two cubic scalar interactions listed above would imply that $h_2$ and $h_3$ are both CP-even.
Consequently, the simultaneous observation of the three interactions in each of the above four cases would unambiguously point to the presence
of P-even CP violation.

In a more general extended Higgs sector (e.g. with additional neutral scalar states), 
the simultaneous observation of the three interactions in each of the following six cases: 
\beqa
&& \hspace{-0.5in}   5.~~h_i H^+ H^-\,, \quad h_j H^+ H^-\quad \mathrm{and} \quad Z  h_i h_j\,,\quad \text{(for $i\neq j$ and $i$, $j\neq 1$)}, \label{eq:CPVobs1p}  \\
&& \hspace{-0.5in}  6.~~h_i h_k h_k\,, \quad h_j H^+ H^-\quad \mathrm{and} \quad Z  h_i h_j\,,\quad \text{(for $i\neq j$ and $i$, $j$, $k\neq 1$)}, \label{eq:CPVobs2p}  \\
&& \hspace{-0.5in}  7.~~h_i h_j h_k\,,\quad h_k  H^+ H^- \quad  \mathrm{and} \quad Z  h_i h_j\,,\quad \text{(for $i\neq j\neq k$ and $i$, $j$, $k\neq 1$)},\label{eq:CPVobs2pp} \\
&& \hspace{-0.5in}  8.~~h_i h_k h_k\,, \quad h_j h_\ell h_\ell\quad \mathrm{and} \quad Z  h_i h_j\,,\quad \text{(for $i\neq j$ and $i$, $j$, $k$, $\ell\neq 1$)},\label{eq:CPVobs5} \\
&& \hspace{-0.5in}  9.~~h_k h_\ell h_\ell\,, \quad h_i h_j h_k\quad \mathrm{and} \quad Z  h_i h_j\,,\quad \text{(for $i\neq j\neq k$ and $i$, $j$, $k$, $\ell\neq 1$)},\label{eq:CPVobs2} \\
&& \hspace{-0.5in}  10.~~h_i h_j Z\,, \quad h_i h_k Z\quad \mathrm{and} \quad h_j h_k Z\,,\quad \text{(for $i\neq j\neq k$ and $i$, $j$, $k$, $\neq 1$)}, \label{eq:CPVobs3} 
\eeqa
would be a signal of P-even CP violation.  

In models with inert scalar doublets in which all physical scalars (with the exception of~$h_1$) are $\mathbb{Z}_2$ odd, the couplings of $h_1$ are identical to those of the SM Higgs boson.   Hence, 
only the set of couplings exhibited in \eq{eq:CPVobs3} survive in the inert limit and provide a signature for P-even CP violation.\footnote{However, inert models based on larger symmetries as in Refs.~\cite{Ivanov:2012hc,Keus:2013hya,Aranda:2019vda,Khater:2021wcx} can yield signals of P-even CP violation in some of the other channels exhibited above.}
However, the identification of a CP-violating signal in the
direct detection of such processes will be quite difficult, since the lightest inert scalar is necessarily stable and thus will result in events with missing energy.  On the other hand, the indirect detection of P-even CP violation via loop effects may be viable as discussed in Section~\ref{sec:ZZZ}.

For a practical application of the scenarios listed in \eqst{s1}{s4}, we will omit cases that involve the cubic self-interaction of a given scalar.
It is noteworthy that the scenario corresponding to \eq{s4} involves the discovery of just two neutral states.   In the next section, we shall first focus on processes in which $h_2$, $h_3$ can be observed via an $s$-channel $Z$ exchange.   Once $h_2$ and $h_3$ are established, one can search for $h_2 h_2 h_3$ and $h_2 h_3 h_3$ production (which are also mediated via an $s$-channel $Z$ exchange).  Simultaneous observation of both production channels would signal the presence of P-even CP violation. 
The other scenarios corresponding to \eqst{s1}{s3} also require the observation of a charged Higgs boson.   
Thus, in the next section we also examine a suite of processes in which $h_2$, $h_3$ and $H^\pm$ can be observed.  

For certain scalar mass spectra, it is possible to access all three vertices of a given scenario from one production process.   For example, consider $h_2 h_3$ production via $s$-channel $Z$ exchange.   If kinematically accessible, both $h_2$ and $h_3$ could have observable branching ratios into $H^+ H^-$, in which case
the observation of $h_2 h_3$ production via their respective decays into $H^+ H^-$ would yield a simultaneous observation of all three processes listed in \eq{s1}. 

The vertices listed in \eqst{s1}{s4} are unsuppressed in the Higgs alignment limit, and none involve quartic scalar couplings.  Any contributions of C-even, CP-violating interactions via the Yukawa couplings are loop suppressed.
Thus, to a good approximation, the simultaneous observation of the three processes  in any of the scenarios listed in \eqst{s1}{s4} would constitute an unambiguous signal of P-even CP violation.

The detection of P-even, CP-violating signals at colliders requires the simultaneous observation of processes typically governed by three distinct bosonic interactions.
In some cases, the relevant observable involves an on-shell two-body scalar.  In other cases, the relevant bosonic interaction can only be probed
when at least one of the bosonic states is off shell.  Typically, the on-shell decay processes are the easiest to observe and interpret, although the kinematic availability of
such processes will depend on a favorable spectrum of scalar masses.   A more comprehensive search for P-even, CP-violating signals will require an experimental probe of
bosonic interactions involving off-shell scalars.  Such processes will require higher statistical data samples, subtraction of backgrounds, and more sophisticated methods
to interpret the relevant signals.

\section{Observation of P-even, CP-violating scalar processes at lepton colliders}
\label{sec:collider}

We begin with the assumption that the neutral scalars (beyond the SM-like $h_1$) and the charged scalars of the extended Higgs sector have been discovered in future runs of the LHC and/or at some future lepton collider.  For example, neutral heavy scalars can typically be produced via gluon-gluon fusion and discovered through their decays into heavy fermion pairs at the LHC over a significant part of the scalar sector parameter space (e.g., see Refs.~\cite{Kling:2020hmi,Accomando:2022nfc}).  
At an $e^+e^-$ collider operating at a center of mass (CM) energy of 500 GeV to 1 TeV, new scalars of an extended Higgs sector can be discovered, as shown in Ref.~\cite{Desch:2004yb},
if kinematically accessible and if their main decay modes can be probed (with 
decays to $b$-quarks typically the most relevant for scalar masses below $2m_t$).  For example, the process $e^+ e^- \to H A$ was examined in Ref.~\cite{Desch:2004yb} in the framework of the 
minimal supersymmetric extension of the Standard Model (MSSM) close to the decoupling limit (where the scalars $H$ and $A$ are roughly mass degenerate) and the cross section reaches its maximum value. The authors concluded that for a linear collider with $\sqrt{s}$ = 800 GeV with an integrated luminosity of 500~fb$^{-1}$, it would be possible to observe $e^+ e^- \to H A$ (at a level of $5\sigma$ statistical significance) for scalar masses of around 385 GeV.
In this case, the tree-level cross section $\sigma({e^+ e^- \to H A})$ is about 2 fb, which yields 1000 signal events before cuts.   

Higher energy $e^+ e^-$ colliders with $\sqrt{s}\geq 1$~TeV can significantly extend the discovery reach of heavy scalars.   For example,  Ref.~\cite{No:2018fev} considered
the production of a heavy scalar~$H$ via the process $e^+ e^- \to H \nu \bar \nu$ with
the subsequent decay $H \to hh \to 4b$. This study was performed for the proposed Compact Linear Collider (CLIC) with a CM energy of 3 TeV assuming an integrated luminosity
of 2 ab$^{-1}$.  The authors concluded that the sensitivity for discovery in the mass range
250 GeV~$\leq m_H\leq 1$~TeV provides up to two orders of magnitude improvement in sensitivity compared to the discovery potential at the High Luminosity (HL)~LHC.


%

The unambiguous observation 
P-even CP-violating phenomena via any one of the four sets of three interaction vertices exhibited in \eqst{s1}{s4} at the LHC of is a challenging task.
In contrast, at a lepton or photon collider, the corresponding backgrounds that must be subtracted to reveal the signal are significantly easier to handle. 
The corresponding signals at a lepton or photon collider provide direct access to the interaction vertices exhibited in \eqst{s1}{s4} at tree level that can be directly identified as
P-even and CP-violating (once the three interaction vertices of a given set are confirmed).   Moreover, potential P-odd, CP-violating effects that would involve scalar couplings to fermions only arise at the one-loop level and are hence subdominant.
Of course, the Yukawa couplings will enter when considering the decays of the produced neutral and charged Higgs bosons, which we shall address at the end of this section.

\subsection{Discovery potential at future lepton (and photon) colliders}

\begin{table}[b!]
\centering
\begin{tabular}{|c|c|c|}\hline
$\mbox{Accelerator}$ & $\sqrt{s} \, (\mbox{TeV})$ & $\mbox{Integrated luminosity}\,  ({\rm ab}^{-1})$ \Tstrut \\ [2mm]
\hline 
CLIC & $1.5$ & $ 2.5$\Tstrut \\[2mm]
CLIC & $3$ & $ 5$ \\[2mm]
Muon Collider & $3$ & $ 1$ \\[2mm]
Muon Collider  & $10$ & $ 10$ \\ [2mm]
Muon Collider  & $14$ & $20$ \\ [2mm]
\hline
\end{tabular}
\caption{\small Accelerators used in the analysis with different CM energies proposed and the corresponding total integrated luminosity after the completion of a multiyear program (typically of order 10 years).  \label{coll}}
\label{colliders}
\end{table}

Consider the 
discovery potential of final states related to the P-even, CP-violating observables at future lepton colliders listed in Table~\ref{coll}. 
CLIC~\cite{deBlas:2018mhx} is an electron-positron collider that has been proposed to run at CM energies of 1.5 and 3 TeV with total integrated luminosities of 2.5 and 5 ab$^{-1}$, respectively, after the completion of a multiyear program  (typically of order 10 years).  We also consider the possibility of 
a muon collider~\cite{AlAli:2021let} with CM energies of 3, 10, and 14 TeV and with total integrated luminosities of 1, 10, and 20 ab$^{-1}$, respectively. 
In addition, we shall show results for a photon-photon collider of CM energies of 1 and 2 TeV that could be achieved via the Compton backscattering of laser light on high energy 
electrons at CLIC~\cite{Burkhardt:2002vh}.  Other 
lepton colliders now under development such as the Circular Electron Positron Collider in China~\cite{CEPCStudyGroup:2018ghi} ($\sqrt{s}_{max}  \sim 250$ GeV), the International Linear Collider in Japan~\cite{Baer:2013cma} ($\sqrt{s}_{max} \sim 250$ GeV) 
and the FCC-ee at CERN ($\sqrt{s}_{max}  \sim 365$ GeV)~\cite{FCCee} have energies well below the production threshold of our final states, and thus are not considered here.

Although lepton colliders provide a very clean environment for the final states of the processes under consideration,
a proper analysis would still have to take into account both the efficiencies and the main background processes.   Nevertheless, having assumed above that the scalars of the extended Higgs sector have already been discovered, either at the HL-LHC or at an $e^+e^-$ collider of sufficient CM energy,
one can consider smaller samples of signal events (compared, say, to the 1000 events before cuts employed in Ref.~\cite{Desch:2004yb}) in searches for P-even CP-violating phenomena using
di-Higgs and tri-Higgs final states via the processes specified in \eqst{s1}{s4}. 
In this work, we are considering lepton collider luminosities that are roughly one order of magnitude larger than employed in earlier studies that focused on lower energy colliders. 
Thus, to ensure at least 100 events before cuts, signals of interest should have cross sections that are above 10--100 ab.   



 \subsection{Coupling parameters that govern the three-scalar interactions}

As previously indicated, the observed SM-like Higgs boson is denoted by $h_1(125)$. In the context of the 2HDM, one needs to  simultaneously detect the other two scalars $h_2$ and $h_3$ 
either in production or decay processes involving the interaction vertices $Z  h_2 h_3$, $H^+ H^- h_2$ (or alternatively $h_3 h_3 h_2$) 
and $H^+ H^- h_3$ (or alternatively $h_2 h_2 h_3$).  In principle, $h_2$ and $h_3$ can be either heavier or lighter than the SM-like Higgs boson. As we will be working in the exact Higgs alignment limit, the tree-level couplings $\lambda_{Z Z h_2} $  and $\lambda_{Z Z h_3} $ as well
as $\lambda_{Z h_2 h_1} $ and $\lambda_{Z h_3 h_1} $ vanish, precluding the possibility of using a combination of these decays to signal CP violation as discussed in Refs.~\cite{Mendez:1991gp,Fontes:2015xva,Keus:2015hva}.  
In the exact Higgs alignment limit, the interaction vertices that contribute to the P-even, CP-violating observables corresponding to \eqst{s1}{s4} can be obtained from Table~\ref{trialign} of Appendix~\ref{2hdm},
\beq
\lambda_{H^+ H^- h_1} =  vZ_3\,,\quad  \lambda_{H^+ H^- h_2} = \lambda_{h_3 h_3 h_2}  =   v\,\Re\zvii\,,\quad
\lambda_{H^+ H^- h_3} = \lambda_{h_2 h_2 h_3} =  -v\,\Im\zvii\,,
\eeq
where $\zvii\equiv  Z_7 e^{-i\eta}$ is defined in \eq{zees}.
Thus, the production rate for each process governed by the $\varphi^*\varphi h_i$ interaction vertices,
where $\varphi= H^+$ or $h_j$ ($j\neq i$, $j\neq 1$), is proportional to $\Lambda_i^2$ defined by
\beq \label{Lambdas}
\Lambda_1 \equiv \lambda_{H^+ H^- h_1}/v, \quad   \Lambda_2 \equiv \lambda_{H^+ H^- h_2}/v = \lambda_{h_3 h_3 h_2}  /v, \quad \Lambda_3 \equiv - \lambda_{H^+ H^- h_3}/v =- \lambda_{h_2 h_2 h_3}/v\,.
\eeq 
In particular,
\beq \label{Z377}
\Lambda_1 = Z_3,
\qquad
\Lambda_2 =  \Re\zvii,
\qquad
\Lambda_3 = \Im\zvii.
\eeq

The detection of scalars in decay processes with fermionic final states depends on the Yukawa couplings for which there is some freedom in the 2HDM. 
In order to maximize this freedom while avoiding tree-level flavor changing neutral currents, we shall employ the Yukawa sector of the flavor-aligned 2HDM (A2HDM)~\cite{Pich:2009sp}
as a benchmark in the Higgs alignment limit [cf.~\eq{A2HDMalign}].
With this choice, the Yukawa couplings are independent of the gauge couplings and of the parameters of the scalar potential, which in turn weakens the constraints on the scalar masses.  

\subsection{Allowed mass range for non-SM Higgs states of the 2HDM}

The processes that we propose as probes of P-even CP violation involve the production of neutral Higgs bosons [e.g., $h_2$ and $h_3$, but excluding the $h_1(125)$] and charged Higgs bosons.  Experimental searches for these scalar states have already yielded some constraints on their masses and couplings.
The production of charged Higgs bosons at the LHC is governed mainly by the Yukawa interactions via $pp \to tb H^\pm$ 
with subsequent decay channels,
$H^\pm \to tb$, $H^\pm \to \tau \nu$ and $H^\pm \to c s$. 
Charged Higgs bosons can also be produced via scalar or gauge boson exchange, as in the $s$-channel (Drell-Yan type) processes $\bar u (\bar c) d (s,b) \to W^- \to H^- h_i$ (and its conjugate process)  and also via $\bar f f \to \gamma (Z) \to H^+ H^-$.\footnote{See, e.g.,  Ref.~\cite{Aoki:2011wd} for a list of the main charged Higgs production processes at the LHC.}

The LHC searches performed so far rely on the charged Higgs couplings to fermions. 
For example, in a CP-conserving 2HDM with Type-I Yukawa couplings in the exact Higgs alignment limit, the Yukawa coupling of  $h_1$ coincides with that of the SM, whereas 
the Yukawa couplings of all other scalar states (both neutral and charged) are proportional to a common factor of $1/\tan \beta$. 
That is, the couplings of the charged Higgs boson to fermions are suppressed at large $\tan\beta\gsim 10$, in which case 
a charged Higgs mass of the order of 100 GeV \cite{Sanyal:2019xcp,Bahl:2021str,Cheung:2022ndq}
is still allowed.
Indeed, other constraints on the parameter space, such as those derived from $pp \to A \to Z h_i$ searches are also weakened if $\tan \beta$ is large. 
In order to circumvent the smallness of the  Yukawa couplings, the authors of Ref.~\cite{Bahl:2021str} proposed employing  the processes $\bar u (\bar c) d (s,b) \to W^- \to H^- h_i \to W^- h_i h_j \, (i,\, j = 2,\, 3)$
 to search for charged Higgs bosons at the LHC. The corresponding production rates only depend on the Yukawa couplings in the decays of the neutral scalars. The cross sections can reach a few hundred femtobarns  
 if both scalars are relatively light (i.e., not much heavier than 100 GeV). 
 
The Yukawa sector of the A2HDM is less constrained than that of a Type I 2HDM in light of the recent global fit of Ref.~\cite{Eberhardt:2020dat}
(which assumed no new sources of CP violation 
beyond the CKM phase). Hence, light charged Higgs bosons are also not excluded
in the A2HDM.
As for the neutral scalars $h_2$ and $h_3$, these scalars do not couple to gauge bosons in the exact Higgs alignment limit. Their couplings to fermions can also be significantly smaller than those of the SM-like Higgs boson $h_1$.  
In particular, the couplings of $h_2$ and $h_3$ to top quarks are suppressed by $1/\tan \beta$ in the 2HDM with Type I, II, X or Y Yukawa couplings when $\tan\beta\gg 1$.
Therefore, it is quite likely that $h_2$ and $h_3$ with masses as low as 100 GeV could have so far evaded detection in a non-negligible region of the 2HDM parameter space.

\subsection{Cross sections for P-even, CP-violating scalar processes}
\label{xsections}

The  cross sections presented in this section are calculated using {\tt SARAH~4.14.4}~\cite{Staub:2009bi,Staub:2013tta}
to generate the Feynman rules for the model and  \texttt{MadGraph5\_aMC@NLO}~\cite{Alwall:2011uj} to evaluate the cross section for the different scenarios. 
The Equivalent Photon Approximation is used~\cite{Budnev:1974de, Frixione:1993yw} by \texttt{MadGraph5}  when calculating the cross sections
$\ell^+ \ell^- \to  \ell^+ \ell^- H^+ H^- h_i$ (see also the recent work dedicated to muon colliders~\cite{Ruiz:2021tdt}).

\begin{figure}[ht!]
    \centering
    \hspace{-2cm}
    \includegraphics[width=0.7\textwidth]{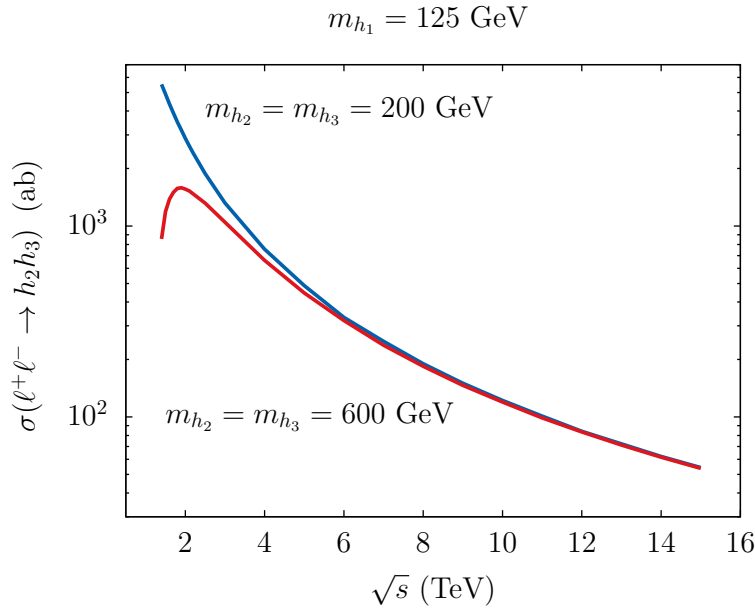}
	\caption{\small $\sigma (\ell^+ \ell^-\to h_2 h_3)$ as a function of the CM energy for $m_{h_2}= m_{h_3}= 200$ GeV and
	 $m_{h_2}= m_{h_3}= 600$ GeV. }
	\label{fig:fig1}
\end{figure}

Let us first consider the production process $\ell^+ \ell^-\ \to h_2 h_3$, for which the main contribution comes from a tree-level $s$-channel process with $Z$ boson exchange. 
In the Higgs alignment limit, this is purely an electroweak process and the number of Higgs bosons produced depends only on their masses. 
In Fig.~\ref{fig:fig1} we show the total cross section $\sigma (\ell^+ \ell^-\to h_2 h_3)$ as a function of the 
CM energy for $m_{h_2}= m_{h_3}= 200$ GeV and for $m_{h_2}= m_{h_3}= 600$ GeV. As expected, the cross section falls with the CM energy but is still above
1000 ab for both scenarios for $\sqrt{s} = 3$ TeV. 
This means that there is a good chance of producing and detecting the two scalars even if they are heavy. For even higher CM energies the number of events will
decrease but the dependence on the masses of the scalars will become milder.

These scalars still have to be detected in some particular final state.  In the exact Higgs alignment limit $h_2$ and $h_3$ cannot decay to gauge bosons.
To simplify the discussion of the possible final states let us assume $m_{h_2} \leq  m_{h_3}$.  
The most relevant $h_2$ and $h_3$ decay modes (if kinematically allowed) are $h_3 \to h_2 Z$, $h_3 \to h_2 h_2$, $h_{2,3} \to H^\pm W^\mp$ , $h_{2,3} \to H^+ H^-$, $h_{2,3} \to \bar t t$, $h_{2,3} \to \bar bb$ and $h_{2,3} \to \tau^+ \tau^-$.
Because the couplings of $h_2$ and $h_3$ to other scalars can be large enough to allow
the decays to charged scalars to be dominant, this process alone could signal P-even CP violation in the exact Higgs alignment limit (e.g., by considering 
$h_3 \to h_2 h_2$ and $h_2 \to H^+ H^-$). Clearly all the masses would have to be fully reconstructed via the hadronic decays
of the charged Higgs boson, which can be carried out at a lepton collider (where the cross sections for the relevant background processes are
of the same order of magnitude as the signal process).

If the two-body decays of $h_2$ and $h_3$ into bosonic final states are kinematically forbidden, then it is necessary to consider separately the three production processes governed by
one of the sets of bosonic interactions listed in \eqst{s1}{s4}. 
This strategy has the advantage of being constrained only
by the collider energy but the disadvantage of requiring the observation of three-body processes with smaller cross sections.

\begin{figure}[t!]
    \hspace{-0.4cm}
    \includegraphics[width=0.5\textwidth]{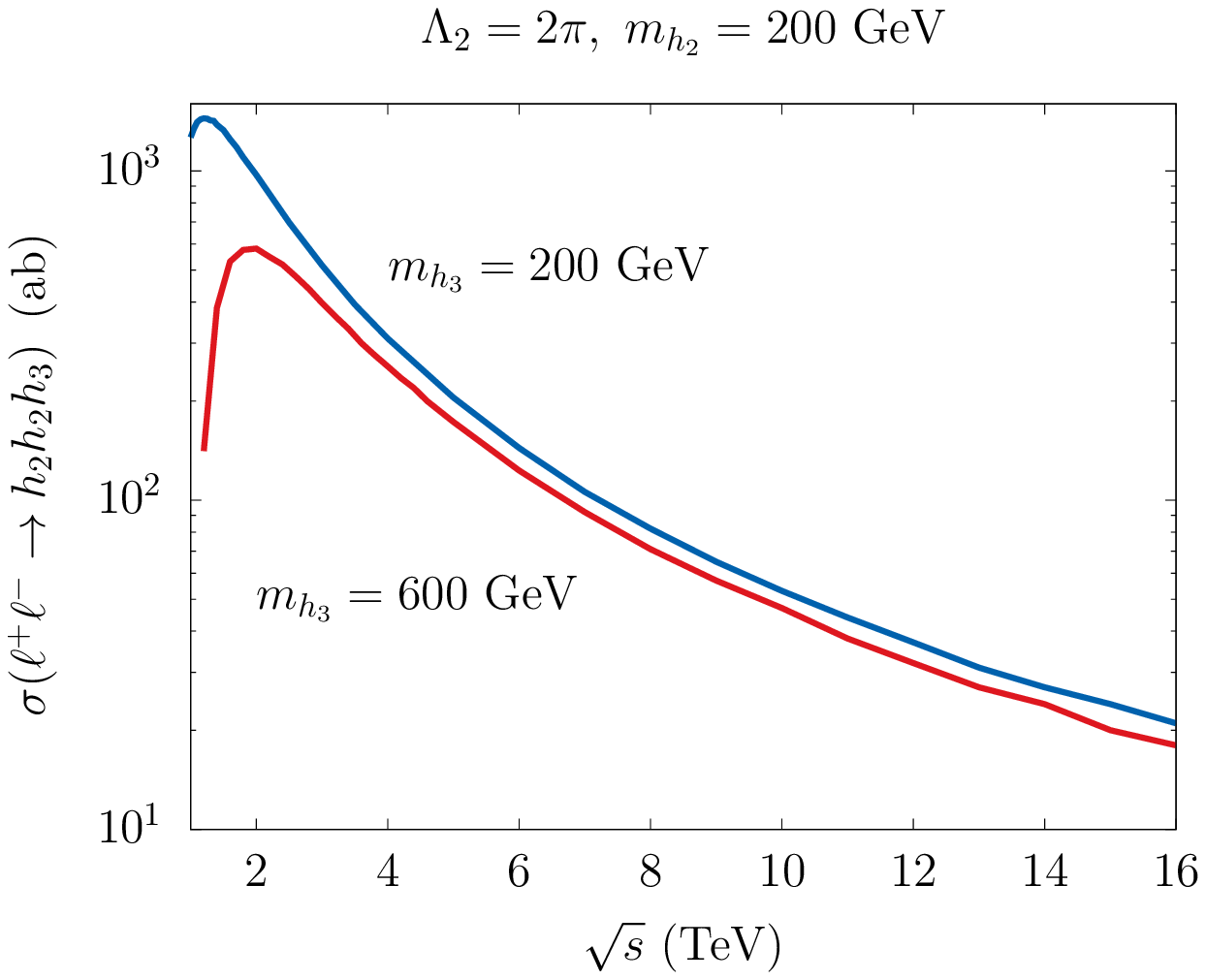} \hspace{-0.4cm}
    \includegraphics[width=0.5\textwidth]{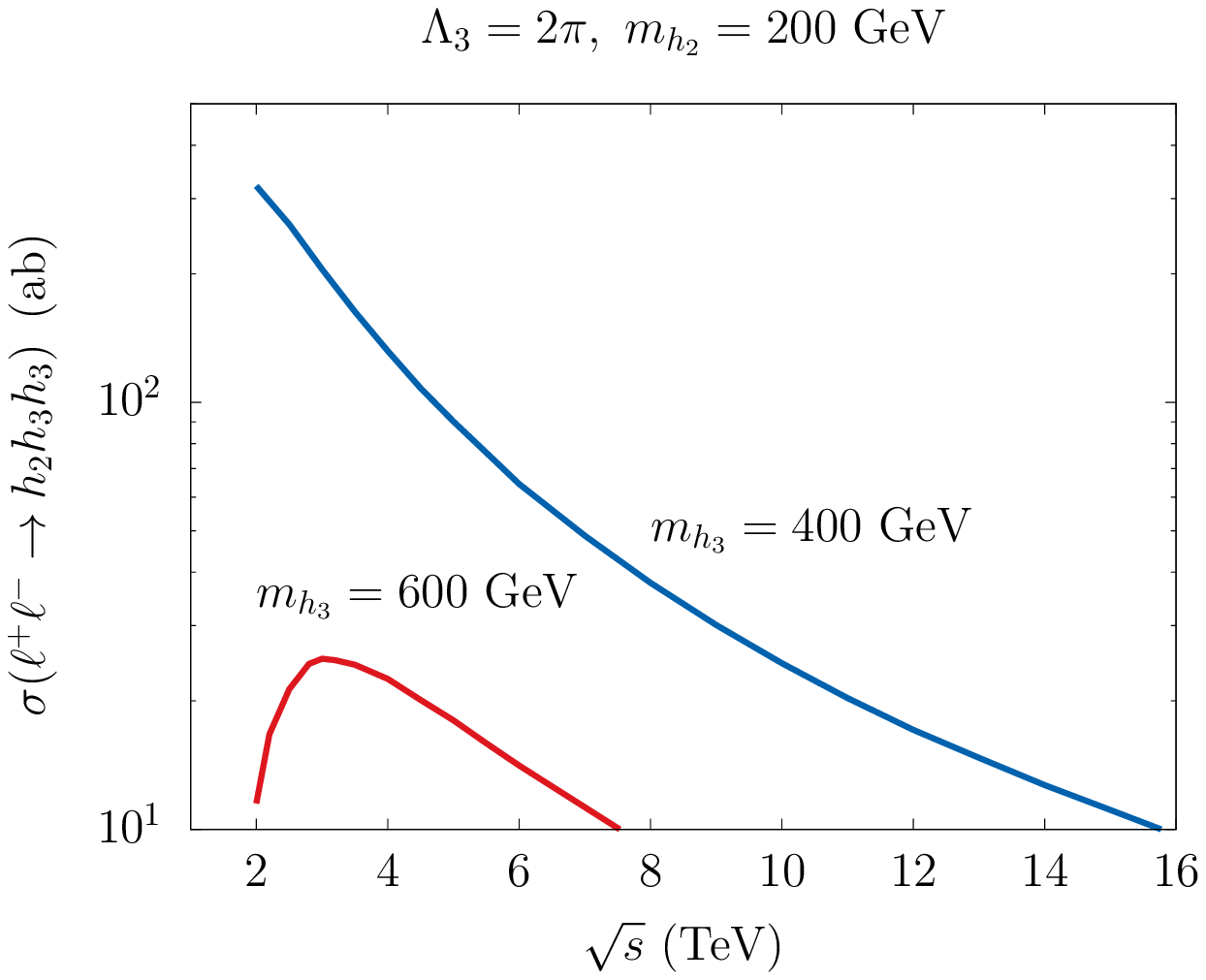}
	\caption{\small $\sigma (\ell^+ \ell^-\to h_2 h_2 h_3)$ (left) and $\sigma (\ell^+ \ell^-\to h_2 h_3 h_3)$ (right)  as a function of the CM energy, with $m_{h_2} =200$ GeV. }
	\label{fig:fig0}
\end{figure}

We begin with the $s$-channel three-body process with the exchange of a $Z$ boson.  In Fig.~\ref{fig:fig0}, we fix the value of $m_{h_2}=200$~GeV and plot the total cross sections
for $\ell^+\ell^-\to h_i  h_j h_j$ (for $i\neq j=2,3$)
as a function of the CM energy.   In the left panel of Fig.~\ref{fig:fig0}, we exhibit
$\sigma (\ell^+ \ell^-\to h_2 h_2 h_3)$ with $\Lambda_2 = 2 \pi$ 
for two choices of $m_{h_3}=200$ and 600 GeV, and in the right panel we exhibit $\sigma (\ell^+ \ell^-\to h_2 h_3 h_3)$ with $\Lambda_3 = 2 \pi$ for two choices of $m_{h_3}=400$ and 600 GeV. 
The cross section for $\ell^+ \ell^-\to h_2 h_2 h_3$ is dominated by the value $\Lambda_2$ because of the relation
$\lambda_{h_2 h_2 h_2} = 3 \lambda_{h_3 h_3 h_2} =  \Lambda_2/v $ (cf.~Table~\ref{trialign}). All diagrams except for the ones with two $Z h_2 h_3$ vertices are proportional to  $\Lambda_2$, and 
in the region relevant for our analysis where $\Lambda_2 > 1$,
all other contributions are negligible. The same can be said for the relation between $\sigma$($\ell^+ \ell^-\to h_3 h_3 h_2)$ and the value of $\Lambda_3$ because  
$\lambda_{h_3 h_3 h_3} = 3 \lambda_{h_2 h_2 h_3} =  - \Lambda_3/v $ (cf.~Table~\ref{trialign}).  The results shown in Fig.~\ref{fig:fig0} suggest that if the masses of $h_2$ and $h_3$ are not significantly heavier than the scale of electroweak symmetry breaking then the 
observation of $\ell^+\ell^-\to h_i h_j h_j$ will provide an opportunity for detecting evidence
for P-even CP violation (if present), if the CM energy of the lepton collider is in the range of 1--3~TeV.

Consider next the $t$-channel processes, which 
are dominated by $\gamma \gamma$ fusion with a cross section that
is proportional to $\ln^2 (s/m_\ell^2)$. There are also $Z$ fusion diagrams contributing but the corresponding cross sections are proportional to $\ln^2 (s/m_Z^2)$\cite{Dawson:1984gx}
 and are thus subdominant. 
 In light of \eq{Lambdas}, the cross section for any final state of the type $ H^+ H^- h_i$ (for $i=1,2,3$) is proportional to~$\Lambda_i^2$. 
That is,  by choosing ${\Lambda}_1 ={\Lambda}_2={\Lambda}_3 =2 \pi$, the cross sections exhibited in this section are applicable to any of the neutral scalars.  

\begin{figure}[t!]
    \hspace{-0.4cm}
    \includegraphics[width=0.5\textwidth]{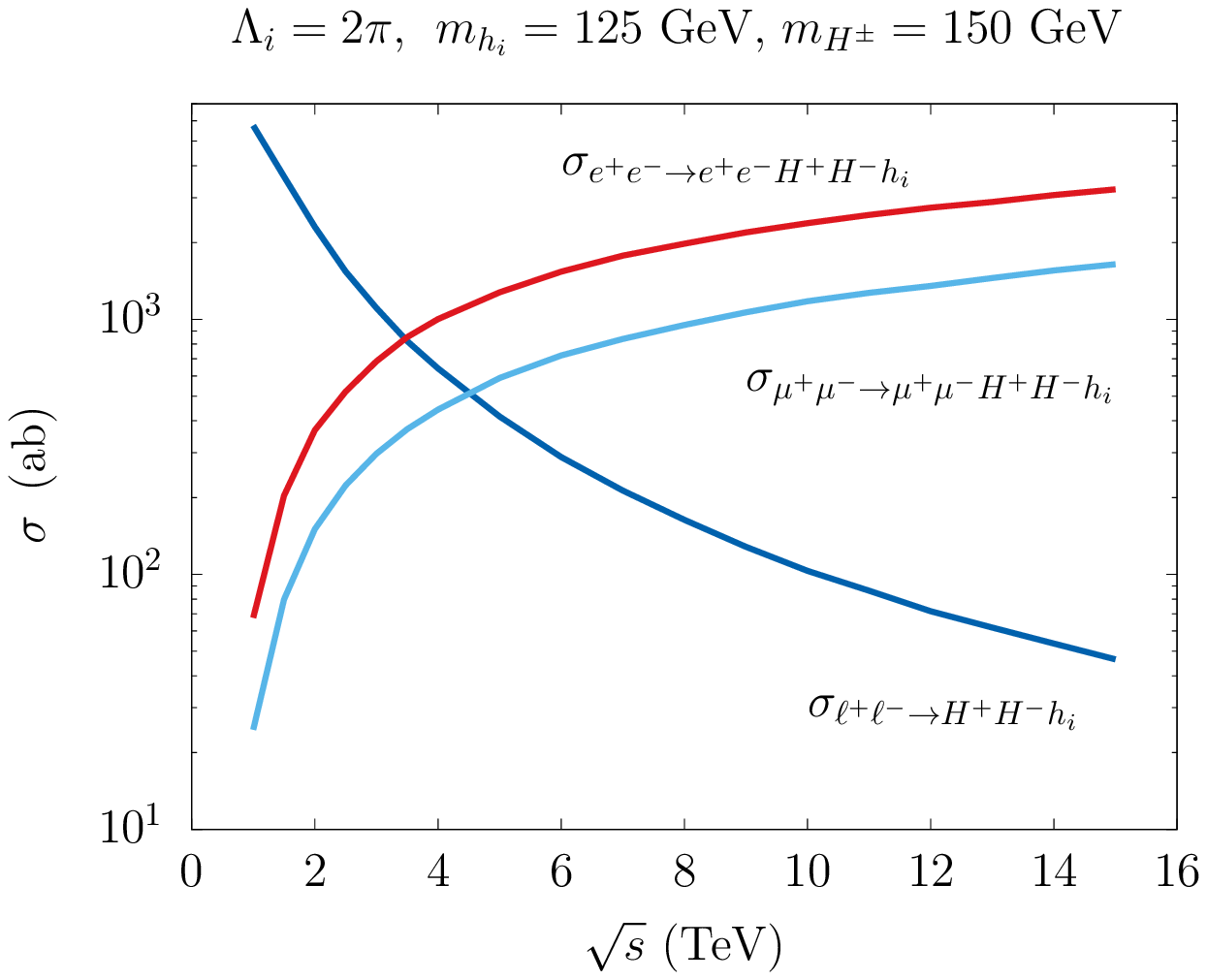} \hspace{-0.4cm}
    \includegraphics[width=0.5\textwidth]{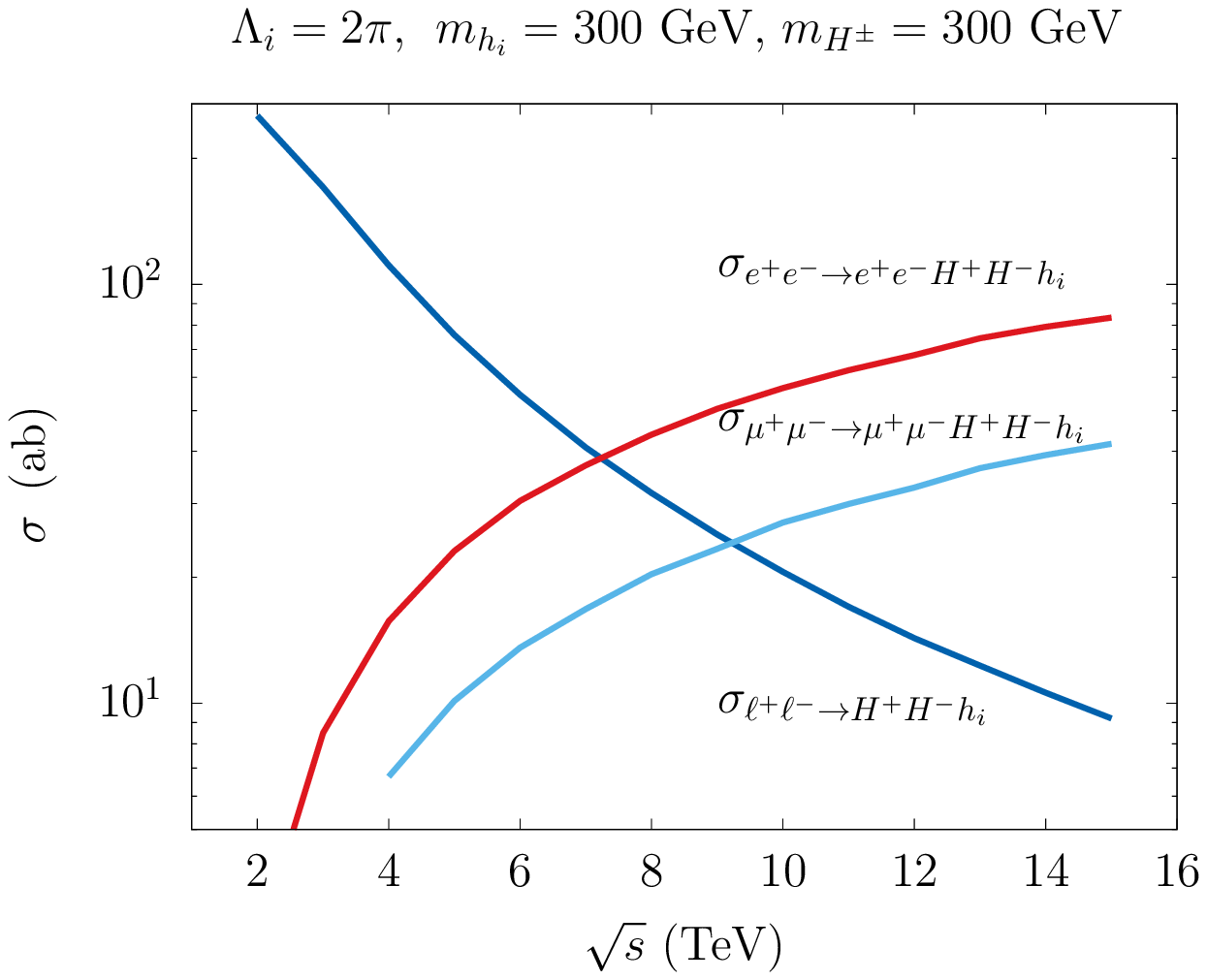}
	\caption{\small $\sigma({e^+ e^- \to  e^+ e^-  H^+ H^- h_i})$,  $\sigma({\mu^+ \mu^- \to  \mu^+ \mu^-  H^+ H^- h_i})$ and  $\sigma({\ell^+ \ell^- \to  H^+ H^- h_i})$
	as a function of the CM energy. In the left panel $m_{h_i} = 125$ GeV and $m_{H^\pm}= 150$ GeV, and in the right panel
  $m_{h_i} = m_{H^\pm} = 300$ GeV. 
 The scalar potential parameters are chosen such that ${\Lambda}_i = 2 \pi$.}
	\label{fig:fig3}
\end{figure}

\begin{figure}[t!]
    \hspace{-1.4cm}
    \includegraphics[width=0.57\textwidth]{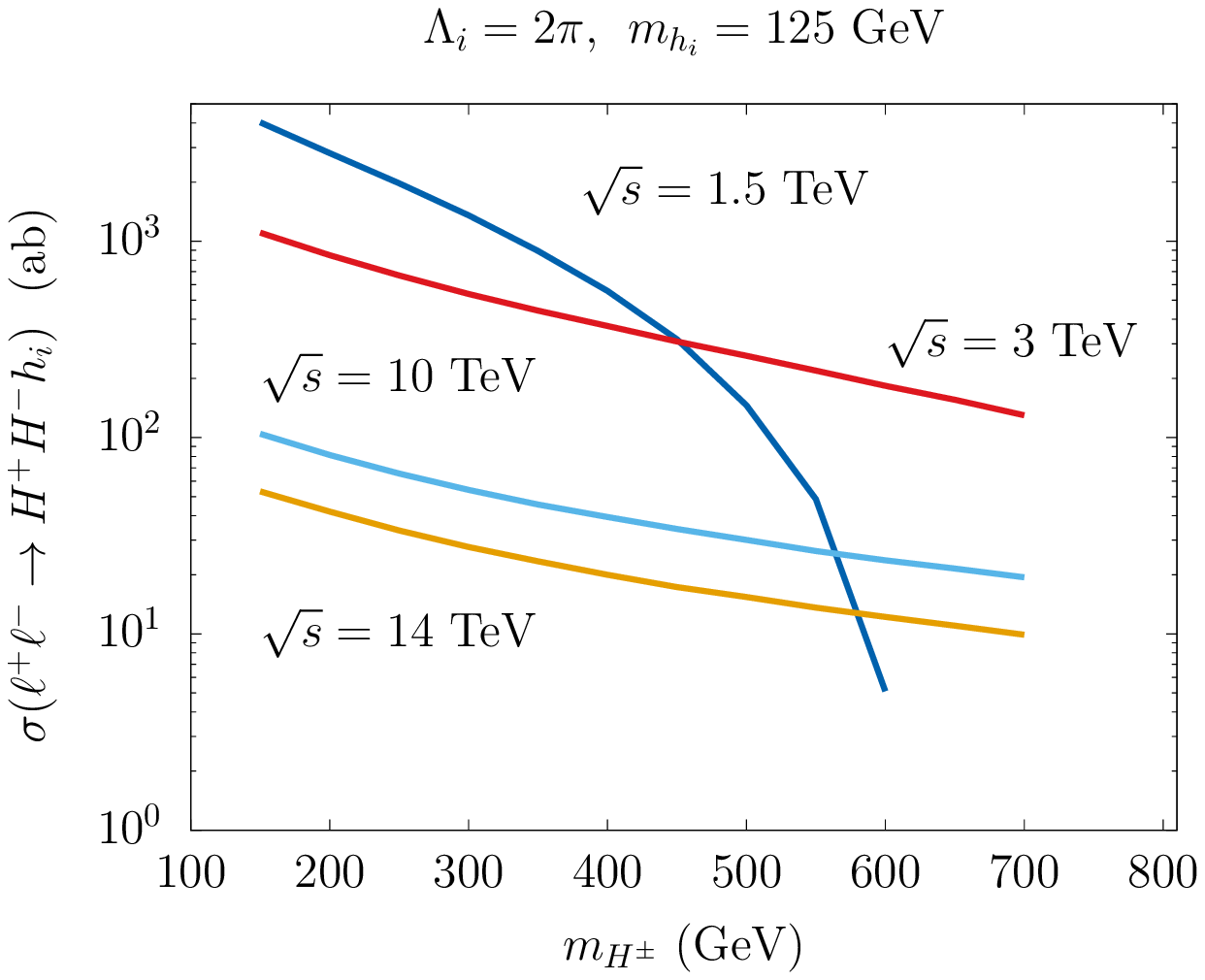} \hspace{-1.4cm}
    \includegraphics[width=0.57\textwidth]{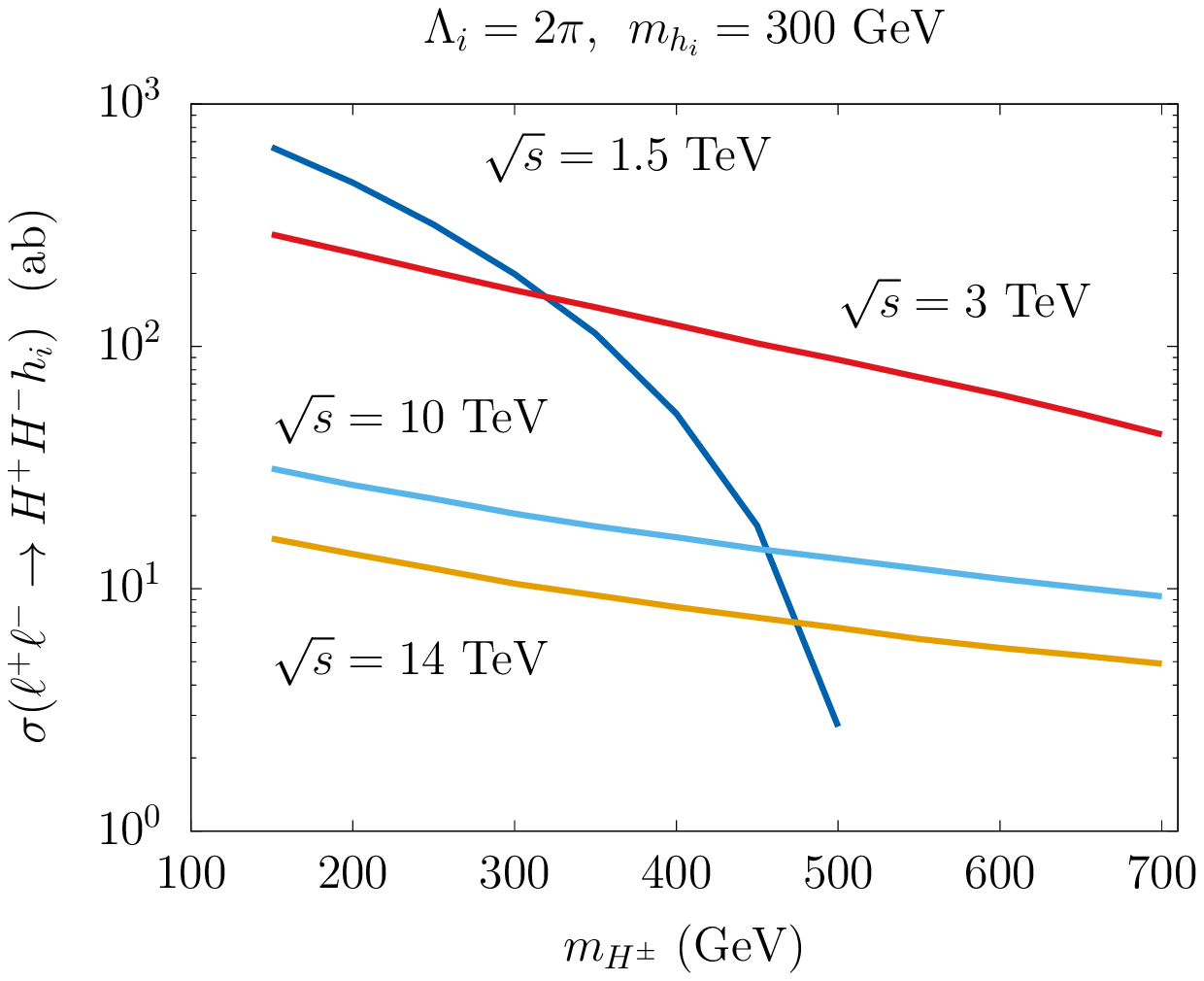}
	\caption{\small$\sigma (\ell^+ \ell^- \to  H^+ H^- h_i)$ as a function of the charged Higgs mass for four CM energies of $\sqrt{s} = 1.5$, $3$, $10$ and $14$ TeV.
	In the left 	panel $m_{h_i}= 125$ GeV, and in the right panel $m_{h_i}= 300$ GeV. The scalar potential parameters are chosen such that ${\Lambda}_i = 2 \pi$.}
	\label{fig:fig4}
\end{figure}

\begin{figure}[ht!]
    \hspace{-1.4cm}
    \includegraphics[width=0.57\textwidth]{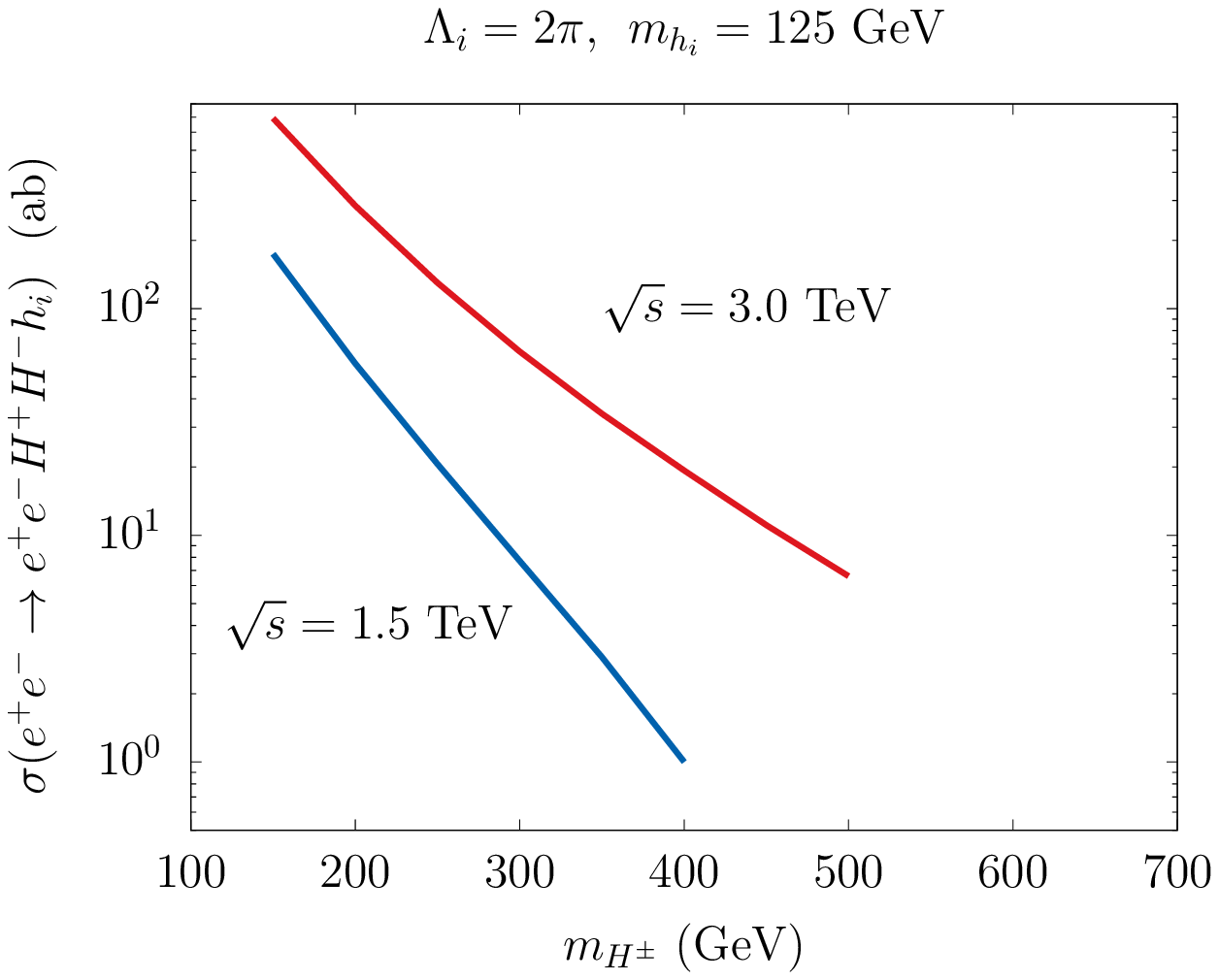} \hspace{-1.4cm}
    \includegraphics[width=0.57\textwidth]{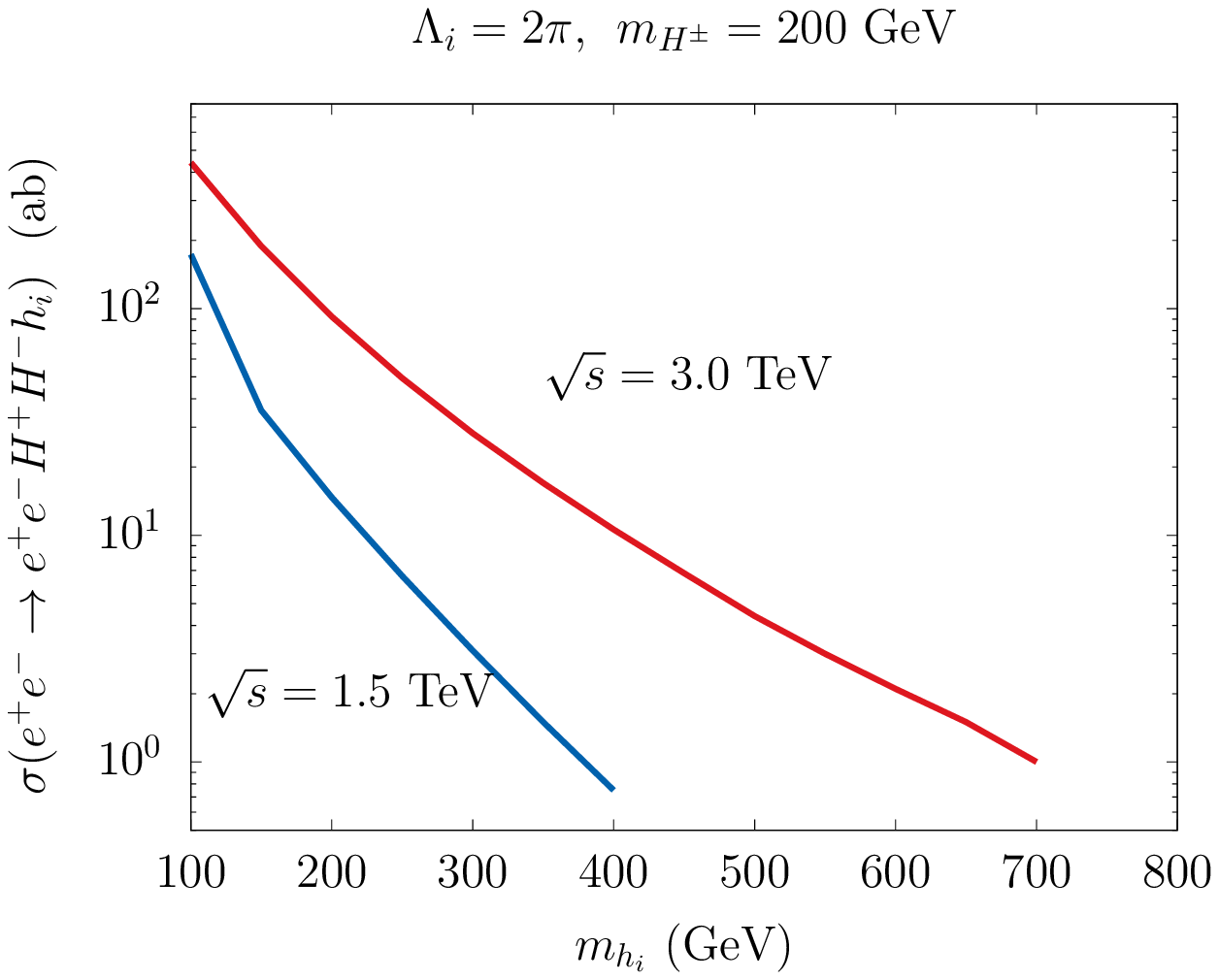}
	\caption{\small $\sigma (e^+ e^- \to e^+ e^-  H^+ H^- h_i)$ via $\gamma \gamma$ fusion as a function of the charged Higgs mass (left) and of the neutral Higgs mass (right) 
	for two CM energies of $\sqrt{s} = 1.5$ TeV and $\sqrt{s} = 3$ TeV.  The scalar potential parameters are chosen such that ${\Lambda}_i = 2 \pi$.}
	\label{fig:fig5}
\end{figure}

\begin{figure}[t!]
    \hspace{-1.8cm}
    \includegraphics[width=0.6\textwidth]{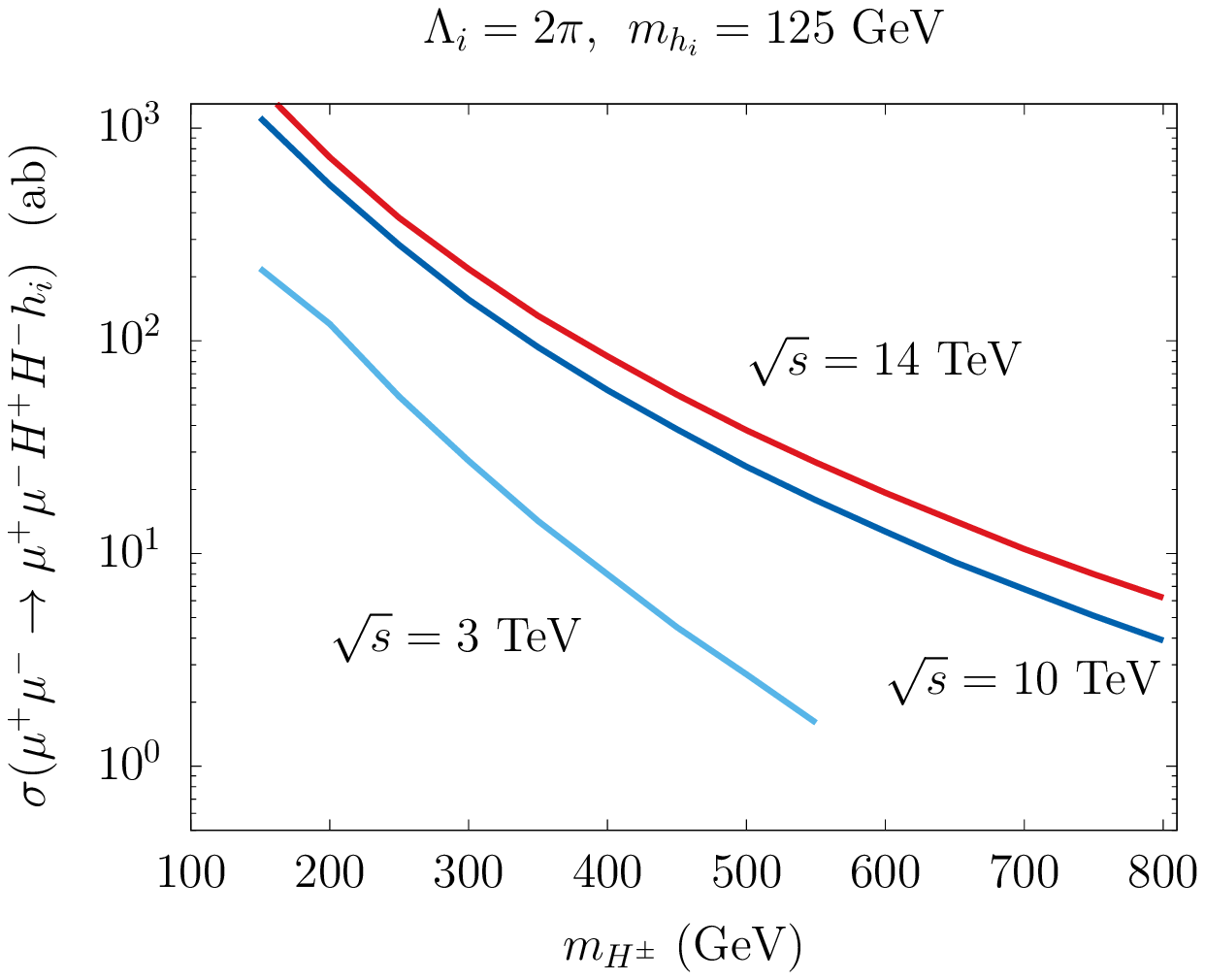} \hspace{-1.8cm}
    \includegraphics[width=0.6\textwidth]{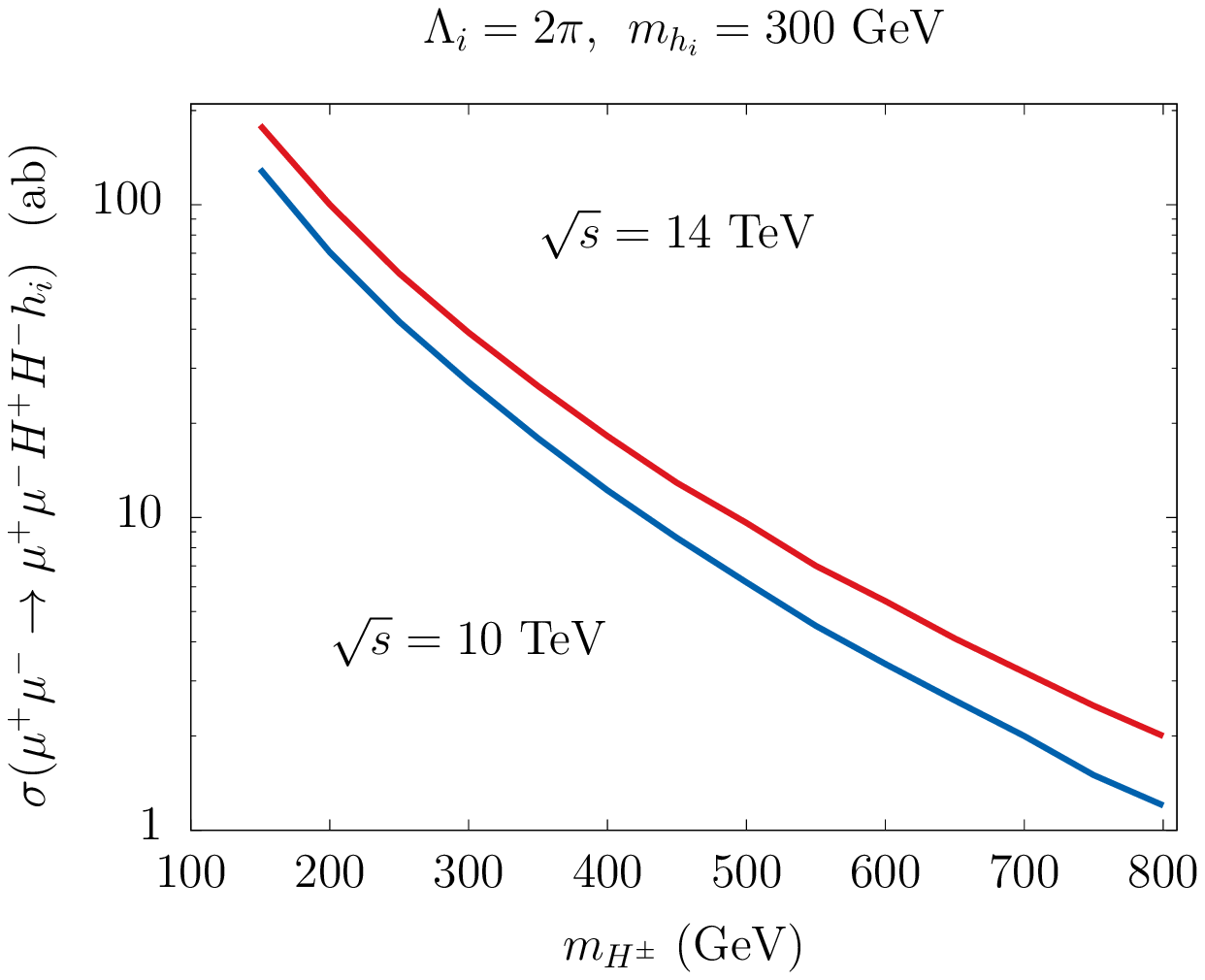}
	\caption{\small $\sigma (\mu^+ \mu^- \to \mu^+ \mu^- H^+ H^- h_i)$ as a function of the charged Higgs mass for three CM energies of $\sqrt{s} = 3$, $10$ and $14$ TeV.  In the left
	panel $m_{h_i}= 125$ GeV, and in the right panel $m_{h_i}= 300$ GeV.  The scalar potential parameters are chosen such that ${\Lambda}_i = 2 \pi$.}
	\label{fig:fig6}
\end{figure}

\begin{figure}[t!]
   \hspace{-1.4cm}
    \includegraphics[width=0.55\textwidth]{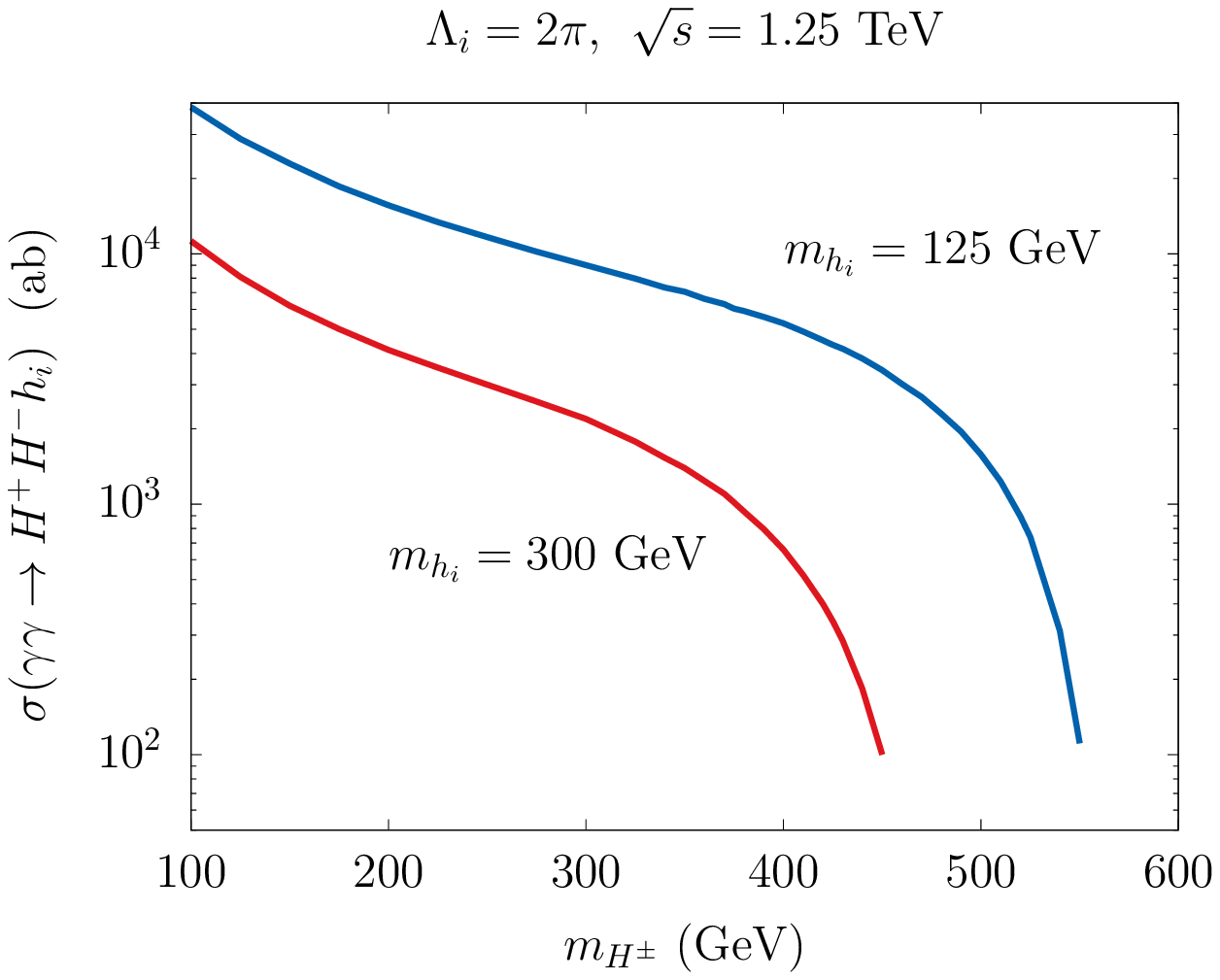} \hspace{-1.cm}
    \includegraphics[width=0.55\textwidth]{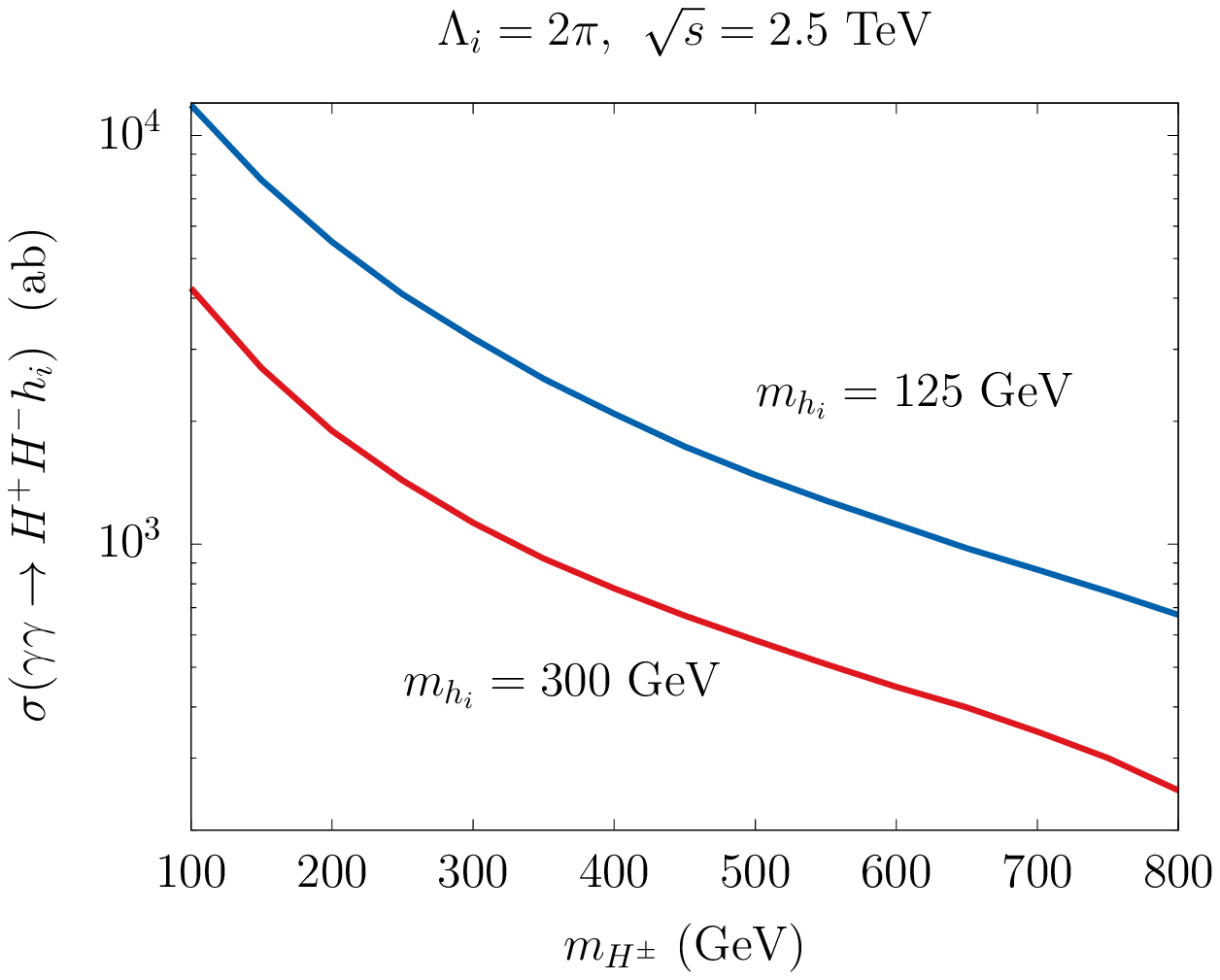}
	\caption{\small The cross section $\sigma (\gamma \gamma \to H^+ H^- h_i)$ as a function of the charged Higgs mass for two CM energies,   $\sqrt{s} = 1.25$ TeV (left panel)
	and  $\sqrt{s} = 2.5$ TeV (right panel), for two values for the neutral Higgs masses, $m_{h_i} = 125$ GeV and $m_{h_i} = 300$ GeV.  
 The scalar potential parameters are chosen such that ${\Lambda}_i = 2 \pi$.}
	\label{fig:fig2}
\end{figure}

In Figs.~\ref{fig:fig3}--\ref{fig:fig2}, we present cross sections for the production of  $H^+ H^- h_i$ final states.   In order to 
confirm the existence of P-even, CP-violating phenomena (if present), we shall focus primarily on processes that include $h_2$ or $h_3$ in the final state.
If such channels are detected, then it will also be possible to observe the $H^+ H^- h_1$ final state. 
Note that the production cross section for $h_1$ is proportional to the factor ${\Lambda}_1$, 
which provides us with a benchmark cross section for a final state with at least one known particle.  

In Fig.~\ref{fig:fig3}, we plot the cross sections, $\sigma (e^+ e^- \to  e^+ e^-  H^+ H^- h_i)$, $\sigma (\mu^+ \mu^- \to  \mu^+ \mu^-  H^+ H^- h_i)$ and $\sigma (\ell^+ \ell^- \to  H^+ H^- h_i)$,
as a function of the CM energy. In the left panel we have chosen a neutral scalar boson with $m_{h_i} = 125$ GeV and a charged Higgs boson with $m_{H^\pm}= 150$~GeV.
 For $i=1$, the corresponding plot refers to the production of the SM-like Higgs boson.
For $i=2$ and 3, the same plot refers to the production of the scalar $h_i$ of mass 125 GeV, assuming that $\Lambda_i=2\pi$.  Although we do not expect either $h_2$ and $h_3$ to
be (approximately) degenerate in mass with $h_1$,\footnote{Indeed, this possibility of an approximate mass degeneracy is either excluded based on present LHC Higgs data or will be excluded by the time
the higher energy lepton colliders are operational~\cite{Gunion:2012he,Ferreira:2012nv,Heikkila:2015xzp,Bian:2017gxg}.} we exhibit these figures to provide the reader with a sense of how large the cross sections of interest may be.
 In the right panel, the masses
are chosen to be $m_{h_i} = m_{H^\pm} = 300$ GeV. The parameters of the potential are ${\Lambda}_i = 2 \pi$. As expected, the
first two cross sections that occur mainly via $\gamma \gamma$ fusion, grow with the collider energy as  $\ln^2 (s/m_\ell^2)$. 
Taking into account only the leading term in the Equivalent Photon Approximation, which scales as $\ln^2 (s/m_\ell^2)$, the ratio of the electron to muon cross section yields 2.5
for $\sqrt{s} = 1$ TeV and 2.1 for $\sqrt{s} = 10$ TeV.  
The $t$-channel and $s$-channel cross sections are complementary to each other giving us access to the final state $ H^+ H^- h_i$ at both the low and high energy ends. 
Note that even with the coupling constants as large as ${\Lambda}_i = 2 \pi$,  the maximum value for the  $s$-channel cross section for
$m_{h_i} = m_{H^\pm} = 300$ GeV is roughly 200 ab and the corresponding maximum value for $\gamma \gamma$ fusion cross section is below 100 ab for $e^+ e^-$
and below 50 ab for $\mu^+ \mu^-$ processes. 
Hence, if both the neutral and the charged Higgs bosons are simultaneously heavy, it is unlikely that we will be able to detect these final sates. In the next plots we present in more
detail how the different cross sections vary with the scalar masses.

In Fig.~\ref{fig:fig4} we exhibit the cross section $\sigma (\ell^+ \ell^- \to  H^+ H^- h_i)$  as a function of the charged Higgs mass for four CM energies of $\sqrt{s} = 1.5$, $3$, $10$ and $14$ TeV. 
This covers the energy ranges of
both CLIC and the muon collider. Note that for the $s$-channel the $e^+e^-$ and $\mu^+ \mu^-$ cross sections have the same values. In the left panel we have set $m_{h_i}= 125$ GeV, and in the right panel $m_{h_i}= 300$ GeV.
Clearly there is a wide range of charged Higgs masses that can be probed for all collider energies. 

In Fig.~\ref{fig:fig5} we present the cross section $ \sigma (e^+ e^- \to e^+ e^-  H^+ H^- h_i)$ via $\gamma \gamma$ fusion at CLIC as a function of the charged Higgs mass in the left panel and as a function of the neutral Higgs mass in the right panel, 
for two CM energies of $\sqrt{s} = 1.5$ and 3 TeV. 
Again we choose  ${\Lambda}_i = 2 \pi$. 
We shall insist that the corresponding cross sections must exceed (roughly) 10~ab in order that the SM-like Higgs boson  
can be detected in association with a charged Higgs boson pair at CLIC.
In light of the results exhibited in the left panel of Fig.~\ref{fig:fig5}, it follows that $m_{H^\pm}$ can be at most $300$~GeV for $\sqrt{s}=1.5$~TeV and
500 GeV for $\sqrt{s}= 3$~TeV.  This behavior is expected
because at these CM energy ranges the $s$-channel production process provides the dominant contribution.  In the right panel of Fig.~\ref{fig:fig5}, 
 if the charged scalar has a mass of 200 GeV, then the neutral scalar masses need to be less than
about 200 GeV for $\sqrt{s}=1.5$~TeV and less than about 400 GeV for $\sqrt{s}= 3$~TeV.

In Fig.~\ref{fig:fig6} we show the cross section $\sigma (\mu^+ \mu^- \to \mu^+ \mu^- H^+ H^- h_i)$ as a function of the charged Higgs mass for three CM energies of $\sqrt{s} = 3$, $10$ and $14$ TeV.  
In the left panel $m_{h_i}= 125$ GeV, and in the right panel $m_{h_i}= 300$ GeV and the relevant parameters of the potential are set to ${\Lambda}_i =2 \pi$. 
Again we see that a large number of events will be produced if one of the Higgs bosons is light but detection will be difficult if both the neutral and the charged Higgs bosons are heavy.

There are currently no plans to alter the CLIC design to permit it to run in a $\gamma\gamma$ collider mode.  Nevertheless, it is instructive to have some measure of what to expect at such a 
facility.  Here, we assume following Ref.~\cite{Burkhardt:2002vh} that the $\gamma \gamma$ CM energy is
 approximately 80\% of the $e^+e^-$ CM energy, i.e.~roughly 1.25 and 2.5 TeV, with a luminosity of the order of 10\% of the CLIC design luminosity.\footnote{The $\gamma\gamma$ collider luminosity assumed in Ref.~\cite{Burkhardt:2002vh} may be somewhat optimistic.  A more recent study of multi-TeV $\gamma\gamma$ colliders carried out in Ref.~\cite{Barzi:2022rax} examines the possibility of employing a free electron laser in the construction of a high luminosity $\gamma\gamma$ collider with a CM energy in the multi-TeV regime.  For example, Table~4 of Ref.~\cite{Barzi:2022rax} imagines an $e^+ e^-$ collider with $\sqrt{s}=3$~TeV, which yields $\gamma\gamma$ collisions with $E_\gamma=1.38$~TeV and a luminosity of $L_{\gamma\gamma}\leq 0.06\, L_{ee}$, corresponding to a $\gamma\gamma$ collider with a slightly higher CM energy and a slightly lower luminosity, as compared to $\sqrt{s_{\gamma\gamma}}=2.5$~TeV and $L_{\gamma\gamma}= 0.1\, L_{ee}$ discussed in Ref.~\cite{Burkhardt:2002vh}.   An alternative approach has been investigated in 
Ref.~\cite{Ginzburg:2020koq}, which examines a $\gamma\gamma$ collider based on an $e^-e^-$ linear collider with $\sqrt{s}=2$~TeV.   These authors claim a maximal photon energy of $E_\gamma=0.95$~TeV but with a luminosity of $L_{\gamma\gamma}=0.02\, L_{ee}$.  Such a collider would yield an integrated luminosity of roughly 0.1 ab$^{-1}$ over a ten year program, which would be insufficient to study P-even, CP-violating phenomena over most of the scalar mass range of interest.   We thank the referee for alerting us to these two references cited above.}
We show in Fig.~\ref{fig:fig2} the cross section for the scattering of two on-shell photons, $\sigma (\gamma \gamma \to H^+ H^- h_i)$, 
as a function of the charged Higgs mass for two CM energies $\sqrt{s} = 1.25$ TeV (left panel) and  $\sqrt{s} = 2.5$ TeV (right panel), for two
values for the neutral Higgs masses, $m_{h_i} = 125 $ GeV and $m_{h_i} = 300 $ GeV. 
In light of \eq{Lambdas}, it follows that
$\sigma (\gamma \gamma \to H^+ H^- h_i)$ is proportional to $\Lambda_i^2$. 
The cross sections shown in the plots for $\gamma\gamma$ fusion are all above 100 ab, and with the planned luminosity for CLIC the $\gamma\gamma$ collider would provide a larger number of 
signal events than the corresponding signals anticipated at the lepton colliders
discussed in this section.  However, the CM energy of the $\gamma\gamma$ collider will 
restrict the observable scalar masses accessible in the production of 
$H^+ H^- h_i$ to no more than a few hundred GeV.

In light of our considerations of the $\gamma\gamma$ fusion processes at lepton colliders above, a brief comment on $\gamma\gamma$ physics at the LHC is warranted. The ATLAS Collaboration has reported~\cite{Aad:2020lsx} the observation of photon-induced $W^+ W^-$
production in $pp$ collisions at $\sqrt{s} = 13$ TeV with 139 fb$^{-1}$ of data in the process $\gamma \gamma \to W^+ W^- \to e^\pm \mu^\mp \, + \, \slashed{E}_T $,
with an observed cross section of $3.13 \pm 0.31 (\text{stat}) \pm 0.28 (\text{syst})$~fb.
In light of this measurement, it is of interest to consider whether $\gamma \gamma \to H^+ H^- $ is within
the reach of the LHC program. The cross section for $pp \to pp H^+ H^-$ via photon fusion was calculated in Ref.~\cite{Lebiedowicz:2015cea} 
and shown to be  about 0.1 fb for a charged Higgs mass of 150 GeV at $\sqrt{s} = 14$ TeV.
One must also account for the charged Higgs boson decay channels, while contending with a significant
$\gamma \gamma \to W^+ W^-$ irreducible background with a much larger cross section. 
Therefore, although not impossible, it will be very difficult to detect charged Higgs bosons produced via $\gamma \gamma$ fusion at the LHC.
If in addition, one demands that a neutral
Higgs boson is emitted from one of the final state (virtual) charged Higgs bosons, then the resulting $\gamma\gamma\to H^+ H^- h_i$ cross section will be too small to be detected at the LHC (even with the anticipated 2~ab dataset at the HL-LHC).

Finally, we discuss the detection prospects of the final states produced by the processes analyzed in this section. To reiterate, we are assuming the exact Higgs alignment limit where the tree-level properties of $h_1$ coincide precisely with those of the observed SM-like Higgs boson.
In the A2HDM, the couplings of the up- and down-type quarks to $h_2$, $h_3$ and $H^\pm$ are exhibited in \eq{YUK4}.  In the Higgs alignment limit the
couplings to the neutral scalars, $h_2$ and $h_3$, reduce to the form shown in \eq{A2HDMalign} [whereas the corresponding couplings to $H^\pm$ shown in \eq{YUK4} are unchanged].  
The corresponding couplings of the leptons are obtained by replacing the down-type [up-type] quark fields with the corresponding charged lepton [neutrino] fields,
replacing $M_D$ by the diagonal lepton mass matrix and replacing $M_U$ with the zero matrix.
The charged Higgs boson can be detected via one of its decay channels: $W^\pm h_2$, $W^\pm h_3$, or a fermion pair, as discussed below.\footnote{Note that the $H^\mp W^\pm h_1$ coupling vanishes in the Higgs alignment limit.}

Considering that all Yukawa couplings can be independently chosen in flavor-aligned extended Higgs sectors, one must take into account all final states with quark pairs of different generations. Assuming a good $b$ and $c$ 
tagging efficiency and denoting light quark ($u$, $d$, $s$) jets collectively by $j$, the final states resulting from the $H^\pm$ decay to a fermion pair will be combinations of $tb$ (or $tj)$, $cb$ (or $cj$ or $jb)$, or $\tau^\pm$ and missing transverse energy. 
At a lepton collider, one can employ the hadronic decay modes of
 the charged Higgs boson to reconstruct its mass.
At tree level, the other
important decay channels are $H^\pm \to W^\pm h_2$ and $H^\pm \to W^\pm h_3$ with rates proportional to square of the SU(2)$_L$ electroweak coupling, $g^2$. 
At one loop, the subdominant decay mode $H^\pm \to W^\pm \gamma$ could eventually be utilized if a sufficiently large data sample were available.

The
neutral Higgs bosons $h_2$ and $h_3$ do not couple to pairs of gauge bosons in the exact Higgs alignment limit and hence do not decay (at tree level) to $W^+ W^-$ and $ZZ$. 
If kinematically allowed, the decays $h_2 \to W^\pm H^\mp$, $Z h_3$ and $h_3 \to W^\pm H^\mp$, $Z h_2$ have to be considered particularly because their rates are proportional to $g^2$. 
The tree-level decay rates of the neutral scalars $h_2$ and $h_3$ to fermion pairs will be dominated by third-generation final states, $t\bar{t}$, $b\bar{b}$ and $\tau^+\tau^-$.
In addition, the one-loop neutral Higgs decay modes to a pair of gluons may also be competitive.  
The subdominant one-loop neutral Higgs decay modes to $\gamma \gamma$, $Z\gamma$ could eventually be utilized if a sufficiently large data sample were available.

\section{P-even, CP-violating signals via loop effects}
\label{sec:ZZZ}

An indirect way to examine the set of Higgs couplings exhibited in \eq{eq:CPVobs3}
that yield P-even, CP-violating signals considered in Sections~\ref{section3} and \ref{sec:collider} is to 
probe the same set of couplings via the Higgs loop contributions to the $ZZZ$ and $ZW^+ W^-$form factors.\footnote{The $ZZ\gamma$ and $Z\gamma\gamma$ form factors cannot be used for this purpose, as the only scalar loop contributions involve the charged Higgs boson, which does not contribute to the P-even, CP-violating form factor.}
CP-violating contributions to the triple gauge boson vertices $ZZZ$ and $ZW^+W^-$ vertices were first studied 
in Refs.~\cite{Hagiwara:1986vm,Gounaris:1999kf,Gounaris:2000dn,Baur:2000ae, Grzadkowski:2016lpv}.
For example, the general Lorentz structure of the $Z(P)W^+(q) W^-(\bar{q})$ vertex (with all momenta pointing into the vertex) given in Ref.~\cite{Hagiwara:1986vm} is
\beqa
\Gamma_V^{\alpha \beta \mu}(q, \bar{q}, P)&=& f_{1}^V (\bar{q}-q)^{\mu} g^{\alpha \beta}-\frac{f_{2}^V}{m_{W}^{2}}(\bar{q}-q)^{\mu} P^{\alpha} P^{\beta}+f_{3}^V\left(P^{\alpha} g^{\mu \beta}-P^{\beta} g^{\mu \alpha}\right) \nonumber \\
&&+i f_{4}^V\left(P^{\alpha} g^{\mu \beta}+P^{\beta} g^{\mu \alpha}\right)+i f_{5}^V \epsilon^{\mu \alpha \beta \rho}(\bar{q}-q)_{\rho} \nonumber  \\
&&-f_{6}^V \epsilon^{\mu \alpha \beta \rho} P_{\rho}-\frac{f_{7}^V}{m_{W}^{2}}(\bar{q}-q)^{\mu} \epsilon^{\alpha \beta \rho \sigma} P_{\rho}(\bar{q}-q)_{\sigma}.\label{f}
\eeqa
The three form factors proportional to the Levi-Civita tensor are P-violating.  
One can easily
verify that $f_{1,2,3}^V$ separately conserve C, P and T, whereas $f_4^V$ is the unique form factor that is P-conserving and CP-violating.  This form factor is generated by the bosonic sector of the extended Higgs sector provided that the scalar potential and/or vacuum is CP-violating.   In light of \eq{eq:CPVobs3}, it follows that in the exact Higgs alignment limit, a nonzero scalar contribution to $f_4$ requires at least three neutral scalars beyond the SM-like Higgs boson.\footnote{Note that there are two triangle diagrams with internal scalars that contribute at one-loop order to the $ZW^+W^-$ form factors, consisting of an $H^+ H^- h_j$ and an $h_j h_k H^+$ loop, with corresponding  $ZH^+ H^-$ and $Zh_j h_k$ vertices, respectively.  Only the latter can contribute to the P-even, CP-violating form factor $f_4$.}

There are several calculations in the literature of $f_4$
in the complex 2HDM~\cite{Grzadkowski:2016lpv, Belusca-Maito:2017iob}, in a CP-violating 3HDM with two inert doublets~\cite{Cordero-Cid:2016krd}, and in a model with two Higgs doublets and a singlet~\cite{Azevedo:2018fmj}.   Note that in the 2HDM, the contribution of the neutral scalars vanishes in the exact Higgs alignment limit. Hence, an
extended Higgs sector with at least three neutral scalars beyond the SM-like Higgs boson $h_1$ is required to yield a nonzero value of $f_4$ as noted above.
In practice, the maximal values
for $f_4$ are, however, still an order of magnitude\footnote{The contributions to $f_4$ in the model with two Higgs doublets and a singlet are suppressed by an order of magnitude as compared to the CP-violating 3HDM with two inert doublets, due to the presence of 
fewer number of inert states contributing to the $ZZZ$ loop, along with diluted $Zh_ih_j$ couplings (since the singlet has no direct couplings to the SM gauge bosons).} away from the experimentally measured values~\cite{Aaboud:2017rwm, Sirunyan:2017zjc}.
Future projection for the measurement $f_4$ can be found in Ref.~\cite{CMS:2018qgz} for the HL-LHC and in Ref.~\cite{Ogawa:2018ssv} for the International Linear Collider (ILC).

The observation of a nonzero value of $f_4$ would provide unambiguous evidence for the existence of P-even CP violation.   The main disadvantages of this observable is that it is loop suppressed and is  subject to significant systematic uncertainties that must be overcome in any realistic experiment.  On the other hand, $f_4$ is sensitive to a wider class of Higgs sectors than those that can be probed by observables examined in Section~\ref{sec:collider}.   For example, in models with two or more inert doublets,
the corresponding scalar potential can contain CP-violating terms that generate neutral scalars of indefinite CP, as was shown initially in Ref.~\cite{Cordero-Cid:2016krd} and examined further in Refs.~\cite{Cordero-Cid:2018man,Azevedo:2018fmj,Keus:2019szx,Cordero-Cid:2020yba}.  The P-even, CP-violating signals arising at tree level from such models cannot be probed by the methods discussed in Sections~\ref{section3} and~\ref{sec:collider}  as noted below \eq{eq:CPVobs3}.  Nevertheless,
these inert scalars can appear in the loops that contribute to the $ZZZ$ and $ZW^+ W^-$ form factor (since every vertex of the loop diagram involves a coupling of a pair of $\mathbb{Z}_2$-odd scalars) thereby yielding a nonzero result for~$f_4$.  Further details will be left to a future study.

\section{Final state photons as a diagnostic for CP violation}
\label{sec:photons}

In our study of P-even, CP-violating phenomena at colliders, we relied on tree-level production processes at lepton colliders involving the Higgs bosons and gauge bosons.  
In such cases, any P-odd, CP-violating contribution to the production process arising from the Yukawa sector would be loop suppressed and hence unlikely to have a significant impact on the interpretation of the source of CP violation.  In some cases, the Higgs bosons produced could decay into either two lighter Higgs bosons or into a Higgs boson and a gauge boson. In such cases, any P-odd, CP-violating contribution to the decay process arising from the Yukawa sector would again be loop suppressed.  Ultimately the Higgs bosons produced, either directly in the initial production process and/or in the decay chain of the produced Higgs boson, are identified experimentally via their fermionic decay channels.  However, the presence of these fermionic channels does not obscure the original tree-level bosonic couplings involved in the original Higgs production and the bosonic Higgs decay processes. Thus the interpretation of the P-even, CP-violating signal based on the bosonic Higgs vertices is not compromised.

If one is willing to tolerate branching ratio suppressions associated with the rarer one-loop mediated Higgs decay processes, then in principle one could have additional channels to probe the presence of P-even CP violation in the scalar sector.   In such cases, P-odd CP violation due to a fermion loop would compete at the same order with the P-even CP violation due to a bosonic loop.  In this section, we explore this possibility in more detail by examining the impact of Higgs decay processes involving final state photons and $Z$ bosons.

\subsection{Higgs boson decays to two photons}
\label{multiphoton}

In the absence of Yukawa couplings of the Higgs bosons, the simultaneous observations of the interactions ($i\neq j$, $i$, $j\neq 1$)  
\beq \label{eq:CPVobs3p}
h_i \to \gamma\gamma\,\,{\rm or}\,\,Z\gamma\,, \quad h_j \to  \gamma\gamma\,\,{\rm or}\,\,Z\gamma\,, \quad \text{and the $Z h_ih_j$ vertex}\,,
\eeq
would also constitute a signal of P-even CP violation.   Once again, by taking all scalars to be P-even, 
the existence of the
$Z h_i h_j$ vertex in a CP-conserving theory
would imply  that $h_i$ and $h_j$ have opposite C quantum numbers, whereas the observation of a $\gamma\gamma$ or $Z\gamma$ decay 
would imply that the corresponding neutral scalar is C even.    Due to the suppressed rate for the decay of a scalar to $Z\gamma$ relative to $\gamma\gamma$,
we henceforth focus only on the $\gamma\gamma$  decay mode of $h_i$ (although the corresponding analysis for the $Z\gamma$ decay mode is similar).

In the exact Higgs alignment limit, the $\gamma\gamma$ decay is mediated by a charged Higgs loop (since the $h_i W^+ W^-$ vertex is absent for $i\neq 1$).   
Phenomenologically, the $h_{i,j} \gamma\gamma$ interactions may have a better discovery potential than the 
signals discussed in Section~\ref{sec:collider},
despite the loop-suppression of its coupling strength. In particular, the decays $h_{i,j} \to \gamma\gamma$ are not kinematically suppressed, and the two photon signal can easily be identified experimentally as a resonance in the diphoton-invariant mass spectrum. In addition, one can probe these interactions via photon-photon scattering in a future $e^+e^-$ linear collider or a photon collider.
The observation of the three processes listed in \eq{eq:CPVobs3p} requires the discovery of two new neutral Higgs bosons $h_i$ and $h_j$ ($i\neq j\neq 1$) in addition to the discovered SM-like Higgs boson~$h_1$. Neither $h_i$ nor $h_j$ can be identified as $h_1$ since in the exact Higgs alignment limit the interaction $Z h_i h_1$ ($i\neq 1$) vanishes.

It appears that the utility of \eq{eq:CPVobs3p} as a signal of CP violation is spoiled once the effects of the Yukawa couplings are included.   For example, in the CP-conserving 2HDM where $h_2$ and $h_3$ are CP-even and CP-odd respectively, both $h_2\gamma\gamma$ and $h_3\gamma\gamma$ couplings are generated at one loop mediated by fermions.   The effective Lagrangian that governs these couplings is,
\beq \label{Leff}
\mathscr{L}_{\rm eff}=g_2 h_2 F_{\mu\nu}F^{\mu\nu}+g_3h_3 \epsilon_{\mu\nu\alpha\beta}F^{\mu\nu}F^{\alpha\beta}\,.
\eeq
The two operators 
above can be experimentally distinguished.  For example, the corresponding decay amplitudes have difference dependences on the photon polarizations 
$\boldsymbol{\varepsilon_1}$ and $\boldsymbol{\varepsilon_2}$,
\beq \label{pol}
\mathcal{A}(h_2\to\gamma\gamma)\sim \boldsymbol{\varepsilon_1\cdot\varepsilon_2}\,,\qquad\quad
\mathcal{A}(h_3\to\gamma\gamma)\sim \boldsymbol{\varepsilon_1\times\varepsilon_2}\,.
\eeq
In contrast, the one-loop couplings of a scalar $\phi$ to $\gamma\gamma$ mediated by a charged Higgs boson can only produce an effective coupling $\phi F_{\mu\nu}F^{\mu\nu}$.
Consequently, the observation of a $Zh_2 h_3$ coupling in conjunction with evidence for the presence of the effective couplings $h_2 F_{\mu\nu}F^{\mu\nu}$ and $h_3 F_{\mu\nu}F^{\mu\nu}$ would constitute a signal of P-even CP violation.

\subsection{Higgs boson decays to multivector-boson final states}
\label{multivectors}

Consider the implications of the simultaneous observation of the decays,
\beq \label{eq:CPVobs4}
h_i \to  \gamma\gamma, \quad \mathrm{and} \quad h_i \to \gamma\gamma\gamma\,.
\eeq
Since $\gamma\gamma$ is C-even and $\gamma\gamma\gamma$ is C-odd, it follows that, in the absence of Higgs boson couplings to fermions, the
simultaneous observation the loop-induced decays exhibited in \eq{eq:CPVobs4}
 would be a signal for P-even CP violation.  It is noteworthy that this CP-violating observable
 requires the discovery of only one new Higgs boson $h_i$ (in contrast to the signals
 of Section~\ref{section3} that rely on the $Z h_i h_j$ coupling).
However, an analysis presented in Appendix~\ref {3gamma} reveals that the $h_i \to \gamma\gamma\gamma$ decay amplitude vanishes at the one-loop level (independently of the presence or absence of Higgs boson couplings to fermions).   Thus, the decay rate for $h_i\to\gamma\gamma\gamma$ relative to $h_i\to\gamma\gamma$ is suppressed by both a loop factor and the three-body phase space suppression. 
Thus in practice, the detection of CP violation via this scenario is not viable.

\begin{figure}[t!]
\centering
\scalebox{0.45}{
  \begin{picture}(320,0)(320,140)
    \SetWidth{1.5}
        \SetArrowScale{1.5}
    \DashArrowLine(320,-17)(400,31){8}
    \DashArrowLine(400,31)(400,-65){8}
    \DashArrowLine(400,-65)(320,-17){8}
    \Photon(400,127)(256,63){7.5}{5}
    \DashLine(320,-17)(256,63){8}
    \DashLine(160,63)(256,63){8}
    \Photon(400,31)(515,31){7.5}{5}
    \Photon(400,-65)(515,-65){7.5}{5}
    \Text(130,55)[lb]{\huge{$h_2$}}
    \Text(355,-25)[lb]{\huge{$H^+$}}    
    \Text(415,120)[lb]{\huge{$Z$}}    
    \Text(255,10)[lb]{\huge{$h_3$}}        
    \Text(535,20)[lb]{\huge{$\gamma$}}   
    \Text(535,-75)[lb]{\huge{$\gamma$}}
    \Text(190,-90)[lb]{\Huge{(a)}}
  \end{picture}
  }
\scalebox{0.45}{
  \begin{picture}(180,0)(180,140)
    \SetWidth{1.5}
        \SetArrowScale{1.5}
    \DashArrowLine(320,-17)(400,31){8}
    \DashArrowLine(400,31)(400,-65){8}
    \DashArrowLine(400,-65)(320,-17){8}
  \Photon(400,127)(256,63){7.5}{5}
    \DashLine(320,-17)(256,63){8}
    \DashLine(160,63)(256,63){8}
    \Photon(400,31)(515,31){7.5}{5}
    \Photon(400,-65)(515,-65){7.5}{5}
    \Text(130,55)[lb]{\huge{$h_3$}}
    \Text(355,-25)[lb]{\huge{$H^+$}}    
    \Text(415,120)[lb]{\huge{$Z$}}    
    \Text(255,10)[lb]{\huge{$h_2$}}        
    \Text(535,20)[lb]{\huge{$\gamma$}}   
    \Text(535,-75)[lb]{\huge{$\gamma$}}
    \Text(190,-90)[lb]{\Huge{(b)}}
  \end{picture}
  }
\vspace{4cm}
\caption{\small Sample one-loop diagrams contributing to the decay process $h_{2,3}\to Z\gamma\gamma$.}
\label{fig:h3togagaZ}
\end{figure}
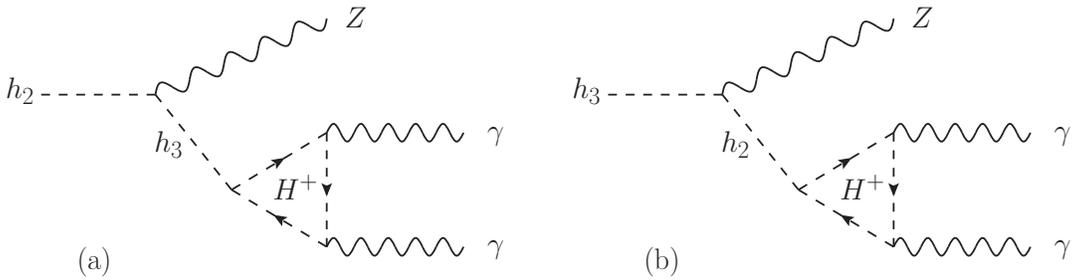

One can improve matters by replacing one or two of the photons with $Z$ bosons.  For example, consider  the implications of the simultaneous observation of the decays,
\beq \label{eq:CPVobs4p}
h_i \to \gamma\gamma~~{\rm or}~~Z \gamma, \quad \mathrm{and} \quad h_i \to Z\gamma\gamma\,.
\eeq
First suppose that the couplings of $h_i$ to fermions are absent.   In this case, we can again make use of the analysis presented in Appendix~\ref {3gamma}, where one of the final state photons in Fig.~\ref{fig:hitogagaga} of Appendix~\ref{3gamma} is replaced by a $Z$ boson.  We again conclude that the contributions to the one loop matrix element for $h_i\to Z\gamma\gamma$ due to the sum of the diagrams analogous to those of Fig.~\ref{fig:hitogagaga} (as well as the corresponding diagrams where the charged Higgs loop is replaced by a charged $W$ loop) must exactly vanish for any choice of Higgs boson state $h_i$.    However, in contrast to $h_i\to\gamma\gamma\gamma$, additional one-loop diagrams shown in Fig.~\ref{fig:h3togagaZ} contribute to $h_i\to Z\gamma\gamma$.
These diagrams do not vanish.   Indeed, if P and C are separately conserved, then the diagrams shown in Fig.~\ref{fig:h3togagaZ}(b) contribute to the C and P conserving decay $h_3 \to Z\gamma\gamma$ while the one-loop decay $h_3\to \gamma\gamma/Z\gamma$ is forbidden, whereas the diagram for $h_2\to Z\gamma\gamma$  shown in Fig.~\ref{fig:h3togagaZ}(a) is forbidden while the one-loop decay $h_2\to \gamma\gamma/Z\gamma$ is allowed.  These results follow in light of the absence of an $h_3 H^+ H^-$ vertex when CP is conserved.   
Thus, the simultaneous observation of the decays of \eq{eq:CPVobs4p} would constitute a signal for P-even CP violation.  Moreover, both decays arise at one-loop order (although, admittedly, the $Z\gamma\gamma$ final state would be suppressed relative to the $\gamma\gamma$ and $Z\gamma$ final states due to a coupling constant and the three-body phase space suppression).

If the couplings of $h_i$ to fermions are present, then one has to reconsider the analysis in which the charged Higgs boson loop in Figs.~\ref{fig:h3togagaZ}(a) and (b) is replaced by a fermion loop.
In this case CP conservation does not forbid the simultaneous decay channels, $h_2\to Z\gamma$ or $h_3\to Z\gamma$.  Thus, isolating any P-even, CP-violating contribution to
the simultaneous observation of \eq{eq:CPVobs4p} would require a separation of the CP-even and CP-odd components of the $Z\gamma$ system analogous to the discussion below \eq{Leff}.

Finally, a similar analysis applies to 
\beq \label{eq:CPVobs4pp}
h_i \to \gamma\gamma~~{\rm or}~~Z \gamma,  \quad \mathrm{and} \quad h_i \to ZZ\gamma\,.
\eeq
In particular, one simply replaces one of outgoing photons in Fig.~\ref{fig:h3togagaZ} by a $Z$ boson.   Once again, this scenario is 
less useful than \eq{eq:CPVobs4p} due to the suppression of the effective $h_i Z\gamma$ vertex relative to that of $h_i\gamma\gamma$.

\section{Beyond the exact Higgs alignment limit}
\label{sec:beyond}

In previous sections, we focused on P-even, CP-violating signals in the bosonic sector of the 2HDM in the exact Higgs alignment limit.  In this section, we shall extend our considerations to the case where there is a departure from exact Higgs alignment.  The present Higgs data require that any departure from exact Higgs alignment should be small.  We shall identify new P-even, CP-violating signals that, although suppressed due to approximate Higgs alignment, would provide new observables that could be probed in future experiments.
Several observables for scalar sector CP violation that appear at tree level or one-loop level have been discussed in Refs.~\cite{Mendez:1991gp,Keus:2015hva,Fontes:2015xva} and Ref.~\cite{Grzadkowski:2016lpv}, respectively.  In contrast to the observables introduced in Section~\ref{section3}, these observables vanish in the exact Higgs alignment limit. 

It is noteworthy that in the exact Higgs alignment limit,  the observation of two tree-level processes is not sufficient
to identify the presence of P-even CP violation.  Indeed, in light of \eqst{s1}{eq:CPVobs3}, it is necessary to observe at least three tree-level processes to provide evidence of 
P-even CP violation.\footnote{Note that the processes exhibited in \eqst{eq:CPVobs4}{eq:CPVobs4pp} are loop induced and therefore exempt from the three processes requirement.} In contrast, if one also allows for Higgs alignment suppressed processes, then there are cases in which the observation of only two tree-level processes is enough to claim the detection of P-even CP violation. Examples of such cases are shown below. 

Given that $h_1$ is the SM-like Higgs boson (which is a CP-even scalar), our list of interactions in \eqst{s1}{eq:CPVobs3} expands to include the following.
\beqa
&& \hspace{-0.75in} \text{One Higgs alignment suppressed observables:}  \nonumber \\
&& \hspace{-0.5in}  1.~~h_i H^+ H^-\,,\, Z h_1 h_i\,, \label{onesup}
\quad \text{(for $i \neq 1 $)}, \label{s1-1-beyond} \\
&& \hspace{-0.5in}  2.~~h_i h_j h_j\,,\quad Z h_1 h_i \,,\quad \text{(for $i,j\neq 1$)}, \label{s2-1-beyond} \\
&& \hspace{-0.5in}  3.~~ h_1 h_i h_j  \,,\quad Z h_i h_j \,,\quad \text{(for $i,j\neq 1$ and $i\neq j$)}. \label{s3-1-beyond} \\
&& \hspace{-0.5in}  4.~~h_i h_1 h_1\,,\quad h_j H^+ H^- \,,\quad Zh_i h_j\,,\quad\text{(for $i,j\neq 1$ and $i\neq j$)}, \label{s4-1-beyond} \\
&& \hspace{-0.5in}  5.~~h_k h_1 h_1\,,\quad h_i h_j h_k \,,\quad Zh_i h_j\,,\quad\text{(for $i,j,k\neq 1$ and $i\neq j$)}, \label{s5-1-beyond} \\
&& \hspace{-0.5in}  6.~~h_i h_1 h_1\,,\quad h_j h_k h_k \,,\quad Zh_i h_j\,,\quad\text{(for $i,j,k\neq 1$ and $i\neq j$)}. \label{s6-1-beyond} \\
&& \hspace{-0.75in} \text{Two Higgs alignment suppressed observables:} \nonumber  \\
&& \hspace{-0.5in}  1.~~h_i h_1 h_1\,,\quad Z h_1 h_i \,,\quad \text{(for $i\neq 1$)}, \label{s1-2-beyond} \\
&& \hspace{-0.5in}  2.~~h_1 h_i h_j\,,\quad Z h_1 h_i \,,\quad h_j H^+ H^-\,,\quad\text{(for $i,j\neq 1$ and $i\neq j$)}, \label{s2-2-beyond} \\
&& \hspace{-0.5in}  3.~~h_i h_1 h_1\,,\quad h_j h_1 h_1\,, \quad Zh_i h_j\,,\quad\text{(for $i,j\neq 1$ and $i \neq j$)}, \label{s3-2-beyond}  \\
&& \hspace{-0.5in}  4.~~h_1 h_i Z\,,\quad h_1 h_j Z\,, \quad Zh_i h_j\,,\quad\text{(for $i,j\neq 1$ and $i \neq j$)},  \label{s4-2-beyond} 
\\
&& \hspace{-0.5in}  5.~~h_1 h_i h_j\,,\quad Z h_1 h_i \,,\quad h_j h_k h_k \,,\quad\text{(for $i,j,k\neq 1$ and $i \neq j$)}. \label{s5-2-beyond}  \\
&& \hspace{-0.75in} \text{Three Higgs alignment suppressed observables:} \nonumber  \\
&& \hspace{-0.5in}  1.~~h_i h_1 h_1 \,,\quad h_1 h_i h_j \,,\quad Zh_1 h_j\,,\quad\text{(for $i,j\neq 1$ and $i\neq j$)}. \label{s1-3-beyond}
\eeqa
The observation of any one of the combinations of observables listed in \eqst{onesup}{s1-3-beyond} would constitute a signal of P-even CP violation,
under the assumption that the one-loop corrections to the above observables due to fermion loops are subdominant to the 
tree-level contributions (after taking into account any suppressions due to approximate Higgs alignment).   If the fermion-loop contributions to the above
processes are competitive with the corresponding tree-level contributions, then the distinction between P-even and P-odd CP violation
becomes less clear, although the CP violation interpretation of the signal is still maintained.

If the $h_iZZ$ vertex is included then the presence or absence of fermion-loop contributions becomes critical to the interpretation of the signal.
In the absence of fermion-loop contributions, 
the C and~P quantum number listed in Table~\ref{candp} can be employed to conclude that the coupling of a CP-odd scalar to $ZZ$ would constitute evidence for
CP violation.   However, such a conclusion is no longer tenable if fermion-loop contributions are present, since dimension-five operators induced by
fermion loop contributions exhibited in \eq{Leff} demonstrate that both CP-even and CP-odd scalars can couple to $ZZ$ in a CP-conserving
theory.

For completeness, we list below additional combination of observables that can be interpreted as evidence of 
P-even CP violation in the absence of fermion-loop contributions:
\beqa
&& \hspace{-0.75in} \text{One Higgs alignment suppressed observables:}  \nonumber \\
&& \hspace{-0.5in}  1.~~h_i ZZ\,,\quad h_j H^+ H^-\,, \quad\text{(for $i,j\neq 1$ and $i \neq j$) \quad [only in the 2HDM]}, \label{s1p-1-beyond} \\
&& \hspace{-0.5in}  2.~~h_i ZZ\,,\quad h_j H^+ H^-\,, \quad Z h_i h_j\,, \quad\text{(for $i,j\neq 1$ and $i \neq j$)}, \label{s2p-1-beyond} \\
&& \hspace{-0.5in}  3.~~ h_i ZZ\,,\quad h_j h_k h_k\,,\quad  Zh_i h_j\,, \quad\text{(for $i,j,k\neq 1$ and $i \neq j$)}, \label{s3p-1-beyond}\\
&& \hspace{-0.5in}  4.~~h_k ZZ\,, \quad h_i h_j h_k\,,\quad Zh_i h_j \,, \quad \text{(for $i,j,k\neq 1$ and $i \neq j\neq k$)}. \label{s4p-1-beyond} \\
&& \hspace{-0.75in} \text{Two Higgs alignment suppressed observables:} \nonumber  \\
&& \hspace{-0.5in}  1.~~h_i ZZ\,, \quad Z h_1 h_i\,,  \quad \text{(for $i \neq 1 $)}, \label{s1p-2-beyond} \\
&& \hspace{-0.5in}  2.~~h_i ZZ\,,\quad  h_jh_1 h_1 \,,\quad\text{(for $i,j\neq 1$ and $i\neq j$) \quad [only in the 2HDM]}, \label{s2p-2-beyond} \\
&& \hspace{-0.5in}  3.~~h_i ZZ\,, \quad h_j ZZ \,, \quad \text{(for $i,j \neq 1$ and $i \neq  j$) \quad [only in the 2HDM]}, \label{s3p-2-beyond} 
\\
&& \hspace{-0.5in}  4.~~h_i ZZ\,,\quad h_j ZZ\,, \quad Z h_i h_j\,, \quad\text{(for $i,j\neq 1$ and $i \neq j$)}, \label{s4p-2-beyond} 
\\
&& \hspace{-0.5in}  5.~~h_i ZZ\,,\quad  h_jh_1 h_1\,,\quad Z h_i h_j \,,\quad\text{(for $i,j\neq 1$ and $i\neq j$)}. \label{s5p-2-beyond} \\
&& \hspace{-0.75in} \text{Three Higgs alignment suppressed observables:} \nonumber  \\
&& \hspace{-0.5in}  1.~~h_i ZZ\,, \quad h_1 h_i h_j\,,\quad Zh_1 h_j \,, \quad \text{(for $i,j\neq 1$ and $i \neq j$)}. \label{s1p-3-beyond}
\eeqa
Note that the observables in \eq{s4-2-beyond}, \eqst{s3p-2-beyond}{s4p-2-beyond} and \eq{s1p-3-beyond} appear in Refs.~\cite{Mendez:1991gp,Keus:2015hva,Fontes:2015xva} where decay processes are studied.

In principle, one could replace $ZZ$ with $\gamma\gamma$ (or $Z\gamma$) in \eqst{s1p-1-beyond}{s1p-3-beyond}.
If fermion loops are absent, then the observation of any of the corresponding ten sets of observables could be interpreted as evidence of P-even CP violation.   The $h_i\gamma\gamma$ one-loop amplitude in each case would be mediated by a $W$ or charged Higgs boson loop.  For $i\neq 1$, the $W$-loop contribution would be Higgs alignment suppressed, whereas the charged Higgs loop contribution would be present in the exact Higgs alignment limit.  However, if $h_i\to ZZ$ is kinematically allowed, then the tree-level decay would dominate the corresponding $h_i\to\gamma\gamma$ decay in which case the observables listed in \eqst{s1p-1-beyond}{s1p-3-beyond} would be the preferred signatures of P-even CP violation.  If fermion loops are present, then the fermion loop contribution to the $h_i\gamma\gamma$ decay amplitude is not Higgs alignment suppressed.  On the other hand, as discussed in Section~\ref{multiphoton}, the observation of the $\gamma\gamma$ decay mode would not constitute evidence for CP violation unless the photon polarizations are measured to distinguish the operators exhibited in \eq{pol}. 

As noted above, if the $h_i$ ($i\neq 1$) couple to fermions, then the presence or absence of the $h_iZZ$ coupling by itself sheds no light on the existence of CP violation in the scalar sector in the exact Higgs alignment limit.  Since the $h_iZZ$ tree-level coupling for $i\neq 1$ is absent 
in the exact Higgs alignment limit, 
this coupling is only generated at the loop level via fermion loop contributions.  Indeed,
as shown in Ref.~\cite{Arhrib:2018pdi} for the CP-conserving 2HDM where $h$ is identified as the observed Higgs boson of mass 125 GeV, the decay width of $A \to ZZ$ is of the same order of magnitude as $H \to ZZ$ in the exact Higgs alignment limit.  
Therefore, when the $h_i ZZ$ couplings ($i\neq 1$) are of the same order, one can no longer use them to probe CP violation. 
Away from the exact Higgs alignment limit, tree-level $h_iZZ$ couplings (for $i\neq 1$) may be present.  If the size of these tree-level couplings is parametrically larger than the corresponding one-loop amplitude generated by fermion loop contributions, then the presence of P-even CP violation can be probed using the observables listed in \eqst{s1p-1-beyond}{s1p-3-beyond} in the approximation that the one-loop contributions can be neglected.  In practice, determining the conditions where the latter approximation is valid is model dependent.  
For example,
in models with additional particles beyond the scalar sector that carry electroweak quantum numbers (e.g., vectorlike quarks or supersymmetric particles), additional contributions to the one-loop amplitude will make it more difficult to distinguish between tree-level effects suppressed in the approximate Higgs alignment limit and loop-level effects.

In the scenarios presented in Section~\ref{xsections}, we focused on observables that depended on tree-level couplings that were not suppressed in the Higgs alignment limit.
Given that the results exhibited in this section were obtained in the exact Higgs alignment limit, one can ask whether the potential signals of P-even CP violation are robust as one moves away from exact Higgs alignment.  For example, 
away from the Higgs alignment limit the triple scalar couplings are modified; in particular, they depend on additional parameters of the scalar potential.\footnote{%
In contrast, in the exact Higgs alignment limit, each $h_i H^+ H^-$ coupling is proportional to $\Lambda_i$
 [cf.~\eq{Z377}].}  Hence, both the production cross section and branching ratios obtained in Section~\ref{xsections} will be modified.  The modifications are to some extent random because the processes of interest
are now controlled by different couplings. Nevertheless, by combining all possible sets of CP-violating observables, the chances of being able to probe P-even CP violation in extended Higgs sectors are very good if the scalars $h_i$ ($i\neq 1$) are not too heavy.

\section{Conclusions and outlook}
\label{sec:conclusion}

There is vast literature (e.g., see Refs.~\cite{Grzadkowski:1995rx,Bernreuther:1997af,Berge:2011ij,Berge:2014sra,Cordero-Cid:2020yba,Gunion:1996xu, Boudjema:2015nda, Mileo:2016mxg, AmorDosSantos:2017ayi,Goncalves:2018agy})
discussing CP-violating observables that derive from the P-violating, C-conserving neutral Higgs--fermion Yukawa couplings in models with an extended Higgs sector beyond the Standard Model. 
However, additional sources of CP violation of a different nature can arise from the pure bosonic sector of the model.  Indeed,
in a theory consisting only of bosonic fields (scalars and gauge bosons), CP-violating phenomena, if present,
must be associated with C-violating observables. This statement is independent of the number and representation of the multiplets of scalar fields and the choice of gauge group.  
Hence, if nature employs an extended Higgs sector, then one needs to identify physical observables that either receive negligible fermionic contributions or are completely insensitive to the presence of the fermions in order to experimentally verify the presence of P-even, CP-violating processes.

A combination of three bosonic decays in the 2HDM that signal the presence of P-even CP violation was proposed in~Refs.~\cite{Mendez:1991gp,Fontes:2015xva,Keus:2015hva}
and  involves the interaction of the $Z$ boson with a pair of neutral scalars, $Zh_i h_j$ ($1\leq i< j\leq 3$).  If CP is conserved then $h_i$ and $h_j$ must
have opposite sign CP quantum numbers.   Thus, the observation of all three interactions (e.g., via the three decays $h_3 \to h_2 Z$, $h_3 \to h_1 Z$ and $h_2 \to h_1 Z$)
necessarily implies that CP must be violated.
However, after identifying $h_1$ with the Higgs boson observed at the LHC, the decay rates for two of the three processes cited above
vanish in the exact Higgs alignment limit (where the tree-level properties of $h_1$ coincide with those of the SM Higgs boson). Since the LHC 
Higgs data show a clear preference for a SM-like Higgs boson, it follows that the CP-violating signal considered above is highly suppressed.

In this work we have found and listed all possible combinations of bosonic interactions that can be employed to identify the presence of P-even, CP-violating phenomena
that are not suppressed in the Higgs alignment limit.   Most of these combinations
rely on the discovery of three new scalars (one of which is charged), except for one case where only two new neutral scalars are needed. 
For example, the simultaneous observation of the vertices $Zh_2 h_3$, $h_2 H^+ H^-$ and  $h_3 H^+ H^-$ would provide unambiguous evidence for P-even CP violation.
Moreover, 
the interactions that we have identified 
can 
be quite large and give rise
to detectable signals at future multi-TeV colliders. For completeness, we have
also listed all possible combinations that include interactions with suppressed couplings in the Higgs alignment limit, classifying them by the number of suppressed interactions. 

Uncovering evidence for P-even CP-violation depends on the detection of new heavy scalar particles.  
At the LHC, the decays of the new scalars provide an opportunity to probe P-even, CP-violating processes if a sufficient number of appropriate final states are accessible. 
When considering the combination of Higgs decays, kinematic constraints due to the scalar masses can preclude probing a specific vertex. 
A more robust exploration of P-even, CP-violating phenomena can be carried out
in an experimentally clean environment such as a lepton collider, assuming that the production of the heavy scalars is kinematically accessible.
At lepton colliders, the observation of P-even, CP-violating processes can make use of a combination of production and decay mechanisms that
do not involve the Higgs-fermion Yukawa couplings at leading order.

In Section~\ref{sec:collider}, we performed a phenomenological study for various future multi-TeV lepton colliders. We demonstrated that given
the CM energies and total integrated luminosities of CLIC and of a planned muon collider, many combinations of production processes and/or decay channels are within the reach of these machines.
The simplest production process is $\ell^+ \ell^- \to h_2 h_3$ via $s$-channel $Z$ exchange, where $\ell$ can be either an electron or a muon. If kinematically allowed, both
$h_2$ and $h_3$ could decay to a pair of charged Higgs bosons, $H^+ H^-$. Alternatively one can combine the former production process with  $\ell^+ \ell^- \to h_2 h_2 h_3$ and $\ell^+ \ell^- \to h_2 h_3 h_3$. 
Using only production processes has the clear advantage of being dependent only on the collider CM energy.  As these are $s$-channel processes, colliders with lower CM energy are better suited to probe them.
However, the production cross sections for very heavy scalars are rather small, especially in the case of a three particle final state.  

There are also $t$-channel $\gamma \gamma$ processes, such as
$\ell^+ \ell^- \to \ell^+\ell^- H^+ H^- h_i$,  whose cross sections grow logarithmically with the collider CM energy and are best suited for higher energy colliders. These $t$-channel cross sections yield viable signals at a muon collider with a CM energy of order 10 TeV. 
However, the best option to probe processes such as
$\gamma \gamma \to  h_i H^+ H^-$ is to make use of a (hypothetical) photon collider option at CLIC.
We concluded the phenomenological study of Section~\ref{sec:collider} with a brief description of the possible final states. In order to find the exact reach of each collider, one needs to establish a set of realistic benchmarks by considering a full experimental analysis
by taking into account the complete set of backgrounds to the final state under consideration together with the detector effects.  This task is left to a future study.

This paper has focused primarily on P-even, CP-violating phenomena associated with tree-level processes.  Nevertheless, there are some cases in which loop-induced processes can 
play an important role.    A measurement of a P-even, CP-violating form factor in the loop-induced $ZZZ$ and $ZW^+ W^-$ vertex provides a promising approach for detecting P-even, CP-violating phenomena. 
In principle, such phenomena could also be probed by
loop-induced decays of scalars into final state photons and $Z$ bosons.   
For example, the observation of $h_k \to \gamma \gamma$ ($\gamma Z$) together with $h_k \to \gamma \gamma \gamma$ ($\gamma\gamma Z$ or $\gamma ZZ$)  would constitute a signal of P-even CP violation if the purely bosonic loop contributions can be isolated.  However, we have shown that 
the amplitude for the $\gamma\gamma\gamma$ decay mode vanishes exactly at one-loop order (whereas the corresponding one-loop amplitudes for the $\gamma\gamma Z$ and $\gamma ZZ$ final states are nonzero).   Moreover, due to extra coupling constant and phase space suppression factors that arise in computing the decay rates for the three-body final states,
we concluded that the detection of P-even CP violation via the multiphoton/$Z$ decay channels
is completely impractical.  

The LHC has just begun to probe the existence of 
C-even, CP-violating interactions of the Higgs boson in the top-quark and tau-lepton Yukawa couplings (which requires the observation of both scalar and pseudoscalar couplings of the neutral Higgs boson to a fermion-antifermion pair).   The observation of such interactions would be strongly suggestive of an extended Higgs sector in which the SM-like Higgs boson is not quite an eigenstate of CP due to the mixing with additional scalar degrees of freedom.   In such a framework, one would also expect the existence of P-even, CP-violating interactions.  However, such interactions arise solely from the bosonic sector of the theory, and as such cannot be associated 
with an experimental observable that involves a single scalar state.  Thus, the observation of P-even CP violation is far more challenging and is possible at future runs of the LHC only in very special circumstances.  Ultimately, to achieve a more robust probe of P-even, CP-violating phenomena, the relatively clean environment of a future multi-TeV lepton collider will be required, where the cross sections for multiple heavy scalar production are large enough to yield a viable experimental signal.
Indeed, the detection of the presence or absence of a P-even, CP-violating signal would provide crucial information concerning the fundamental nature of the scalar sector.

\section*{Acknowledgments}
H.E.H.~and V.K.~are grateful for many valuable conversations with Tim Stefaniak and Scott Thomas, who collaborated with us on the initial phase of this project.
R.S.~thanks Richard Ruiz and Jonas Wittbrodt for discussions.   We also acknowledge discussions with Michael Peskin concerning futuristic very high energy $\gamma\gamma$ colliders.

H.E.H. is supported in part by the U.S. Department of Energy Grant
No.~\uppercase{DE-SC}0010107.   V.K.~acknowledges financial support from the Academy of Finland projects ``Particle cosmology and gravitational waves,'' Grant No.~320123 and
 ``Particle cosmology beyond the Standard Model,'' Grant No.~310130, and from the Science Foundation Ireland Grant 21/PATH-S/9475 (MOREHIGGS) under the SFI-IRC Pathway Programme.
Both H.E.H.~and V.K.~acknowledge the support of Grant H$2020$-MSCA-RISE-$2014$
Grant No.~$645722$ (NonMinimalHiggs), which provided funds for two separate month long visits by V.K.~to the Santa Cruz Institute for Particle Physics (SCIPP)
and for the travel of H.E.H. 
to the $2019$ NonMinimalHiggs conference at the University of Helsinki.  Both visits were highly productive in advancing this work, and H.E.H.~and V.K.~are grateful for the hospitality furnished by SCIPP and the University of Helsinki.
R.S.~is supported by FCT under Contracts UIDB/00618/2020, UIDP/00618/2020, PTDC/FIS-PAR/31000/2017, CERN/FIS-PAR/0002/2017, and CERN/
FIS-PAR/0014/2019.
H.E.H., V.K. and R.S.~also benefited from discussions that took place at the University of Warsaw during visits supported by the HARMONIA  project of  the National Science Centre,  Poland,  under Contract No.~UMO-$2015$/$18$/M/ST$2$/$00518$ ($2016$--$2021$).

\vskip 0.5in

\begin{appendices}

\section{Theoretical aspects of the 2HDM}
\label{2hdm}
\renewcommand{\theequation}{A.\arabic{equation}}
\setcounter{equation}{0}

In this Appendix, we briefly introduce the 2HDM and identify a convenient set of parameters that will provide the basis for the parameter choices employed in this work.
We survey the various couplings of the Higgs scalars (to gauge bosons, Higgs bosons and fermions) and discuss the Higgs alignment limit, in which the tree-level properties of one of the neutral scalars coincides with those of the Standard Model Higgs boson.

The fields of the 2HDM consist of two
identical complex hypercharge-one, SU(2) doublet scalar fields
$\Phi_a(x)\equiv (\Phi^+_a(x)\,,\,\Phi^0_a(x))$, 
where the ``Higgs flavor'' index $a=1,2$ labels the two Higgs doublet fields.
The most general renormalizable SU(2)$_L\times$U(1)$_Y$-invariant scalar potential is given in the $\Phi$-basis
by
\beqa  \label{pot}
\mathcal{V}&=& m_{11}^2\Phi_1^\dagger\Phi_1+m_{22}^2\Phi_2^\dagger\Phi_2
-[m_{12}^2\Phi_1^\dagger\Phi_2+{\rm h.c.}]+\half\lambda_1(\Phi_1^\dagger\Phi_1)^2
+\half\lambda_2(\Phi_2^\dagger\Phi_2)^2
+\lambda_3(\Phi_1^\dagger\Phi_1)(\Phi_2^\dagger\Phi_2)\nonumber\\[8pt]
&&\quad 
+\lambda_4(\Phi_1^\dagger\Phi_2)(\Phi_2^\dagger\Phi_1)
+\left\{\half\lambda_5(\Phi_1^\dagger\Phi_2)^2
+\big[\lambda_6(\Phi_1^\dagger\Phi_1)
+\lambda_7(\Phi_2^\dagger\Phi_2)\big]
\Phi_1^\dagger\Phi_2+{\rm h.c.}\right\}\,,
\eeqa
where $m_{11}^2$, $m_{22}^2$, and $\lam_1,\cdots,\lam_4$ are real parameters
and $m_{12}^2$, $\lambda_5$, $\lambda_6$ and $\lambda_7$ are
potentially complex parameters.  We assume that the
parameters of the scalar potential are chosen such that
the minimum of the scalar potential respects the
U(1)$\ls{\rm EM}$ gauge symmetry.\footnote{As shown in Refs.~\cite{Ferreira:2004yd, Barroso:2005sm}, given a generic 2HDM tree-level scalar potential that
has a minimum which respects U(1)$\ls{\rm EM}$, then any competing stationary point that 
breaks the U(1)$\ls{\rm EM}$ symmetry is a saddle point that lies above the symmetry conserving vacuum.}
Then, the scalar field vevs are of the form
\beq \label{vhat}
\langle\Phi_a\rangle=
\frac{v}{\sqrt{2}}\begin{pmatrix} 0\\ \widehat{v}_a\end{pmatrix}\,,
\eeq
where $ \widehat{v}$ is a complex vector of unit norm,  
\beq \label{veedef}
\widehat{v}=(\widehat{v}_1\,,\,\widehat{v}_2) =(c_\beta\,,\, s_\beta e^{i\xi}), 
\eeq
and
$c_\beta\equiv \cos\beta$ and $s_\beta\equiv\sin\beta$, $0\leq\beta\leq\half\pi$, $0\leq \xi< 2\pi$, and $v$ is determined by the Fermi constant,
\beq \label{v246}
v\equiv \frac{2\mw}{g}=(\sqrt{2}G_F)^{-1/2}\simeq 246~{\rm GeV}\,.
\eeq

\subsection{Higgs basis}

It is convenient to introduce the Higgs basis as follows.   Starting from a generic $\Phi$-basis, the Higgs basis fields $\mathcal{H}_1$ and $\mathcal{H}_2$ are defined 
by the linear combinations of $\Phi_1$ and $\Phi_2$ such that $\langle \mathcal{H}_1^0\rangle=v/\sqrt{2}$ and $\langle \mathcal{H}_2^0\rangle=0$.  That is,
\beq \label{invhiggs}
\mathcal{H}_1\equiv c_\beta\Phi_1+s_\beta e^{-i\xi}\Phi_2\,,\qquad\quad \mathcal{H}_2=e^{i\eta}\bigl[-s_\beta e^{i\xi}\Phi_1+c_\beta \Phi_2\bigr]\,,
\eeq
where we have introduced (following Ref.~\cite{Boto:2020wyf}) the complex phase factor $e^{i\eta}$ to account for the nonuniqueness of the Higgs basis, since one is always free to rephase the Higgs basis field whose vacuum expectation value vanishes.  In particular,  $e^{i\eta}$ is a pseudoinvariant quantity that is rephased under the unitary basis transformation,
$\Phi_a\to U_{a\bbar}\Phi_b$, as\footnote{The use of unbarred and barred indices follows the conventions introduced in Ref.~\cite{Davidson:2005cw}.}
\beq \label{shift}
e^{i\eta}\to (\det~U)^{-1} e^{i\eta}\,,
\eeq
where $\det U$ is a complex number of unit modulus.
Equivalently, one can write,
\beq \label{hbasisdef}
\mathcal{H}_1=(\mathcal{H}_1^+\,,\,\mathcal{H}_1^0)\equiv \widehat v_{\abar}^{\,\ast}\Phi_a\,,\qquad\qquad
\mathcal{H}_2=(\mathcal{H}_2^+\,,\,\mathcal{H}_2^0)\equiv e^{i\eta} \widehat  w_{\abar}^{\,\ast}\Phi_a\,,
\eeq
where there is an implicit sum over unbarred/barred index pairs and
\beq \label{what}
\widehat{w}_b=\widehat{v}_{\abar}^{\,\ast}\epsilon_{ab}\qquad \quad (\text{$\epsilon_{12}=-\epsilon_{21}=1$ and $\epsilon_{11}=\epsilon_{22}=0$})
\eeq
is a unit vector that is orthogonal
to $\widehat{v}$ (i.e., $\widehat{v}_{\bbar}^{\,\ast}\widehat{w}_b=0$).
Under a U(2) basis transformation, 
\beq \label{wtrans}
\widehat{v}_a\to U_{a\bbar}\,\widehat{v}_b, \quad \text{which implies that \quad
$\widehat{w}_a\to (\det U)^{-1}U_{a\bbar}\,\widehat{w}_b$}.
\eeq
In light of \eqst{shift}{wtrans},
it follows that both $\mathcal{H}_1$ and $\mathcal{H}_2$ are invariant fields with respect to U(2) basis transformations.

In terms of the Higgs basis fields defined in \eq{invhiggs}, the scalar potential is given by, 
 \beqa
 \mathcal{V}&=& Y_1 \mathcal{H}_1^\dagger \mathcal{H}_1+ Y_2 \mathcal{H}_2^\dagger \mathcal{H}_2 +[Y_3 e^{-i\eta}
\mathcal{H}_1^\dagger \mathcal{H}_2+{\rm h.c.}]
\nn\\
&&\quad 
+\half Z_1(\mathcal{H}_1^\dagger \mathcal{H}_1)^2+\half Z_2(\mathcal{H}_2^\dagger \mathcal{H}_2)^2
+Z_3(\mathcal{H}_1^\dagger \mathcal{H}_1)(\mathcal{H}_2^\dagger \mathcal{H}_2)
+Z_4(\mathcal{H}_1^\dagger \mathcal{H}_2)(\mathcal{H}_2^\dagger \mathcal{H}_1) \nn \\
&&\quad
+\left\{\half Z_5 e^{-2i\eta}(\mathcal{H}_1^\dagger \mathcal{H}_2)^2 +\big[Z_6 e^{-i\eta} (\mathcal{H}_1^\dagger
\mathcal{H}_1) +Z_7 e^{-i\eta} (\mathcal{H}_2^\dagger \mathcal{H}_2)\big] \mathcal{H}_1^\dagger \mathcal{H}_2+{\rm
h.c.}\right\}.\label{higgspot}
\eeqa
The coefficients of the quadratic and quartic terms of the scalar potential in \eq{higgspot} are independent of the initial choice of the $\Phi$-basis.  It then follows that $Y_3$, $Z_5$, $Z_6$ and $Z_7$ are also pseudoinvariant quantities that are rephased under $\Phi_a\to U_{a\bbar}\Phi_b$ as follows:
\beq \label{rephasing}
 [Y_3, Z_6, Z_7]\to (\det~U)^{-1}[Y_3, Z_6, Z_7] \quad{\rm and}\quad
Z_5\to  (\det~U)^{-2} Z_5\,.
\eeq
The minimization of the scalar potential in the Higgs basis yields 
\beq \label{minconds}
Y_1=-\half Z_1 v^2\,,\qquad\quad Y_3=-\half Z_6 v^2\,.
\eeq

\subsection{Physical mass eigenstates}

Given the scalar potential and its minimization conditions, one can determine the masses of the neutral scalars.  
After removing the massless Goldstone boson, $G^0=\sqrt{2}\,\Im~\!\mathcal{H}_1^0$ from the $4\times 4$ neutral scalar squared-mass matrix, 
the physical neutral scalar mass-eigenstate fields are obtained by diagonalizing the resulting $3\times 3$ neutral scalar squared-mass matrix,
\beq  \label{matrix33}
\mathcal{M}^2=v^2\left( \begin{array}{ccc}
Z_1&\, \Re(Z_6 e^{-i\eta}) &\, -\Im(Z_6 e^{-i\eta})\\
\Re(Z_6 e^{-i\eta})  &\, \half\bigl[Z_{34}+\Re(Z_5 e^{-2i\eta})\bigr]+Y_2/v^2 & \,
- \half \Im(Z_5  e^{-2i\eta})\\ -\Im(Z_6  e^{-i\eta}) &\, - \half \Im(Z_5  e^{-2i\eta}) &\,
\half\bigl[Z_{34}-\Re(Z_5 e^{-2i\eta})\bigr]+Y_2/v^2\end{array}\right)\!, 
\eeq
with respect to the $\{\sqrt{2}\,\Re \mathcal{H}_1^0-v,\sqrt{2}\,\Re \mathcal{H}_2^0,\sqrt{2}\,\Im \mathcal{H}_2^0\}$ basis, where $Z_{34}\equiv Z_3+Z_4$.
The squared masses of the physical neutral scalars, denoted by $m_k^2$ ($k=1,2,3$) with no implied mass ordering, are the eigenvalues of $\mathcal{M}^2$, which are independent of the choice of~$\eta$.

The real symmetric squared-mass matrix $\mathcal{M}^2$ can be diagonalized by
a real orthogonal transformation of unit determinant,
\beq \label{rmrt}
R\mathcal{M}^2 R^{\T}= {\rm diag}~(m_1^2\,,\,m_2^2\,,\,m_3^2)\,,
\eeq
where $R\equiv R_{12}R_{13}R_{23}$   is the product of three rotation matrices parametrized by $\theta_{12}$, $\theta_{13}$ and $\theta_{23}$, respectively~\cite{Haber:2006ue}. 
Since the matrix elements of $\mathcal{M}^2$ are independent of the scalar field basis, it follows that the mixing angles $\theta_{ij}$ are invariant parameters.
The physical neutral mass-eigenstate scalar fields are 
\beq \label{hsubkay}
h_k=q_{k1}\bigl(\sqrt{2}\,\Re \mathcal{H}_1^0-v\bigr)+\frac{1}{\sqrt{2}}\bigl(q_{k2}^*\mathcal{H}_2^0 e^{i\theta_{23}}+{\rm h.c.}\bigr)\,,
\eeq
where $q_{k1}$ and $q_{k2}$ are exhibited in Table~\ref{tabinv}.  The charged scalar mass eigenstates are defined by,
\beq
G^\pm=\mathcal{H}_1^\pm\,,\qquad\quad H^\pm\equiv e^{\pm i\theta_{23}}\mathcal{H}_2^\pm\,,
\eeq
where we have rephased the charged Higgs fields as a matter of convenience.
The mass of the charged Higgs scalar is given by
\beq \label{plusmass}
m_{H^\pm}^2=Y_2+\half Z_3 v^2\,.
\eeq

 \begin{table}[t!]
\centering
\begin{tabular}{|c||c|c|}\hline
$\phaa k\phaa $ &\phaa $q_{k1}\phaa $ & \phaa $q_{k2} \phaa $ \\ \hline
$1$ & $c_{12} c_{13}$ & $-s_{12}-ic_{12}s_{13}$ \\
$2$ & $s_{12} c_{13}$ & $c_{12}-is_{12}s_{13}$ \\
$3$ & $s_{13}$ & $ic_{13}$ \\
\hline
\end{tabular}
\caption{\small The U(2)-invariant quantities $q_{k\ell}$ are functions of 
the neutral Higgs mixing angles $\theta_{12}$ and $\theta_{13}$, where
$c_{ij}\equiv\cos\theta_{ij}$ and $s_{ij}\equiv\sin\theta_{ij}$.   The angles $\theta_{12}$ and $\theta_{23}$ are defined modulo $\pi$.
By convention, we take $0\leq c_{12}, c_{13}\leq 1$.
\label{tabinv}}
\end{table}

Inverting \eq{hsubkay}, one can now express the Higgs basis fields in terms of the mass eigenstate fields,
\beq \label{Hbasismassbasis}
\mathcal{H}_1=\begin{pmatrix} G^+ \\[3pt] \displaystyle\frac{1}{\sqrt{2}}\left(v+iG+\sum_{k=1}^3 q_{k1}h_k\right)\end{pmatrix},
\qquad\quad
e^{i\theta_{23}}\mathcal{H}_2=\begin{pmatrix} H^+ \\[3pt] \displaystyle\frac{1}{\sqrt{2}}\sum_{k=1}^3 q_{k2}h_k\end{pmatrix}.
\eeq
In light of \eq{Hbasismassbasis},
the parameter $\theta_{23}$ can be eliminated by rephasing $\mathcal{H}_2\to e^{-i\theta_{23}}\mathcal{H}_2$.  Thus, without loss of generality, we shall henceforth 
set $\theta_{23}=0$.

If we denote the physical neutral scalar masses by $m_k$ ($k=1,2,3$), then the equation for the diagonalization of the neutral scalar squared-mass matrix
yields the following squared mass sum rules (e.g., see eqs.~(47)--(50) of Ref.~\cite{Boto:2020wyf}):
\beqa
Z_1  &=& \frac{1}{v^2}\sum_{k=1}^3 m_k^2 q_{k1}^2\,,\label{zee1id} \\
Z_4  &=& \frac{1}{v^2}\left[\sum_{k=1}^3 m_k^2 |q_{k2}|^2 -2m_{H^\pm}^2\right]\,, \label{zeefour}  \\
Z_5  e^{-2i\eta} &=& \frac{1}{v^2}\sum_{k=1}^3 m_k^2 (q_{k2}^*)^2\,, \label{zee5id}\\
Z_6  e^{-i\eta} &=& \frac{1}{v^2}\sum_{k=1}^3 m_k^2 \,q_{k1} q_{k2}^*\,.\label{zee6id}
\eeqa
Due to \eqs{shift}{rephasing}, the quantities $Z_5  e^{-2i\eta}$ and $Z_6  e^{-i\eta} $ are basis-invariant quantities.

More explicitly, \eqst{zee1id}{zee6id} yield the following expressions:
\beqa
Z_1 v^2&=&m_1^2 c_{12}^2 c_{13}^2+m_2^2 s_{12}^2 c_{13}^2 + m_3^2
s_{13}^2\,, \\
Z_4 v^2&=&  m_1^2+m_2^2-c_{13}^2(c_{12}^2 m_1^2+s_{12}^2 m_2^2-m_3^2)-2m_{H^\pm}^2\,, \label{rez4} \\
\Re(Z_6\,e^{-i\eta})\,v^2 &=& c_{13}s_{12}c_{12}(m_2^2-m_1^2)\,,\label{rez6}
  \\
\Im(Z_6\,e^{-i\eta})\,v^2 &=& s_{13}c_{13}(c_{12}^2 m_1^2+s_{12}^2
m_2^2-m_3^2) \,, \label{imz6}  \\
\Re(Z_5\,e^{-2i\eta})\,v^2 &=& (c_{12}^2-s_{12}^2)(m_2^2-m_1^2)+c_{13}^2(c_{12}^2 m_1^2+s_{12}^2
m_2^2-m_3^2)\,, \label{rez5} \\
\Im(Z_5\,e^{-2i\eta})\,v^2 &=& 2s_{12}c_{12}s_{13}(m_2^2-m_1^2)\,.\label{imz5}
\eeqa
Finally, one additional relation of significance is the trace condition,
\beq
\Tr\,\mathcal{M}^2=\sum_{k=1}^3 m_k^2=2Y^2+(Z_1+Z_{34})v^2=2m_{H^\pm}^2+(Z_1+Z_4)v^2\,.
\eeq

\subsection{Parameter set of the bosonic sector of the 2HDM}

Let us now count the parameters that govern the most general 2HDM.    In light of \eq{minconds}, it follows that the 2HDM is governed by
six real parameters, $v$, $Y_2$, $Z_1$, $Z_2$, $Z_3$ and $Z_4$ and three complex parameters $Z_5$, $Z_6$ and $Z_7$ for a total
of 12 real parameters, where $v=246~{\rm GeV}$.  However, due to the presence of $\eta$ in \eq{higgspot}, we are free to rephase the Higgs basis fields. 
In particular, consider what happens if one transforms between two Higgs bases.   That is, suppose that $\vev{\Phi_1^0}=v/\sqrt{2}$ and $\vev{\Phi_2^0}=0$.   To transform to another Higgs basis, one can employ the U(2) transformation $\Phi_a\to U_{a\bbar}\Phi_b$, where $U={\rm diag}(1,e^{i\chi})$.   Then, \eq{shift} implies that $\eta\to\eta-\chi$.  
It then follows that
\beq \label{rephasing2}
[Y_3, Z_6, Z_7]\to e^{-i\chi}[Y_3, Z_6, Z_7] \quad \text{and} \quad
Z_5\to  e^{-2i\chi} Z_5\,.
\eeq
In contrast, $Y_1$, $Y_2$ and $Z_{1,2,3,4}$ are invariant when transforming between two Higgs bases.  This means that among the three complex parameters,
$Z_5$, $Z_6$ and $Z_7$, there are only five independent real physical degrees of freedom.   Thus, in total there are 11 real parameters that govern the most general 2HDM.

However, it is more convenient to choose a different set of parameters to define the most general 2HDM.   Here is the list:
\beq \label{list}
v\,,\, m_1\,,\, m_2\,,\, m_3\,,\, m_{H^\pm}\,,\, \theta_{12}\,,\, \theta_{13}\,,\, Z_2\,,\, Z_3\,,\, Z_7 e^{-i\eta}\,.
\eeq
This list includes nine real parameters and one complex parameter $Z_7 e^{-i\eta}$ for a total of 11 real parameters that fix the 2HDM model.   Note that
all parameters on this list are basis-invariant quantities.  In particular, one cannot ``rephase'' the complex parameter $Z_7 e^{-i\eta}$ to remove one degree of
freedom.

It is often assumed that the scalar potential exhibits a $\mathbb{Z}_2$ symmetry, where
$\Phi_1$ is unchanged and $\Phi_2\to -\Phi_2$
in some scalar field basis.
We will allow this symmetry to be softly broken by the dimension-two squared-mass terms of the scalar potential, in which case there exists a $\Phi$-basis where $\lambda_6=\lambda_7=0$ in the notation of \eq{pot}.   Such a basis will be called the $\mathbb{Z}_2$ basis.  
As shown in Ref.~\cite{Boto:2020wyf},  a $\mathbb{Z}_2$-basis exists if and only if the following relation is satisfied,
\beq \label{finalcond}
(Z_1-Z_2)\bigl[Z_{34}Z^*_{67}-Z_1 Z^*_7-Z_2 Z^*_6+Z^*_5 Z_{67}\bigr]-2Z^*_{67}\bigl(|Z_6|^2-|Z_7|^2\bigr)=0\,,
\eeq
where $Z_{34}\equiv Z_3+Z_4$ and $Z_{67}\equiv Z_6+Z_7$.
Since \eq{finalcond} is linear in $Z_3$, we will use this equation to solve for $Z_3$ and remove it from the list given in \eq{list}.
In the special case of $Z_1=Z_2$, $Z_5\neq 0$ and $Z_{67}\neq 0$,  \eq{finalcond} must be replaced by the two conditions,
\beq \label{finalcond2}
\Im(Z_5^* Z^2_{67})=0\,,\qquad\quad |Z_6|=|Z_7|\,,\qquad \text{if $Z_1=Z_2$, $Z_5\neq 0$ and $Z_{67}\neq 0$}.
\eeq
We can specify the $\mathbb{Z}_2$ basis by providing expressions for $\beta$ and $\xi$ [which are defined in \eq{veedef}]. Assuming that $Z_{67}\neq 0$, 
\beq \label{sincos}
s_{2\beta}=\frac{2|Z_{67}|}{\sqrt{(Z_2-Z_1)^2+4|Z_{67}|^2}}\,,\qquad\quad
c_{2\beta}=\frac{\pm(Z_2-Z_1)}{\sqrt{(Z_2-Z_1)^2+4|Z_{67}|^2}}\,,
\eeq
where by convention, $0\leq\beta\leq\half\pi$.  In particular,
\beq \label{tanbeta}
\tan\beta=\sqrt{\frac{1-c_{2\beta}}{1+c_{2\beta}}}\,,\qquad\quad e^{i(\xi+\eta)}=\left(\frac{Z_2-Z_1}{2Z_{67}e^{-i\eta}}\right)\frac{s_{2\beta}}{c_{2\beta}}\,.
\eeq
The twofold ambiguity in the choice of the sign of $c_{2\beta}$ corresponds to the fact that the conditions, $\lambda_6=\lambda_7=0$ are preserved under the interchange $\Phi_1\leftrightarrow\Phi_2$.

Given the list of parameters of \eq{list}, we can use $\theta_{12}$ and $\theta_{13}$ to determine the $q_{k\ell}$ in Table~\ref{tabinv}.  Then, we can determine
the real parameters $Z_1$ and $Z_4$ and the complex parameters $Z_5  e^{-2i\eta}$ and $Z_6  e^{-i\eta}$ using \eqst{zee1id}{zee6id}.  Finally, $Y_2$ is fixed by \eq{plusmass}.

\subsection{2HDM bosonic interactions}

\begingroup
\allowdisplaybreaks
The couplings of Goldstone bosons and Higgs bosons to gauge bosons depend only on the $q_{kj}$ and
the electroweak SU(2)$_L$ and U(1)$_Y$ gauge coupling parameters $g$ and $g'$ as 
\beqa
\mathscr{L}_{VVH}&=&\left(gm_W W_\mu^+W^{\mu\,-}+\frac{g}{2c_W} m_Z
Z_\mu Z^\mu\right)q_{k1} h_k \nonumber \\[5pt]
&&
+em_WA^\mu(W_\mu^+G^-+W_\mu^-G^+)
-gm_Zs_W^2 Z^\mu(W_\mu^+G^-+W_\mu^-G^+)
\,,\label{VVH}\\[8pt]
\mathscr{L}_{VVHH}&=&\left[\quarter g^2 W_\mu^+W^{\mu\,-}
+\frac{g^2}{8c_W^2}Z_\mu Z^\mu\right](G^0 G^0 +h_k h_k) 
 +\biggl[\half g^2 W_\mu^+ W^{\mu\,-} +e^2A_\mu A^\mu \nn \\
&&  \left.  \qquad\qquad +\frac{g^2}{c_W^2}\left(\half
-s_W^2\right)^2Z_\mu Z^\mu +\frac{2ge}{c_W}\left(\half
-s_W^2\right)A_\mu Z^\mu\right](H^+H^- + G^+ G^-)
\nn \\
&& +\biggl\{ \left(\half eg A^\mu W_\mu^+
-\frac{g^2s_W^2}{2c_W}Z^\mu W_\mu^+\right)
(q_{k1}G^- + q_{k2}H^-)h_k +{\rm h.c.}\biggr\} \nn \\
&& +\biggl\{
\half ieg A^\mu W_\mu^+ G^- G^0
-\frac{ig^2s_W^2}{2c_W}Z^\mu W_\mu^+
G^-G^0 +{\rm h.c.}\!\biggr\}
\,,\label{VVHH} \\[8pt]
\mathscr{L}_{VHH}&=&-\frac{g}{4c_W}\,\epsilon_{jk\ell}q_{\ell 1}
Z^\mu h_j\ddel_\mu h_k -\half g\biggl\{iW_\mu^+\left[q_{k1} G^-\ddel\lsup{\,\mu} h_k+
q_{k2}H^-\ddel\lsup{\,\mu} h_k\right]
+{\rm h.c.}\biggr\}\nonumber \\
&& +\frac{g}{2c_W} q_{k1} Z^\mu G^0\ddel_\mu h_k
+\half g\left(W_\mu^+G^-\ddel\lsup{\,\mu}G^0+W_\mu^-G^+\ddel\lsup{\,\mu}G^0
\right)\nn \\
&& +\left[ieA^\mu+\frac{ig}{c_W}\left(\half -s_W^2\right)
Z^\mu\right](G^+\ddel_\mu G^-+H^+\ddel_\mu H^-)\,, \label{VHH}
\eeqa
where $s_W\equiv \sin\theta_W$, $c_W\equiv\cos\theta_W$,
and the sum over pairs of repeated indices $j,k=1,2,3$ is implied.  
\endgroup

The cubic and quartic scalar self-couplings can be expressed in terms of $v$, the $q_{kj}$, $Z_1,\ldots Z_4$, $Z_5  e^{-2i\eta}$, $Z_6  e^{-i\eta}$
and $Z_7  e^{-i\eta}$ as shown in Ref.~\cite{Haber:2006ue}.   For convenience of the presentation, we introduce the following notation:
\beq \label{zees}
\overbar{Z}_5\equiv Z_5 e^{-2i\eta}\,,\qquad\quad  \overbar{Z}_6\equiv Z_6 e^{-i\eta}\,,\qquad\quad  \overbar{Z}_7\equiv Z_7 e^{-i\eta}\,.
\eeq
 The complete list of cubic scalar couplings is exhibited in Table~\ref{threescalar}.

\begin{table}[t!]
\begingroup
\setlength{\tabcolsep}{10pt} 
\renewcommand{\arraystretch}{1.5} 
\begin{tabular}{|c|c|}
\hline
\pht  vertex  \pht  & self-coupling \\
\hline
$G h_j h_k$ & $\ v\bigl[\Im( q_{j2} q_{k2}\zv)+q_{j1}\Im\left(q_{k2} \zvi\right)+q_{k1}\,\Im\left(q_{j2} \zvi\right)\bigr]=\epsilon_{jk\ell}(m_j^2-m_k^2)q_{\ell 1}/v$  \\
$GG h_j$ & $v\bigl[q_{j1}Z_1+\Re(q_{j2}\zvi)\bigr]=q_{j1}m_j^2/v$ \\
$G^+G^- h_j$ &   $v\bigl[q_{j1}Z_1+\Re(q_{j2}\zvi)\bigr]=q_{j1}m_j^2/v$\\
$G^+ H^- h_j$ &  $\half v\bigl[q_{j2} Z_4+q^*_{j2}\zv\llsup{*}+2q_{j1}\zvi\llsup{*}\bigr]=q_{j2}(m_j^2-m_{H^\pm}^2)/v$ \\
$G^- H^+ h_j$ & $\half v\bigl[q^*_{j2} Z_4+q_{j2}\zv+2q_{j1}\zvi\bigr]=q^*_{j2}(m_j^2-m_{H^\pm}^2)/v$  \\
$H^+ H^- h_j$ &   $v\bigl[q_{j1}Z_3+\Re(q_{j2}\zvii)\bigr]$\\
$h_j h_j h_j$ &  $3v\bigl[q^3_{j1}Z_1+q_{j1}|q_{j2}|^2 Z_{34} +q_{j1}\Re(q^2_{j2}\zv)+3q_{j1}^2\Re(q_{j2} \zvi)+|q_{j2}|^2\Re (q_{j2}\zvii)\bigr]$\phantom{xxxxx}  \\[6pt]
$h_j h_j h_k$ & $v\biggl\{3q_{j1}^2 q_{k1}Z_1+ \bigl[q_{k1}|q_{j2}|^2+2q_{j1}\Re(q_{j2}q_{k2}^*)\bigr]Z_{34}+2q_{j1}\Re(q_{j2}q_{k2}\zv)+q_{k1}\Re(q_{j2}^2\zv)$ \\
&  $+3q_{j1}^2\Re(q_{k2}\zvi)+6q_{j1}q_{k1}\Re(q_{j2}\zvi)+2|q_{j2}|^2\Re(q_{k2}\zvii)+\Re(q_{k2}^* q^2_{j2}\zvii)\biggr\}$\\
$h_j h_k h_\ell$ &  $v\biggl\{3q_{j1}q_{k1}q_{\ell 1}Z_1+\bigl[q_{j1}\Re(q_{k2}q_{\ell 2}^*)+q_{k1}\Re(q_{j2}q_{\ell 2}^*)+q_{\ell 1}\Re(q_{j2}q_{k2}^*)\bigr]Z_{34}$\phantom{xxxxxxx} \\
& $+q_{j1}\Re(q_{k2}q_{\ell 2}\zv)+q_{k1}\Re(q_{j2}q_{\ell 2}\zv)+q_{\ell 1}\Re(q_{j2}q_{k2}\zv)$ \\
& $+3q_{j1}q_{k1}\Re(q_{\ell 2}\zvi)+3q_{j1}q_{\ell 1}\Re(q_{k2}\zvi)+3q_{k1}q_{\ell 1}\Re(q_{j2}\zvi)\phantom{xxxxx}$ \\
& $+\Re(q_{j2}^* q_{k2}q_{\ell 2}\zvii)+\Re( q_{j2}q_{k2}^*q_{\ell 2}\zvii)+\Re( q_{j2}q_{k2}q_{\ell 2}^*\zvii)\biggr\}$ \phantom{xxx} \\
\hline
\end{tabular}
\endgroup
\caption{\small Nonvanishing cubic scalar self-couplings of the most general 2HDM.    The indices $j$, $k$ and $\ell$ are distinct integers $\!\in\!\{1,2,3\}$.  To obtain the corresponding Feynman rules, multiply the
self-couplings listed above by $-i$.  Charged fields point \textit{into} the vertex. The cubic 
self-couplings not listed above,
such as $GGG$, $Gh_j h_j$, $GG^+G^-$, $GH^+H^-$, $GG^\pm H^{\mp}$ vanish exactly.
\label{threescalar}}
\end{table}

In the case of a CP-conserving Higgs scalar potential and vacuum, \eq{CPconditions} is satisfied
and the above results simplify significantly.
In particular, one can fix the Higgs basis up to a potential sign ambiguity by choosing the phase $e^{-i\eta}$ such that $Y_3$, $Z_5$, $Z_6$ and $Z_7$ are all simultaneously real, which yields the so-called real Higgs basis (after absorbing the phase into the definition of the Higgs basis fields).  
 The remaining ambiguity in defining the real Higgs basis is  due to the possibility of transforming $\mathcal{H}_2\to -\mathcal{H}_2$, in which case $Y_3$, $Z_6$ and $Z_7$ change sign (whereas all other scalar potential parameters in the real Higgs basis, including $Z_5$ are unchanged.   Thus it is convenient to define $\varepsilon\equiv e^{i\eta}$, where $\varepsilon$ changes sign under $\mathcal{H}_2\to -\mathcal{H}_2$. 
Following Refs.~\cite{Haber:2006ue,Gori:2017qwg}, we set $s_{13}=0$, $c_{13}=1$ and\footnote{If $Z_6=Z_7=0$, then the sign of $Z_5$ is no longer invariant with respect to transformations that preserve the real Higgs basis (since the sign of $Z_5$ changes under $\mathcal{H}_2\to \pm i\mathcal{H}_2$).   In this case, it would be more appropriate to define $\varepsilon\equiv e^{2i\eta}=\sgn Z_5$.}
 \beq \label{twosigns}
\varepsilon\equiv e^{i\eta}=\begin{cases} \sgn Z_6\,, & \quad \text{if $Z_6\neq 0$},\\   \sgn Z_7\,,& \quad \text{if $Z_6=0$ and $Z_7\neq 0$}.\end{cases}
\eeq

In the standard notation of the CP-conserving 2HDM, one chooses a real $\Phi$-basis and defines $\tan\beta\equiv \langle\Phi_2^0\rangle/\langle\Phi_1^0\rangle$.   The corresponding mixing angle that diagonalizes the CP-even Higgs squared-mass matrix is denoted by~$\alpha$.  The 
CP-even scalar mass-eigenstates, $h$ and $H$ (with $m_{h}\leq m_{H}$), and the CP-odd scalar $A$  are related to the neutral fields of the Higgs basis via
\beq \label{hH}
\begin{pmatrix} H\\ h\end{pmatrix}=\begin{pmatrix} \cbma & \,\,\, -\sbma \\
\sbma & \,\,\,\phantom{-}\cbma\end{pmatrix}\,\begin{pmatrix} \sqrt{2}\,\,{\rm Re}~H_1^0-v \\ 
\sqrt{2}\,{\rm Re}~H_2^0
\end{pmatrix}\,,\qquad\quad A=\sqrt{2}\,\Im H_2^0\,.
\eeq
where $\cbma\equiv\cos(\beta-\alpha)$ and $\sbma\equiv\sin(\beta-\alpha)$.  

Under the assumption that the lighter of the two neutral CP-even Higgs bosons is SM-like, it is convenient to make the following identifications,
\beq\label{eq:epsilon6}
h=h_1\,,\qquad H=-\varepsilon h_2\,,\qquad A=\varepsilon h_3\,.
\eeq
In light of \eq{hH}, one can then identify,
\beq \label{anglelimit}
c_{12}=\sbma\,,\qquad\quad s_{12}=-\varepsilon\,\cbma\,,
\eeq
where $0\leq\sbma\leq 1$, and the
$q_{kj}$ of Table~\ref{tabinv}  simplify to the results given in 
Table~\ref{cptabinv}(a).\footnote{Note that the signs of the fields $H$ and $A$ and the sign of $\cbma$ all flip under the redefinition of the Higgs basis field $\mathcal{H}_2\to -\mathcal{H}_2$.   In the CP-conserving 2HDM literature, in models in which the choice of the $\Phi_1$--$\Phi_2$ basis is physically meaningful (e.g., due to the presence of a discrete $\mathbb{Z}_2$ symmetry of the scalar potential), it is traditional to impose one further restriction that $\tan\beta$ is real and positive.  This removes the final sign ambiguity in defining the real Higgs basis.}
\begin{table}[t!]
\begin{subtable}[c]{0.5\textwidth}
\centering
\begin{tabular}{|c|| c ||c|c|}\hline
\multicolumn{4}{|c|}{\text{(a)~~$h$ is SM-like when $|\cbma|\ll 1$}\TBstrut} \\ \hline
$\phm k \phm $ & $\phaa h_k$\phaa\ & $q_{k1}\phaa $ & \phaa $q_{k2} \phaa $ \\
\hline
$\phm 1 \phm $ & $h$ & $\,\,\,s_{\beta-\alpha}$ & $\varepsilon\, c_{\beta-\alpha}$\Tstrut \\
$2$ & $\!\!\!-\varepsilon\, H$ & $-\varepsilon\, c_{\beta-\alpha}$ & $s_{\beta-\alpha}$ \\
$3$ & $\varepsilon\, A$ & $0$ & $i$ \\ \hline
\end{tabular}
\end{subtable}
\begin{subtable}[c]{0.5\textwidth}
\centering
\begin{tabular}{|c|| c ||c|c|}\hline
\multicolumn{4}{|c|}{\text{(b)~~$H$ is SM-like when $|\sbma|\ll 1$}\TBstrut} \\ \hline
$\phm k\phm $ & $\phaa h_k$\phaa\ & $q_{k1}\phaa $ & \phaa $q_{k2} \phaa $ \\
\hline
$\phm 1 \phm $ & $H$ & $\,\,\,c_{\beta-\alpha}$ & $-\varepsilon\, s_{\beta-\alpha}$\Tstrut \\
$2$ & $\varepsilon\, h$ & $\varepsilon\, s_{\beta-\alpha}$ & $c_{\beta-\alpha}$ \\
$3$ & $\varepsilon\, A$ & $0$ & $i$ \\ \hline
\end{tabular}
\end{subtable}
\caption{\small Invariant combinations $q_{kj}$ defined in Table~\ref{tabinv} in the CP-conserving limit, corresponding to a real Higgs basis where $\varepsilon\equiv e^{i\eta}$ is given by \eq{twosigns}.
If the lighter of the two CP-even neutral scalars, $h$, is SM-like, then it is convenient
to identify the $h_k$ as shown in Table~\ref{cptabinv}(a).  
If the heavier of the two CP-even neutral scalars, $H$, is SM-like, then it is convenient
to identify the $h_k$ as shown in Table~\ref{cptabinv}(b).  
  \label{cptabinv}\\[-20pt]}
\end{table}

If we examine the Higgs couplings to vector bosons and the Higgs boson self-couplings in the CP-conserving limit, we find that the following interactions that were originally present are now 
absent:\footnote{Due to Bose symmetry, the couplings $Zh_i h_i$ are absent in \eq{VHH}.  Hence, in the CP-conserving limit, this means that the couplings $Zhh$, $ZHH$ and $ZAA$ are absent.}
\beqa
&& W^+ W^- A\,,\,ZZA\,,\,ZHh\,,\,hhA\,,\,hHA\,,\,HHA\,,\,H^+ H^- A\,,\label{hhhv}\\
&& hhhA\,,\,hhHA\,,\,hHHA\,,\,HHHA\,,\,hAAA\,,\,HAAA\,,\,hAH^+ H^-\,,\,HAH^+ H^-\label{hhhhv}\,.
\eeqa
The cubic scalar couplings that are nonvanishing in the CP conserving limit are exhibited in Tables~\ref{tri1} and \ref{tri2}.

\begin{table}[t!]
\centering
\begingroup
\setlength{\tabcolsep}{10pt} 
\renewcommand{\arraystretch}{1.5} 
\begin{tabular}{|c|c|c|}
\hline
\pht  vertex  \pht  & self-coupling  & in terms of masses \\
\hline
$G h A$ & $v(Z_5 \cbma+Z_6\sbma)$  &$\phm \bigl[(m_h^2-m_A^2)/v\bigl] \cbma$ \\
$G HA$ & $v(-Z_5 \sbma+Z_6\cbma)$ & $-\bigl[ (m_H^2-m_A^2)/v\bigl] \sbma$  \\
$GG h$ & $v(Z_1\sbma+Z_6\cbma)$  & $(m_h^2/v)\sbma$\\
$GG H$ & $v(Z_1\cbma-Z_6\sbma)$ & $(m_H^2/v)\cbma$\\
$G^+G^- h$ &   $v(Z_1\sbma+Z_6\cbma)$  & $(m_h^2/v)\sbma$ \\
$G^+ G^- H$ &  $v(Z_1\cbma-Z_6\sbma)$  & $(m_H^2/v)\cbma$\\
$G^\pm H^\mp h$ &  $v\bigl[\half (Z_4+Z_5)\cbma+Z_6\sbma\bigr]$ & $-\bigl[(m_{H^\pm}^2-m_h^2)/v\bigl]\cbma$ \\
$G^\pm H^\mp H$ & $v\bigl[-\half (Z_4+Z_5)\sbma+Z_6\cbma\bigr]$  & $\phm \bigl[(m_{H^\pm}^2-m_H^2)/v\bigl]\sbma$ \\
$G^\pm H^\mp A$ &  $\mp\half iv(Z_5-Z_4)$ & $\mp i(m_{H^\pm}^2-m_A^2)/v$\\
\hline
\end{tabular}
\endgroup
\caption{\small Nonvanishing cubic scalar self-couplings of the CP-conserving 2HDM involving the neutral and charged Goldstone fields.  
Charged fields point into the vertex. The interactions that do not appear
in this table are zero after invoking the CP symmetry. See caption to Table~\ref{threescalar}.\label{tri1} }
\end{table}
\begin{table}[ht!]
\centering
\begingroup
\setlength{\tabcolsep}{10pt} 
\renewcommand{\arraystretch}{1.5} 
\begin{tabular}{|c|c|}
\hline
\pht  vertex  \pht  & self-coupling  \\
\hline
$hAA$ & ${v}\bigl[(Z_{34}-Z_5)\sbma+Z_7\cbma\bigr]$ \\
$HAA$ & ${v}\bigl[(Z_{34}-Z_5)\cbma-Z_7\sbma\bigr]$  \\
$hHH$ & $ {3 v}\bigl[Z_1\sbma\cbmaii
   +Z_{345}\sbma\left(\tfrac{1}{3}-\cbmaii\right)+Z_6\cbma(1-3\sbmaii)
    +Z_7\sbmaii\cbma\bigr]$\\
$Hhh$ & $ {3 v}\bigl[Z_1\cbma\sbmaii
   +Z_{345}\cbma\left(\tfrac{1}{3}-\sbmaii\right)-Z_6\sbma(1-3\cbmaii)
    -Z_7\cbmaii\sbma\bigr]$\\
$hhh$ &   $ {3v}\bigl[Z_1\sbmaiii+Z_{345}\sbma\cbmaii+3Z_6\cbma\sbmaii
    +Z_7 c^3_{\beta-\alpha}\bigr]$ \\
$HHH$ &  $ {3 v}\bigl[Z_1\cbmaiii+Z_{345}\cbma\sbmaii-3Z_6\sbma\cbmaii
    -Z_7 s^3_{\beta-\alpha}\bigr]$\\
$hH^+ H^-$ &  $ {v}\bigl(Z_3\sbma+Z_7\cbma\bigr)$ \\
$HH^+ H^-$ & $  {v}\bigl(Z_3\cbma-Z_7\sbma\bigr)$ \\
\hline
\end{tabular}
\endgroup
\caption{\small Nonvanishing cubic scalar of the CP-conserving 2HDM involving the physical scalar fields.  We denote $Z_{34}\equiv Z_3+Z_4$
and $Z_{345}\equiv Z_{34}+Z_5$.
The interactions that do not appear
in this table are zero after invoking the CP symmetry.   See the caption to Table~\ref{threescalar}. 
 \label{tri2}}
\end{table}

For completeness, we note that under the assumption that the heavier of the two neutral CP-even Higgs bosons is SM-like, it is more convenient to replace the identifications previously made in
\eq{eq:epsilon6} as follows,
\beq\label{eq2:epsilon6}
H=h_1\,,\qquad h=\varepsilon h_2\,,\qquad A=\varepsilon h_3\,.
\eeq
In this case, the identifications given in \eq{anglelimit} are replaced by
\beq \label{anglelimit2}
c_{12}=\cbma\,,\qquad\quad s_{12}=\varepsilon\,\sbma\,,
\eeq
where $0\leq\cbma\leq 1$, and the $q_{kj}$ of Table~\ref{tabinv} simplify to the results given in Table~\ref{cptabinv}(b).
Note that the cubic scalar couplings exhibited in Tables~\ref{tri1} and \ref{tri2} remain unchanged.

\subsection{2HDM Yukawa couplings}
\label{app:Yukawa}

Given the most general Yukawa Lagrangian involving the scalar fields of the 2HDM and the interaction eigenstate quark fields, one can derive expressions for the
$3\times 3$ complex up-type and down-type quark mass matrices by setting the neutral Higgs fields to their vacuum expectation values.   
Each of the two quark mass matrices can then be diagonalized via singular value decomposition, which yields a pair of unitary matrices that are then employed in defining the left-handed and right-handed quark mass-eigenstate fields, respectively.

After determining the quark mass eigenstate fields and the Higgs mass eigenstate fields, the
resulting 2HDM Yukawa couplings in their most general form are (cf.~eq.~(58) of Ref.~\cite{Boto:2020wyf}),
\beqa
 && \hspace{-0.5in} -\mathscr{L}_Y = \frac{1}{v}\overline D
\biggl\{ q_{k1} M_D +\frac{v}{\sqrt{2}}
\left[q_{k2}\,\rho^{D^\dagger} P_R+
q^*_{k2}\,\rho^D P_L\right]\biggr\}Dh_k \nonumber \\
&&  +\frac{1}{v}\overline U \biggl\{ q_{k1}M_U +\frac{v}{\sqrt{2}}\left[
q^*_{k2}\,\rho^U P_R+
q_{k2}\,\rho^{U\dagger} P_L\right]\biggr\}U h_k
\nonumber \\
&&+\biggl\{\overline U\left[K\rho^{D\dagger}
P_R-\rho^{U\dagger} KP_L\right] DH^+  +\frac{\sqrt{2}}{v}\,\overline
U\left[K\mdd P_R-\mud KP_L\right] DG^+ +{\rm
h.c.}\biggr\},\phantom{x} \label{Yukawas}
\eeqa
where there is an implicit sum over the index $k=1,2,3$; $P_{R,L}\equiv \half(1\pm\gamma\ls{5})$;
$K$ is the CKM matrix; the mass-eigenstate down-type and up-type quark fields are 
$D=(d,s,b)^{\T}$ and $U\equiv (u,c,t)^{\T}$, respectively;
and $M_U$ and $M_D$ are the diagonal quark mass matrices,
\beq \label{MQ}
M_U=\frac{v}{\sqrt{2}}\kappa^U={\rm diag}(m_u\,,\,m_c\,,\,m_t)\,,\qquad
M_D=\frac{v}{\sqrt{2}}\kappa^{D\,\dagger}={\rm
diag}(m_d\,,\,m_s\,,\,m_b) \,.
\eeq
The matrices $\rho^U$ and $\rho^D$ are independent complex
$3\times 3$ matrices that are invariant with respect to scalar basis transformations.  

It is convenient to rewrite the Higgs-quark Yukawa couplings in terms of
the following three $3\times 3$ hermitian matrices, 
\beq \label{rhoRI}
\rho^F_R \equiv \frac{v}{2\sqrt{2}}\,M^{-1/2}_F
(\rho^F +
\rho^{F\,^\dagger})M^{-1/2}_F\,,
\qquad\quad
\rho^F_I \equiv \frac{v}{2\sqrt{2}\,i}M^{-1/2}_F
(\rho^F -
\rho^{F\,\dagger})M^{-1/2}_F\,,
\eeq
for $F=U,D$,
where the $M_F$ are the diagonal 
fermion mass matrices [cf.~\eq{MQ}] and the Yukawa
coupling matrices are introduced in \eq{Yukawas}.
Then, the Yukawa couplings take the form:
\beqa
\!\!\!\!\!\!\!\!\!\!
-\mathscr{L}_Y &=& \frac{1}{v}\,\overline U \sum_{k=1}^3 M_U^{1/2}\biggl\{q_{k1}\mathds{1}
+ \Re(q_{k2})\bigl[\rho^U_R+i\gamma\ls{5}\rho^U_I\bigr]+\Im(q_{k2})\bigl[\rho^U_I-i\gamma\ls{5}\rho^U_R\bigr]\biggr\} M_U^{1/2}Uh_k \nonumber \\
&&  +\frac{1}{v}\,\overline D\sum_{k=1}^3 M_D^{1/2}
\biggl\{q_{k1}\mathds{1}  +\Re(q_{k2})\bigl[\rho^D_R-i\gamma\ls{5}\rho^D_I\bigr]+\Im(q_{k2})\bigl[\rho^D_I+i\gamma\ls{5}\rho^D_R\bigr]\biggr\} M_D^{1/2}Dh_k  \nonumber \\
&& +\frac{\sqrt{2}}{v}\biggl\{\overline{U}\bigl[KM_D^{1/2}(\rho^D_R-i\rho^D_I)
M_D^{1/2}P_R-M_U^{1/2}(\rho^U_R-i\rho^U_I) M_U^{1/2}KP_L\bigr] DH^+ +{\rm
h.c.}\biggr\},\nonumber \\
&& \phantom{line} \label{YUK2}
\eeqa
where $\mathds{1}$ is the $3\times 3$ identity matrix.
If the off-diagonal elements of $\rho^F_{R,I}$ are unsuppressed, then tree-level Higgs-mediated flavor changing neutral currents (FCNCs) will be generated
that are incompatible with the strong suppression of FCNCs observed in nature.\footnote{\Eq{YUK2} is easily extended to include the Higgs boson couplings to leptons.   Since neutrinos are massless in the two-Higgs doublet extension of the Standard Model, one simply
replaces $D\to E=(e,\mu,\tau)^{\T}$ and $U\to N=(\nu_e,\nu_\mu,\nu_\tau)^{\T}$, with $M_E={\rm diag}(m_e,m_\mu,m_\tau)$ and $M_N=0$  in \eqs{Yukawas}{YUK2}.} 

The flavor-aligned 2HDM (often denoted by A2HDM) posits that the Yukawa matrices $\kappa^F$ and $\rho^F$ [cf.~\eq{Yukawas}] are proportional at the electroweak 
scale~\cite{Pich:2009sp}.\footnote{Generically,
the flavor-aligned conditions imposed by the A2HDM are not stable under renormalization group running~\cite{Braeuninger:2010td,Gori:2017qwg}, except in special cases where the flavor-aligned Yukawa couplings are a consequence of  a symmetry~\cite{Ferreira:2010xe}.   Indeed, any such special case can be identified as one of the four Type-I, II, X and Y Higgs-fermion Yukawa couplings~\cite{Hall:1981bc,Barger:1989fj,Aoki:2009ha},
whose corresponding symmetries are exhibited in Table~\ref{Tab:type}.}   
In light of \eq{MQ},
$\kappa^F=\sqrt{2}M_F/v$ is diagonal.  Thus in the A2HDM, the $\rho^F$ are likewise diagonal,
which implies that tree-level Higgs-mediated FCNCs are absent.
We define the \textit{alignment parameters} $a^F$ via,
\beq \label{aligned}
\rho^F=a^F \kappa^F\,,\qquad\quad \text{for $F=U,D, E$},
\eeq
where the (potentially) complex numbers $a^F$ are invariant under the rephasing of the Higgs basis field $\mathcal{H}_2\to e^{i\chi}\mathcal{H}_2$. It follows from \eq{rhoRI} that
\beq
\rho_R^F=(\Re a^F)\mathds{1}\,,\qquad\quad \rho_I^F=(\Im a^F)\mathds{1}\,.
\eeq
Inserting the above results into \eq{Yukawas}, the Yukawa couplings take the following form:
\beqa
-\mathscr{L}_Y &=& \frac{1}{v}\,\overline U M_U \sum_{k=1}^3\bigl(q_{k1}
+ q_{k2}^* a^U P_R+q_{k2} a^{U*}P_L\bigr) Uh_k   \nonumber \\
&& +\frac{1}{v}\sum_{F=D,E}\left\{\overline F M_F \sum_{k=1}^3 \bigl(q_{k1}
+ q_{k2} a^{F*} P_R+q^*_{k2}a^{F}P_L\bigr) F h_k\right\}
 \nonumber \\
 && +\frac{\sqrt{2}}{v}\biggl\{\overline U\bigl[a^{D*}KM_DP_R-a^{U*}M_U KP_L\bigr] DH^+  +a^{E*}\overline NM_E P_R EH^+
+{\rm h.c.}\biggr\}. \phantom{xxxxx}\label{YUK4}
\eeqa
This form simplifies further if the neutral Higgs mass eigenstates are also states of definite CP.  In this case, the corresponding Yukawa couplings are given by
\beqa
\!\!\!\!\!\!\!\!\!\!
-\mathscr{L}_Y &=& \frac{1}{v}\sum_{F=U,D,E} \overline F  M_F\biggl\{\sbma
+\varepsilon\,\cbma\bigl[\Re a^F+i\eta^F\Im a^F \gamma\ls{5}\bigr]\biggr\} Fh
 \nonumber \\
 &&
+\frac{1}{v}\sum_{F=U,D,E} \overline F  M_F\biggl\{\cbma
-\varepsilon\,\sbma\bigl[\Re a^F+i\eta^F\Im a^F \gamma\ls{5}\bigr]\biggr\} FH
 \nonumber \\
  &&
+\frac{1}{v}\sum_{F=U,D,E} \overline F  M_F\biggl\{
\varepsilon\,\bigl[\Im a^F-i\eta^F\Re a^F \gamma\ls{5}\bigr]\biggr\} FA
 \nonumber \\
&& +\frac{\sqrt{2}}{v}\,\varepsilon\,\biggl\{\overline U\bigl[a^{D*}KM_DP_R-a^{U*}M_U KP_L\bigr] DH^+  + a^{E*}\overline NM_E P_R EH^+
+{\rm
h.c.}\biggr\}, \label{YUK5}
\eeqa
where $\varepsilon$ is defined in \eq{twosigns} and we have introduced the notation,
\beq
\eta^F\equiv \begin{cases} +1 & \quad \text{for $F=U$}\,,\\
-1 & \quad \text{for $F=D,E$}\,.\end{cases}
\eeq

\begin{table}[t!]
\centering
{\addtolength\tabcolsep{10pt}
\begin{tabular}{|c||c|c|c|c|c|c|}
\hline & $\Phi_1$ & $\Phi_2$ & $U_R$ & $D_R$ & $E_R$ &
 $U_L$, $D_L$, $N_L$, $E_L$\TBstrut \\  \hline
Type I  & $+$ & $-$ & $-$ & $-$ & $-$ & $+$ \\
Type II & $+$ & $-$ & $-$ & $+$ & $+$ & $+$ \\
Type X   & $+$ & $-$ & $-$ & $-$ & $+$ & $+$ \\
Type Y  & $+$ & $-$ & $-$ & $+$ & $-$ & $+$ \\
\hline
\end{tabular}}
\caption{\small Four possible $\mathbb{Z}_2$ charge assignments for scalar and fermion fields.
 The $\mathbb{Z}_2$ symmetry is employed to constrain the Higgs-fermion Yukawa couplings, thereby implementing the conditions
 for the natural absence of tree-level Higgs-mediated FCNCs.
\label{Tab:type}}
\end{table}

Special cases of the A2HDM arise if the flavor alignment is the consequence of a symmetry.  For Yukawa couplings of
Type-I, II, X and Y~\cite{Hall:1981bc,Barger:1989fj,Aoki:2009ha,Ferreira:2010xe}, one imposes a $\mathbb{Z}_2$ symmetry on the dimension-4 terms of the Higgs Lagrangian in the $\Phi$-basis, where the
$\mathbb{Z}_2$ charges are exhibited in Table~\ref{Tab:type}.  
In particular, the Type-I and Type-II 2HDMs (and likewise the Type-X and \mbox{Type-Y} 2HDMs\footnote{In Type-X models, the quarks possess Type-I Yukawa couplings whereas the leptons possess Type-II Yukawa couplings.  
In Type-Y models, the quarks possess Type-II Yukawa couplings whereas the leptons possess Type-I Yukawa couplings.}) 
are special cases of the A2HDM, where one can identify the corresponding complex alignment parameters as follows,
\begin{enumerate}
\item 
Type-I: $a^U=a^D=a^E=e^{i(\xi+\eta)}\cot\beta$.
\item
Type-II: $a^U=e^{i(\xi+\eta)}\cot\beta$ and $a^D=a^E=-e^{i(\xi+\eta)}\tan\beta$.
\end{enumerate}

In the CP-conserving limit, it is conventional to define the scalar potential in the $\Phi$-basis such that $\lambda_6=\lambda_7=0$ and $\xi=0$.  In this convention, the vacuum expectation values are real and $\tan\beta$ is non-negative, in which case we can identify $e^{i(\xi+\eta)}=\varepsilon$.   
Inserting the Type-I or Type-II values of the flavor alignment parameters in \eq{YUK5},
we see that the factors of $\varepsilon\,$ cancel exactly, as they must since there is no remaining twofold ambiguity in defining the real Higgs basis once a convention has been adopted such that $\tan\beta$ is non-negative.

\subsection{Higgs alignment limit}
\label{app:alignment}

In the Higgs alignment limit one of the neutral scalars, which we conventionally choose to be $h_1$,
is identified as the observed SM-like Higgs boson.  Consequently,
\beq
\frac{g_{h_1VV}}{g_{h_\mathrm{SM}VV}}=q_{11}=c_{12} c_{13}\simeq 1\,,\qquad \text{where $V=W$ or $Z$}\,,
\eeq
which implies that $|s_{12}|$, $|s_{13}|\ll 1$.  Thus, \eqs{rez6}{imz6} yield,
\beqa
|s_{12}|&\simeq &
\left|\frac{v^2\Re\zvi}{m_2^2-m_1^2}\right|\ll 1\,, \\
|s_{13}|&\simeq&
\left|\frac{v^2\Im\zvi}{m_3^2-m_1^2}\right|\ll 1\,.
\eeqa
In light of \eq{imz5}, one additional small quantity characterizes the Higgs alignment limit,
\beq \label{imzeefive}
|\Im(\zv)|\simeq \left|\frac{2(m_2^2-m_1^2) s_{12}s_{13}}{v^2}\right|
\simeq \left|\frac{v^2\Im\zvi\llsup{\,2}}{m_3^2-m_1^2}\right|\ll 1\,.
\eeq
Hence, the conditions for approximate Higgs alignment are:
\begin{enumerate}
\item
Higgs alignment via decoupling is achieved if $m_2$, $m_3\gg m_1\simeq 125$~GeV (under the assumption
that $Z_6$ is at most an $\mathcal{O}(1)$ parameter).  That is, $Y_2\gg v^2$.
\item
Approximate Higgs alignment without decoupling is achieved if $|Z_6|\ll 1$, while all Higgs squared masses are of 
$\mathcal{O}(v^2)$.\footnote{More precisely, we require that $|Z_6|\ll\Delta m_{j1}^2/v^2$, where $\Delta m^2_{j1}\equiv m_j^2-m_1^2$ for $j=2,3$.}
\end{enumerate}
\noindent
We also obtain the following approximate mass relations,
\beqa
m_1^2 &\simeq & v^2\bigl[Z_1-s_{12}\Re\zvi+s_{13}\Im\zvi\bigr]\,, \\
m_2^2-m_3^2 & \simeq&  v^2\bigl[\Re\zv+s_{12}\Re\zvi+s_{13}\Im\zvi\bigr]\,,  \\
m_2^2-m_{H^\pm}^2 &\simeq &\half v^2\bigl[Z_4+\Re\zv+2s_{12}\Re\zvi\bigr]\,,
\eeqa
to first order in the deviation from exact Higgs alignment.

In the case of exact Higgs alignment, the SM-like Higgs boson resides entirely in the Higgs basis field $\mathcal{H}_1$, i.e., it is aligned in field space with the direction of
the neutral Higgs vacuum expectation value.   Thus, we can identify $h_1=\sqrt{2}\,\Re\mathcal{H}_1^0-v$.
In this case, $h_1$ does not mix with the neutral scalar fields that reside in $\mathcal{H}_2$.   In light of \eq{matrix33}, it then follows that $Z_6=0$.   However, a second condition arises
in the diagonalization of the neutral scalar squared-mass matrix.
We will demonstrate below that 
the conditions for exact Higgs alignment, where $h_1$ is identified with the SM Higgs boson, are given by
\beq \label{exact}
\Im\zv=0\quad \text{and}\quad Z_6=0\,.
\eeq
 In particular, in the exact Higgs alignment limit where \eq{exact} holds, $\mathcal{M}^2$ is a diagonal matrix, and we can immediately identify
\beq
m_{1}^2=Z_1 v^2\,,\qquad  m^2_{2,3}= Y_2+\half v^2\bigl[Z_{34}\pm\Re\zv\bigr]\,.
\eeq

To understand the origin of the condition that $\Im\zv=0$, note that $c_{12}=c_{13}=1$ in the exact Higgs alignment limit, or equivalently $R_{12}=R_{13}=\mathds{1}_{3\times 3}$.  Hence,  
\beq \label{rmrt23}
R\mathcal{M}_0^2 R^{\T}=R_{23}\mathcal{M}_0^2 R_{23}^{\T}= {\rm diag}~(m_1^2\,,\,m_2^2\,,\,m_3^2)\,,
\eeq
where
\beq
\mathcal{M}_0^2=\left( \begin{array}{ccc}
Z_1 v^2 &\quad 0 &\quad 0\\
0  &\quad Y_2+\half v^2\bigl[Z_{34}+\Re\zv\bigr] & \quad
- \half v^2\Im\zv\\ 0 &\quad - \half v^2 \Im\zv &\quad
Y_2+\half v^2 \bigl[Z_{34}-\Re\zv\bigr]\end{array}\right) 
\eeq
is the neutral scalar squared mass matrix in the exact Higgs alignment limit.
But having rephased $\mathcal{H}_2$ to set $\theta_{23}=0$ as discussed below \eq{Hbasismassbasis},  it follows that $R_{23}=\mathds{1}_{3\times 3}$.
Hence, \eq{rmrt23} implies that $\mathcal{M}_0^2$ is a diagonal matrix, and we conclude that $\Im\zv=0$.

In the exact Higgs alignment limit where $Y_3=Z_6=0$, the only potentially complex parameters of the scalar potential in the Higgs basis are $Z_5$ and $Z_7$.   
With the rephasing freedom exhibited in \eq{rephasing2}, we can assume that $Z_5$ is real without loss of generality.   Thus, the only remaining potentially complex parameter in the scalar potential is $Z_7$.  Since the neutral Higgs squared-mass matrix is independent of $Z_7$, it follows that Higgs boson--gauge boson interactions are separately CP-conserving in the exact Higgs alignment limit.   The scalar mass eigenstates can be identified as eigenstates of the diagonal squared-mass matrix $\mathcal{M}^2$, with squared masses
\beq
m_h^2=Z_1 v^2\,,\qquad\quad m_{H,A}^2= Y_2+\half(Z_3+Z_4\pm Z_5)\,.
\eeq
That is, the neutral scalar mass eigenstates are states of definite CP.
The only potential  source of CP violation resides in the Higgs self-interactions in the case of $\Im(Z_5^*Z_7^2)\neq 0$.

In this paper, we have proposed to make use of the $H^+ H^- h_k$ interactions,
\beq \label{hkhphm}
h_k H^+ H^-: \quad   v\bigl[q_{k1}Z_3+\Re(q_{k2}\zvii)\bigr]\,,
\eeq
where only CP-even states can couple to $H^+ H^-$ if CP is conserved.
In the exact Higgs alignment limit, $q_{11}=1$, $q_{21}=q_{31}=q_{12}=0$, $q_{22}=1$ and $q_{32}=i$.   Thus $h_1$ is CP-even as anticipated.   Assuming that $Z_7\neq 0$, the $H^+ H^- h_2$ coupling is nonvanishing if 
$\Im \zvii=0$, whereas the $H^+ H^- h_3$ coupling is nonvanishing if 
$\Re \zvii=0$.   
Thus, we again conclude (as previously noted in Table~\ref{cpalign}) that if $Z_7\neq 0$ then
\beqa
&&\Im \zvii=0 \quad \Longrightarrow \quad \text{$h_2$ is CP-even and $h_3$ is CP-odd},\nonumber \\
&&\Re \zvii=0 \quad \Longrightarrow \quad \text{$h_2$ is CP-odd and $h_3$ is CP-even}.\label{evenodd}
\eeqa
For more details, see Appendix C of Ref.~\cite{Haber:2010bw}.

\begin{table}[t!]
\centering
\begingroup
\setlength{\tabcolsep}{10pt} 
\renewcommand{\arraystretch}{1.5} 
\begin{tabular}{|c|c|}
\hline
\pht  vertex  \pht  & self-coupling \\
\hline
$H^+ H^- h_1$ &   $v\bigl[Z_3-s_{12}\Re\zvii+s_{13}\Im\zvii\bigr]$\\
$H^+ H^- h_2$ &   $\phm v\bigl[\Re\zvii+s_{12}Z_3\bigr]$\\
$H^+ H^- h_3$ &   $-v\bigl[\Im\zvii-s_{13}Z_3\bigr]$\\
$h_1 h_1 h_1$ & $3vZ_1$ \\
$h_2 h_2 h_2$ & $\phm 3v\bigl[\Re\zvii+s_{12}(Z_{34}+\Re\zv)\bigr]$ \\
$h_3 h_3 h_3$ & $-3v\bigl[\Im \zvii-s_{13}(Z_{34}-\Re\zv)\bigr]$ \\
$h_1 h_2 h_2$ & $v\bigl[Z_{34}+\Re\zv-3s_{12}\Re\zvii+s_{13}\Im\zvii\bigr]$ \\
$h_1 h_3 h_3$ &  $v\bigl[Z_{34}-\Re\zv-s_{12}\Re\zvii+3s_{13}\Im\zvii\bigr]$ \\
$h_2 h_1 h_1$ & $ v\bigl[s_{12}(3Z_1-2Z_{34}-2\Re\zv)+3\Re\zvi\bigr]$ \\
$h_2 h_3 h_3$ & $v\bigl[\Re\zvii+s_{12}(Z_{34}-\Re\zv)\bigr]$ \\
$h_3 h_ 1h_1$ & $v\bigl[s_{13}(3Z_1-2 Z_{34}+2\Re\zv)-3\Im\zvi\bigr]$ \\
$h_3 h_2 h_2$ & $-v\bigl[\Im \zvii-s_{13}(Z_{34}+\Re\zv)\bigr]$ \\
$h_1 h_2 h_3$ & $-v\bigl[s_{13}\Re \zvii-s_{12}\Im\zvii\bigr]$ \\
\hline
\end{tabular}
\endgroup
\caption{\small Cubic self-couplings of the physical Higgs scalars of the 2HDM in the approximate Higgs alignment limit without decoupling (where $|Z_6|\ll 1$).
Charged fields point into the vertex.  The first order corrections to the exact Higgs alignment limit, which are linear in $s_{12}$, $s_{13}$ and $\protect\zvi$, are exhibited.
\label{trialign}}
\end{table}

In light of the precision LHC Higgs data, the departure from the exact Higgs alignment limit is expected to be small.   Thus, it is useful to exhibit the Higgs couplings in the approximate Higgs alignment limit, where only terms that are first order in the small parameters that govern the Higgs alignment limit are kept, as shown in Table~\ref{trialign}.  Here, one must distinguish between the Higgs alignment limit without decoupling, where $|Z_6|\ll 1$ and the decoupling limit where $|s_{12}|$, $|s_{13}|$, $|\Im\zv|\ll 1$ by virtue of the fact that $m_{2}$, $m_{3}\gg v$.   In this paper, our phenomenological considerations are based on the assumption that the masses of $h_2$ and $h_3$ are not significantly larger than $\mathcal{O}(v)$.   As a result, we assume that $\zvi$ is sufficiently suppressed to be consistent with the observed SM-like Higgs boson $h_1$.   In particular, the cubic self-couplings of the physical Higgs scalars of the 2HDM are given in Table~\ref{trialign} in the approximate Higgs alignment limit without decoupling, where $|Z_6|\ll 1$.  The first order corrections to the exact Higgs alignment results shown in Table~\ref{trialign} are linear in $s_{12}$, $s_{13}$ and $\zvi$.   Note that in the approximate Higgs alignment limit without decoupling, $\Im\zv\sim\mathcal{O}(|Z_6|^2)$ is a second order effect and hence can be neglected [cf.~\eq{imzeefive}].

Finally, we record the structure of the neutral Higgs-quark Yukawa couplings of the A2HDM in the exact Higgs alignment limit.   Inserting the values for the $q_{kj}$ given below \eq{hkhphm} into \eq{YUK4}, we end up with
\beqa
-\mathscr{L}&=& \frac{1}{v}\bigl(\overline{U}M_U U +\overline{D} M_D D+\overline{E}M_E E\bigr)h_1 \nonumber \\
&+& \frac{1}{v}\left[\overline{U}M_U\bigl(\Re a^U +i\gamma\ls{5}\Im a^U\bigr)U+\sum_{F=D,E}\overline{F}M_F\bigl(\Re a^F -i\gamma\ls{5}\Im a^F\bigr)F\right]h_2 \nonumber \\
&+& \frac{1}{v}\left[\overline{U}M_U\bigl(\Im a^U -i\gamma\ls{5}\Re a^U\bigr)U+\sum_{F=D,E}\overline{F}M_F\bigl(\Im a^F +i\gamma\ls{5}\Re a^F\bigr)F\right]h_3\,. \phantom{xxxxxx} \label{A2HDMalign}
\eeqa
As expected, in the limit of exact Higgs alignment the Yukawa couplings of $h_1$ coincide with those of the SM Higgs boson.

\section{Processes with an odd number of photons}
\label{3gamma}
\renewcommand{\theequation}{B.\arabic{equation}}
\setcounter{equation}{0}

Consider 2HDM processes involving bosonic external states 
in the limit of exact Higgs alignment
(where the Yukawa interactions, which can contribute via fermionic loops, are neglected).  
If the scalar potential is CP conserving, then $\Re\zvii\Im\zvii=0$ [cf.~\eq{AlignedCPcondition2}]
and the bosonic sector separately conserves C and P, with assigned quantum numbers as indicated in Table~\ref{candp}.   For definiteness, we henceforth assume that 
$\Im\zvii=0$, in which case we can identify $h_2=H$ and $h_3=A$.\footnote{If one were to assume that $\Re\zvii=0$ then simply interchange $h_2$ and $h_3$ in the discussion above (in light of Table~\ref{cpalign}).} 

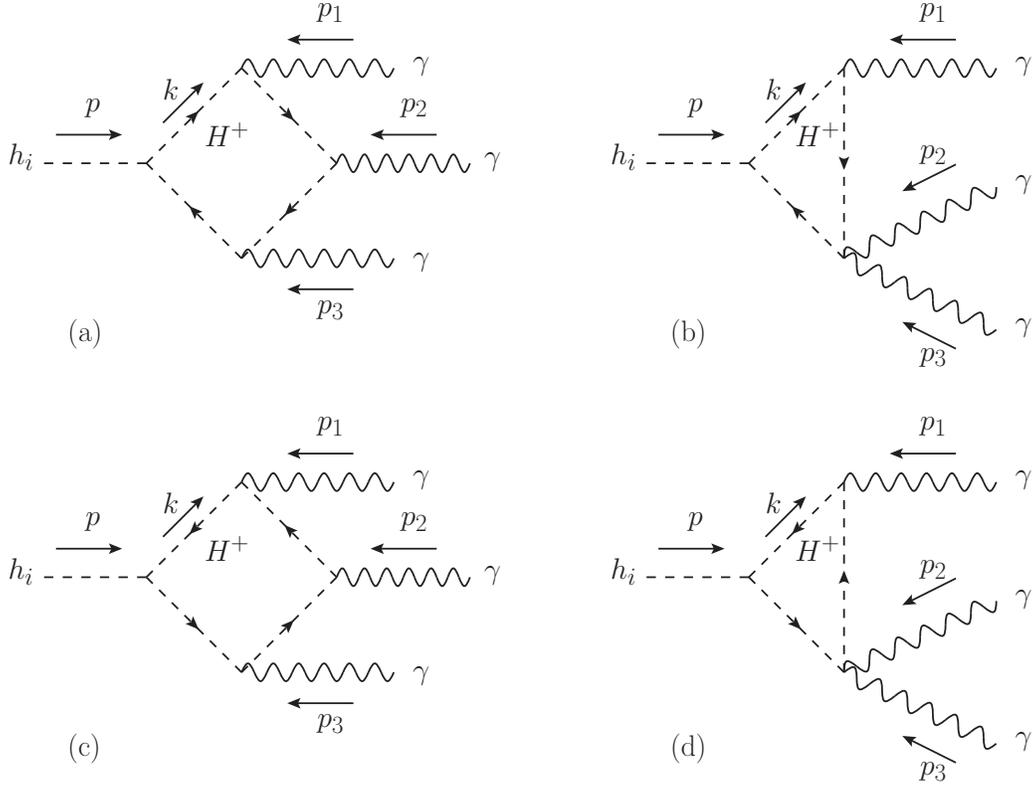
\begin{figure}[t]
\centering
\scalebox{0.45}{
  \begin{picture}(320,0)(320,0)
    \SetWidth{1.5}
        \SetArrowScale{1.5}
    \DashLine(170,-64)(256,-64){8}
    \DashArrowLine(256,-64)(336,16){8}
    \DashArrowLine(336,16)(416,-64){8}
    \DashArrowLine(416,-64)(336,-144){8}
    \DashArrowLine(336,-144)(256,-64){8}
    \Photon(336,16)(464,16){7.5}{6}
    \Photon(416,-64)(528,-64){7.5}{6}
    \Photon(336,-144)(464,-144){7.5}{6}
    \ArrowLine[arrowpos=1](180,-40)(230,-40)
    \ArrowLine[arrowpos=1](270,-30)(300,0)
    \ArrowLine[arrowpos=1](430,40)(380,40)    
    \ArrowLine[arrowpos=1](500,-40)(450,-40)        
    \ArrowLine[arrowpos=1](430,-170)(380,-170)    
    \Text(140,-70)[lb]{\huge{$h_i$}}
    \Text(305,-50)[lb]{\huge{$H^+$}}    
    \Text(270,-10)[lb]{\huge{$k$}}
    \Text(205,-25)[lb]{\huge{$p$}}
    \Text(400,55)[lb]{\huge{$p_1$}}
    \Text(470,-25)[lb]{\huge{$p_2$}}    
    \Text(400,-195)[lb]{\huge{$p_3$}}
   \Text(480,-155)[lb]{\huge{$\gamma$}}    
    \Text(540,-70)[lb]{\huge{$\gamma$}}    
    \Text(480,10)[lb]{\huge{$\gamma$}}

    \Text(190,-220)[lb]{\Huge{(a)}}           
  \end{picture}
}
\scalebox{0.45}{
 \begin{picture}(150,0)(150,0) 
    \SetWidth{1.5}
        \SetArrowScale{1.5}
  \DashLine(170,-64)(256,-64){8}
    \DashArrowLine(256,-64)(336,16){8}
    \DashArrowLine(336,16)(336,-144){8}
    \DashArrowLine(336,-144)(256,-64){8}
    \Photon(336,16)(464,16){7.5}{6}
    \Photon(336,-144)(464,-84){7.5}{6}
    \Photon(336,-144)(464,-204){7.5}{6}
    \ArrowLine[arrowpos=1](180,-40)(230,-40)
  \ArrowLine[arrowpos=1](270,-30)(300,0)
 \ArrowLine[arrowpos=1](430,40)(380,40)    
 \ArrowLine[arrowpos=1](430,-65)(390,-85)        
 \ArrowLine[arrowpos=1](430,-220)(390,-200)    
    \Text(140,-70)[lb]{\huge{$h_i$}}
    \Text(295,-50)[lb]{\huge{$H^+$}}     
    \Text(270,-10)[lb]{\huge{$k$}}
    \Text(205,-25)[lb]{\huge{$p$}}
    \Text(400,55)[lb]{\huge{$p_1$}}
    \Text(400,-65)[lb]{\huge{$p_2$}}    
    \Text(400,-235)[lb]{\huge{$p_3$}}
    \Text(480,-210)[lb]{\huge{$\gamma$}}    
    \Text(480,-90)[lb]{\huge{$\gamma$}}    
    \Text(480,10)[lb]{\huge{$\gamma$}}    
    \Text(190,-220)[lb]{\Huge{(b)}}                       
  \end{picture}
}
\vspace{5cm}

\scalebox{0.45}{
  \begin{picture}(320,0)(320,0)
    \SetWidth{1.5}
    \SetArrowScale{1.5}
    \DashLine(170,-64)(256,-64){8}
    \DashArrowLine(336,16)(256,-64){8}
    \DashArrowLine(416,-64)(336,16){8}
    \DashArrowLine(336,-144)(416,-64){8}
    \DashArrowLine(256,-64)(336,-144){8}
    \Photon(336,16)(464,16){7.5}{6}
    \Photon(416,-64)(528,-64){7.5}{6}
    \Photon(336,-144)(464,-144){7.5}{6}
    \ArrowLine[arrowpos=1](180,-40)(230,-40)
    \ArrowLine[arrowpos=1](270,-30)(300,0)
    \ArrowLine[arrowpos=1](430,40)(380,40)    
    \ArrowLine[arrowpos=1](500,-40)(450,-40)        
    \ArrowLine[arrowpos=1](430,-170)(380,-170)    
    \Text(140,-70)[lb]{\huge{$h_i$}}
    \Text(305,-50)[lb]{\huge{$H^+$}}    
    \Text(270,-10)[lb]{\huge{$k$}}
    \Text(205,-25)[lb]{\huge{$p$}}
    \Text(400,55)[lb]{\huge{$p_1$}}
    \Text(470,-25)[lb]{\huge{$p_2$}}    
    \Text(400,-195)[lb]{\huge{$p_3$}}
    \Text(480,-155)[lb]{\huge{$\gamma$}}    
    \Text(540,-70)[lb]{\huge{$\gamma$}}    
    \Text(480,10)[lb]{\huge{$\gamma$}}
    \Text(190,-220)[lb]{\Huge{(c)}}        
  \end{picture}
}
\scalebox{0.45}{
  \begin{picture}(150,0)(150,0) 
    \SetWidth{1.5}
    \SetArrowScale{1.5}
    \DashLine(170,-64)(256,-64){8}	
    \DashArrowLine(336,16)(256,-64){8}
    \DashArrowLine(336,-144)(336,16){8}
    \DashArrowLine(256,-64)(336,-144){8}
    \Photon(336,16)(464,16){7.5}{6}
    \Photon(336,-144)(464,-84){7.5}{6}
    \Photon(336,-144)(464,-204){7.5}{6}
    \ArrowLine[arrowpos=1](180,-40)(230,-40)
    \ArrowLine[arrowpos=1](270,-30)(300,0)
    \ArrowLine[arrowpos=1](430,40)(380,40)    
    \ArrowLine[arrowpos=1](430,-65)(390,-85)        
    \ArrowLine[arrowpos=1](430,-220)(390,-200)    
    \Text(140,-70)[lb]{\huge{$h_i$}}
    \Text(295,-50)[lb]{\huge{$H^+$}}     
    \Text(270,-10)[lb]{\huge{$k$}}
    \Text(205,-25)[lb]{\huge{$p$}}
    \Text(400,55)[lb]{\huge{$p_1$}}
    \Text(400,-65)[lb]{\huge{$p_2$}}    
    \Text(400,-235)[lb]{\huge{$p_3$}}
    \Text(480,-210)[lb]{\huge{$\gamma$}}    
    \Text(480,-90)[lb]{\huge{$\gamma$}}    
    \Text(480,10)[lb]{\huge{$\gamma$}}
    \Text(190,-220)[lb]{\Huge{(d)}}                       
  \end{picture}
}
\vspace{4cm}
\caption{\small Representative box diagrams [graphs (a) and (c)] and triangle diagrams [graphs (b) and (d)] with an internal charged Higgs boson, $H^\pm$, for the $h_i \to \gamma\gamma\gamma$ ($i=2,3$) process. The arrows on the charged Higgs propagator denote the flow of electric charge.}
\label{fig:hitogagaga}
\end{figure}

Under the conditions elucidated above,
we expect the process $h_2 \to \gamma\gamma\gamma$ to be absent, since the initial state ($h_2$) is C even and the final state ($\gamma\gamma\gamma$) is C odd. In contrast, the process $h_3 \to \gamma\gamma\gamma$ is allowed, because $h_3$ is C odd. At the one-loop level the Feynman diagrams for both processes have identical topologies: (a) the box topology with an internal charged Higgs boson and three $H^+H^-\gamma$ vertices; and (b) the triangle topology with an internal charged Higgs boson that contains a $H^+H^-\gamma$ and a $H^+H^-\gamma\gamma$ vertex. Representative diagrams for these topologies are depicted in Fig.~\ref{fig:hitogagaga} (there are also four additional box diagrams corresponding to the other four possible permutations of the external photon lines). The only difference between the one-loop decay matrix elements for the processes $h_2 \to \gamma\gamma\gamma$ and $h_3 \to \gamma\gamma\gamma$ at the one-loop level is the appearance of the different coupling factors of the $h_2 H^+H^-$ and $h_3 H^+H^-$ vertices, respectively.   
In particular, these coupling factors are just numbers (independent of momenta).   Hence, it follows that if the one-loop matrix element for $h_2 \to \gamma\gamma\gamma$ vanishes due to the C and P-invariance of the 2HDM bosonic sector, then the one-loop matrix element for $h_3 \to \gamma\gamma\gamma$ must vanish as well (despite the fact that $h_3 \to \gamma\gamma\gamma$ is allowed by C and P-invariance).   

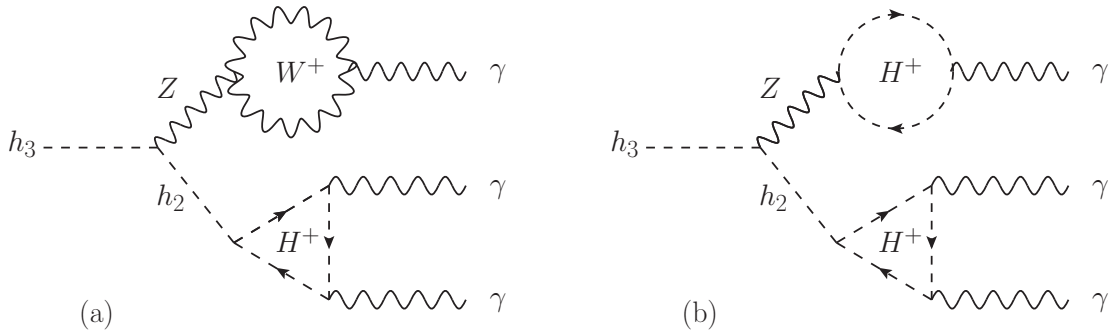
\begin{figure}[b!]
\centering
\scalebox{0.45}{
  \begin{picture}(320,0)(350,150)
    \SetWidth{1.5}
        \SetArrowScale{1.5}
     \PhotonArc(368,127)(48,0,360){7.5}{15}
    \Photon(416,127)(515,127){7.5}{5}
         \DashArrowLine(320,-17)(400,31){8}
     \DashArrowLine(320,-17)(400,31){8}
    \DashArrowLine(400,31)(400,-65){8}
    \DashArrowLine(400,-65)(320,-17){8}
      \DashLine(320,-17)(256,63){8}
    \DashLine(160,63)(256,63){8}
    \Photon(320,127)(256,63){7.5}{5}
    \Photon(400,31)(515,31){7.5}{5}
    \Photon(400,-65)(515,-65){7.5}{5}
    \Text(130,55)[lb]{\huge{$h_3$}}
    \Text(355,-25)[lb]{\huge{$H^+$}}    
    \Text(255,105)[lb]{\huge{$Z$}}    
    \Text(255,10)[lb]{\huge{$h_2$}}        
    \Text(355,118)[lb]{\huge{$W^+$}}
    \Text(535,118)[lb]{\huge{$\gamma$}}    
    \Text(535,20)[lb]{\huge{$\gamma$}}   
    \Text(535,-75)[lb]{\huge{$\gamma$}}
    \Text(190,-90)[lb]{\Huge{(a)}}
  \end{picture}
 }
\scalebox{0.45}{
  \begin{picture}(180,0)(180,150)
    \SetWidth{1.5}
    \SetArrowScale{1.5}
  \DashArrowArcn(370,127)(48,360,180){7}
  \DashArrowArcn(370,127)(48,180,0){7}
    \Photon(416,127)(515,127){7.5}{5}
    \Photon(320,127)(256,63){7.5}{5}
     \DashArrowLine(320,-17)(400,31){8}
    \DashArrowLine(400,31)(400,-65){8}
    \DashArrowLine(400,-65)(320,-17){8}
    \Photon(320,127)(256,63){7.5}{5}
    \DashLine(320,-17)(256,63){8}
    \DashLine(160,63)(256,63){8}
    \Photon(400,31)(515,31){7.5}{5}
    \Photon(400,-65)(515,-65){7.5}{5}
    \Text(130,55)[lb]{\huge{$h_3$}}
    \Text(355,-25)[lb]{\huge{$H^+$}}    
    \Text(255,105)[lb]{\huge{$Z$}}    
    \Text(255,10)[lb]{\huge{$h_2$}}        
    \Text(355,118)[lb]{\huge{$H^+$}}
    \Text(535,118)[lb]{\huge{$\gamma$}}    
    \Text(535,20)[lb]{\huge{$\gamma$}}   
    \Text(535,-75)[lb]{\huge{$\gamma$}}
    \Text(190,-90)[lb]{\Huge{(b)}}
  \end{picture}
  }
\vspace{4.2cm}
\caption{\small Sample diagrams for two-loop contributions to the decay process $h_3\to \gamma\gamma\gamma$.}
\label{fig:h3togagaga_2L}
\end{figure}

Indeed, it is easy to see that the sum of the one-loop diagrams that contribute to $h_i\to\gamma\gamma\gamma$ 
vanishes. In Fig.~\ref{fig:hitogagaga}(a), each $H^+H^-\gamma$ vertex gives rise to a factor $-ie q_j^\mu$, where $j=1,2,3$ enumerates the momenta of the external photons, and the momenta $q_j^\mu$ are
\beqa
q_1^\mu &=& 2 k  + p_1, \nonumber\\
q_2^\mu &=& 2 k  + 2 p_1 + p_2, \nonumber \\
q_3^\mu &=& 2 k  + 2 p_1 + 2 p_2 + p_3.
\eeqa
 In contrast, due to the reversed flow of electric charge in the loop, each vertex in Fig.~\ref{fig:hitogagaga}(c) gives a contribution of $+ie q_j^\mu$, whereas everything else (assignment of momenta, etc.)  is identical to Fig.~\ref{fig:hitogagaga}(a). Thus, with an odd number of $H^+H^-\gamma$ vertices, these two contributions cancel. 
The same can be observed between the diagrams in Fig.~\ref{fig:hitogagaga}(b) and (d), where the $H^+H^-\gamma$ vertex gives $-ie p_j^\mu$ in diagram (b) and $+ie p_j^\mu$ in diagram (d), and the $H^+H^-\gamma\gamma$ vertex in both cases simply gives a factor of $2ie^2$.  

If we now allow for CP-violating interactions in the scalar potential, then $\Re\zvii\Im\zvii\neq 0$ [cf.~\eq{AlignedCPcondition2}], in which case both the $h_2 H^+ H^-$ and $h_3 H^+ H^-$ vertices exist.
Nevertheless the sum of the one-loop diagrams that contribute to $h_i\to\gamma\gamma\gamma$ still vanishes since the cancellation of diagrams
is not affected by the value of the $h_i H^+ H^-$ coupling.
 
The sum of the one-loop contributions in which the internal $H^\pm$ is replaced by the $W^\pm$ must also vanish if C and P are separately conserved, since the $W^+ W^-$ system in a total angular momentum zero state has ${\rm C}={\rm P}=+1$.
Hence, $h_i\to W^+ W^-\to\gamma\gamma\gamma$ (via the one-loop diagrams with topologies exhibited in Fig.~\ref{fig:hitogagaga}) for any scalar $h_i$ is forbidden by the C invariance of the 2HDM bosonic sector.   If $h_2$ and $h_3$ are CP-mixed states, then the vanishing of the corresponding one-loop matrix element is again unchanged.  Once again, this conclusion can be verified diagrammatically by explicitly exhibiting the cancellation among diagrams by repeating the analysis used above for the $H^\pm$ loop.
 
Although the bosonic contributions to the one-loop matrix element  for the C-conserving, P-conserving $h_3\to\gamma\gamma\gamma$ decay vanish, the cancellation does not persist at two loops. 
Two possible diagrams that occur at the two-loop level and contribute to $h_3\to\gamma\gamma\gamma$ are depicted in Fig.~\ref{fig:h3togagaga_2L}. 
This decay can be described by an effective Lagrangian~\cite{Berends:1965ftl,Dolgov2,Basham:1977rj},
\beq
\mathscr{L}_{\rm eff}=\frac{\kappa_3}{\Lambda^7}h_3(\partial^\beta{F}_{\sigma\tau})(\partial_\rho F_{\alpha\beta})(\partial^\rho\partial^\alpha F_{\sigma\tau})\,,
\eeq
where $F_{\mu\nu}$ is the electromagnetic field strength tensor, $\kappa_3$ is dimensionless and $\Lambda$ is a parameter with dimensions of mass.   The corresponding decay amplitude in terms of form factors is given in Ref.~\cite{Dolgov1}.
In contrast, the analogous diagrams for the C-violating, P-conserving $h_2 \to \gamma\gamma\gamma$ decay (obtained by interchanging $h_2$ and $h_3$ in Fig.~\ref{fig:h3togagaga_2L}) require a nonvanishing $h_3 H^+ H^-$ coupling, i.e.~$\Im\zvii\neq 0$. Hence, the matrix element for the process $h_2\to \gamma\gamma\gamma$ is proportional to a bosonic P-even, CP-violating coupling, as expected.
\clearpage

Finally, consider how the above results are modified when the Yukawa couplings are taken into account.
At one-loop order, new diagrams must be considered with the topology of  Figs.~\ref{fig:hitogagaga}(a) and (c), where the $H^+$ is replaced by a fermion.   Thus, to analyze these contributions, we can consider a subset of the 2HDM fields that contains the neutral scalars $h_{1,2,3}$, the fermions and the photon (but with the $W^\pm$, $Z$ and $H^\pm$ removed).  In this truncated theory, if the scalar potential is CP conserving, then C and P are separately conserved.  However, the $J^{\rm PC}$ quantum numbers of the scalars are $h_1(0^{++})$, $h_2(0^{++})$ and $h_3(0^{-+})$.   Note that the
$J^{\rm PC}$ assignment for $h_3$ is different than the $0^{+-}$ assignment that appears in Table~\ref{candp}.   Moreover, if the scalar potential is CP-violating, then the truncated theory remains C conserving since $\bar{\psi}\psi$ and $\bar{\psi}\gamma\ls{5}\psi$ are both ${\rm C}=+1$ bilinears.  It immediately follows that the one-loop amplitude for $h_i\to\gamma\gamma\gamma$ vanishes due to C invariance (otherwise known as Furry's theorem).  

At two-loop order, one can no longer assign a unique C quantum number to $h_3$.
That is, in the 2HDM coupled to the complete electroweak sector,
if the scalar potential is CP-conserving then $h_2$ is CP even and $h_3$ is 
CP odd, but C and P are no longer separately conserved.\footnote{Ultimately, one expects to find the effects of CP violation to leak into the scalar sector due to the CKM phase in the Yukawa interactions.  However, such effects are highly suppressed and will not enter until at least four-loop order~\cite{Fontes:2021znm}.}
For example, if we consider the two loop diagrams of  Fig.~\ref{fig:h3togagaga_2L} where the $H^+$ loop in the triangle subgraph is replaced by a fermion, and the corresponding graph where $h_2$ and $h_3$ are interchanged, 
then the contributions to the two-loop amplitudes for $h_{2,3}\to\gamma\gamma\gamma$ are nonzero.  In other words, there exist both CP-even and CP-odd effective operators involving three photon field strength tensors that can couple to an external scalar field~\cite{Dolgov2,Basham:1977rj}.    The CP-conserving effective Lagrangian that governs the $h_{2,3}\to\gamma\gamma\gamma$ decays is
\beq
\mathscr{L}_{\rm eff}=\frac{1}{\Lambda^7}\bigl[\kappa_2 h_2 \partial^\beta\widetilde{F}_{\sigma\tau}+\kappa_3 h_3\partial^\beta F^{\sigma\tau}\bigr](\partial_\rho F_{\alpha\beta})(\partial^\rho\partial^\alpha F_{\sigma\tau})\,,
\eeq
where $\widetilde{F}_{\sigma\tau}\equiv\half\epsilon_{\mu\nu\sigma\tau}F^{\mu\nu}$ is the dual electromagnetic field strength tensor and $\kappa_{2,3}$ are dimensionless parameters.
In particular, if the scalar sector is CP conserving, then both decays $h_{2,3}\to\gamma\gamma\gamma$ are allowed.  The term in the effective Lagrangian involving
the dual electromagnetic field strength tensor arises due to the $\gamma\ls{5}$ that appears in the $h_3\bar{\psi}\gamma\ls{5}\psi$ coupling.  This is in contrast
to the behavior of the bosonic sector in isolation, where $h_2\to\gamma\gamma\gamma$ can be interpreted as a signal of a P-even, CP-violating interaction, since in this case, there is no way to generate $\kappa_2\neq 0$ via diagrams that involve only bosonic fields.

\end{appendices}


\end{document}